%% file: npeaks1.tex
\documentclass[10pt,aps,nofootinbib,
amsmath,amssymb,twocolumn,
preprintnumbers, 
showpacs,
raggedbottom, 
superscriptaddress,
prd,
floatfix]{revtex4-2}

\usepackage{cases}
\usepackage{pgf,tikz}
\usepackage{graphicx}
\usepackage{booktabs}

\usepackage{slashed} 

\usepackage{hyperref}
\usepackage[nameinlink]{cleveref}
\crefdefaultlabelformat{#2\textup{#1}#3}
\creflabelformat{equation}{#2\textup{#1}#3}

\newcommand{\ud}{\mathrm{d}}

\newcommand{\eps}{\epsilon}

\newcommand{\diff}[2]{\frac{\ud #1}{\ud #2}}

\newcommand{\lrp}[1]{\left(#1\right)} 
\newcommand{\lrb}[1]{\left[#1\right]} 
\newcommand{\lrc}[1]{\left\{#1\right\}} 

\newcommand{\MeV}{\text{MeV}}
\newcommand{\GeV}{\text{GeV}}

\newcommand{\fmic}{\text{fm}^{-3}}

\newcommand{\tov}{\text{TOV}}
\newcommand{\pqcd}{\text{pQCD}}
\newcommand{\cqm}{\mathrm{CQM}}

\newcommand{\ceft}{{\chi\text{EFT}}}

\newcommand{\eden}{\mathcal{E}}

\newcommand{\lmin}{l_{\mathrm{min}}}
\newcommand{\lmax}{l_{\mathrm{max}}}

\newcommand{\cmin}{C_{s,\mathrm{min}}}
\newcommand{\cmax}{C_{s, \mathrm{max}}}
\newcommand{\cmean}{C_{s, \mathrm{mean}}}

\newcommand{\dpmin}{\Delta P_\mathrm{min}}
\newcommand{\dpmax}{\Delta P_\mathrm{max}}
\newcommand{\dpmean}{\Delta P_\mathrm{mean}}

\newcommand{\cmaxns}{C_{s, \mathrm{max}}^\mathrm{NS}}

\newcommand{\cminud}{C_{s,\mathrm{min}}^\mathrm{UD}}
\newcommand{\cmaxud}{C_{s, \mathrm{max}}^\mathrm{UD}}
\newcommand{\cmeanud}{C_{s, \mathrm{mean}}^\mathrm{UD}}

\newcommand{\cmaxl}{\cmaxns}

\newcommand{\cming}{\cminud}
\newcommand{\cmaxg}{\cmaxud}
\newcommand{\cmeang}{\cmeanud}

\newcommand{\maxcmingg}{{\mathrm{max}\lrc{\cmingg}}}

\newcommand{\cmingg}{C_{s,\mathrm{min}}^\mathrm{UD*}} 
\newcommand{\cmaxgg}{C_{s,\mathrm{max}}^\mathrm{UD*}} 

\newcommand{\maxcming}{\mathrm{max}\lrc{\cming}}

\newcommand{\erf}{\mathrm{erf}}

\newcommand{\UCB}{Department of Physics, University of California Berkeley, Berkeley, CA 94720}

\begin{document}

\title{Bayesian evidence for two peaks in the sound speed in cold dense QCD}
\author{Dake Zhou}
\email{dkzhou@berkeley.edu}
\affiliation{Department of Physics, University of Washington, Seattle, WA 98195}
\affiliation{Institute for Nuclear Theory, University of Washington, Seattle, WA 98195}
\affiliation{\UCB}

\date{\today}

\begin{abstract}

I show that in addition to the well-known peak inside massive neutron stars, the sound speed in cold dense QCD matter likely exhibits 
another peak above neutron star densities before 
it asymptotes to $c_s=\sqrt{C_s}=\sqrt{1/3}$.
Based on the 
framework reported in \href{https://arxiv.org/abs/2408.16738}{arxiv:2408.16738}, this approach does not rely on any assumption about
the ultra-dense matter
not realized in nature. 
Current multimessenger observation of neutron stars 
favors the two-peak scenario with a Bayes factor $5.1_{-0.7}^{+0.9}$, 
where the uncertainties are systematics due to models of neutron star inner cores.
This evidence grows to $27_{-8}^{+11}$
if the $2.6M_\odot$ component of GW190814 is a neutron star.
The trough in $C_s$ separating the two peaks is inferred to be 
below $0.05^{+0.04}_{-0.02}$ (at the $50\%$ level) if $2.6M_\odot$ neutron stars exist.
The second peak 
above $1/3$ beyond baryon chemical potential $\mu_B=1.6-1.9$ GeV
most likely signals non-perturbative effects in cold quark matter, for instance color superconductivity.
\end{abstract}

\maketitle

\section{Introduction}

The QCD phase diagram at low temperatures and high baryon density remains elusive because the theoretical and experimental tools needed remain rudimentary. Lattice QCD, the only reliable method to calculate non-perturbative phenomena in QCD, is inapplicable as it suffers from the well-known Fermion sign problem    
~\cite{Troyer:2004ge,deForcrand:2009zkb,Kaplan:2009yg} and heavy-ion collisions cannot simultaneously access low temperatures and high densities. Neutron stars (NSs) provide a unique laboratory to study high-density matter at low-temperature and observation of NS mass, radius, and tidal deformability have provided useful constraints on the equation of state (EOS) at baryon number densities $n_B\simeq2-10n_0$ where $n_0=0.16~\fmic$~(e.g.,\cite{LattimerPrakash:2010,Watts:2016uzu,Abbott:2017aa,Abbott:2018exr,Tews:2018aa,De:2018uhw,Radice:2017lry,Capano:2019eae,Miller:2019cac,Legred:2021hdx,Riley:2019yda,Miller:2021qha,Riley:2021pdl,Salmi:2024aum,Dittmann:2024mbo}).   
Above NS densities
reliable descriptions in the range $n_B\simeq10-30n_0$ are lacking as the strongly interacting cold quark matter (QCM) renders perturbative calculations untractable.
Non-perturbative phenomenon in particular color superconductivity~\cite{Alford:1997zt,Berges:1998rc,Carter:1998ji,Pisarski:1999bf,Son:1998uk,Alford:1998mk,Son:1999cm,Casalbuoni:1999wu,Hong:1999ei,Schafer:1999jg,Rajagopal:2000wf} is 
believed to be active 
in this regime, although estimates of its strength vary~\cite{Alford:2007xm,Braun:2021uua,Braun:2022jme,Geissel:2024nmx,Abbott:2024vhj,Fujimoto:2024pcd}.

An observable of interest is the speed of sound squared $C_s=\ud P/\ud\eden$, where $P$ and $\eden$ are the pressure and energy density respectively.
It influences the hydrostatic equilibrium of NSs~\cite{Tolman:1939jz,Oppenheimer:1939ne} and encodes information about the underlying degrees of freedom and their interactions~\cite{Lattimer:2000nx,McLerran:2007qj,Gandolfi:2011xu,McLerran:2018hbz}.
Earlier work has shown that the existence of two-solar-mass pulsars~\cite{Demorest:2010bx,Antoniadis:2013pzd,Romani:2021xmb,NANOGrav:2019jur,Fonseca:2021wxt} suggests a peak in $C_s$ whose magnitude 
$\cmax>1/3$~\cite{Bedaque:2014sqa,Tews:2018kmu,Drischler:2020fvz,Drischler:2021bup}.
Since the asymptotic value $C_s=1/3$ is approached from below as predicted by perturbative QCD (pQCD)~\cite{Freedman:1976xs,Freedman:1976ub,Vuorinen:2003fs,Kurkela:2009gj,Gorda:2018gpy,Fernandez:2021jfr,Gorda:2021kme,Gorda:2021znl,Gorda:2023mkk,Fernandez:2024ilg}, $C_s>1/3$ inside NSs implies at least one peak and one trough in $C_s$ in cold dense QCD.

Here, I show that the simplest picture of one peak followed by a trough may be inadequate and that $C_s$ is likely required to rise again above $1/3$ 
at high densities.
The second peak in $C_s$ is
 the consequence of thermodynamic concordance between the EOS at NS and CQM densities,
 a paradigm traditionally used to constrain NSs~\cite{Annala:2017llu,Komoltsev:2021jzg,Gorda:2022jvk,Somasundaram:2022ztm,Zhou:2023zrm} or CQM in the perturbative regime~\cite{Zhou:2023zrm,Kurkela:2024xfh} and recently exploited to elucidate the intermediate region under investigation~\cite{Zhou:2024hdi}.
Heuristically, physical matching of the EOS demands a specific arrangement and sizable variance of $C_s$ at intermediate densities.
The astrophysical evidence exceeds naive expectations from phase-space volume arguments, which are taken into account, and is generally substantial despite systematic uncertainties associated with assumptions about NS inner cores.

I achieve this in two steps. 
\Cref{sec:framework} introduces a set of criteria on NS EOSs
that guarantees an additional peak in $C_s$ above the central density of the heaviest NS.
These conditions inform both the magnitude and the order of extrema of $C_s$,
and circumvent the need to model the ultra-dense matter not realized in nature.
Next, \cref{sec:bayesian} reports Bayesian analyses that quantify the preference of the two-peak scenario by current multimessenger observations of NSs.
In this first attempt at inferring the strongly interacting high-density CQM, 
I specifically ignore pairing contributions to the EOS 
and report evidence for features that are difficult to explain by perturbative calculations alone, thus lending credence to the presence of non-perturbative effects.

\section{sufficient condition for two peaks}\label{sec:framework}

Across the zero-temperature QCD phase diagram, two peaks in $C_s$ are guaranteed if  in the ultra-dense phase (i) a minimum (possibly local) 
precedes a maximum (possibly local); 
(ii) this minimum is less than the global maximum within NS densities; and (iii) the maximum following the minimum exceeds the asymptotic value $1/3$~\footnote{Strictly speaking, it only needs to exceed $C_s^\cqm\approx1/3$.}:
\begin{align}
\cming &\text{ precedes } \cmaxg, \label{eq:ordering}\\
\cmaxl &> \cming, \label{eq:cmaxns_cminud}\\
\cmaxg &>1/3. \label{eq:cmax_asymp}
\end{align}
Throughout this letter, superscripts ``NS" denote densities below the highest reached inside NSs. 
Dubbed the TOV point, this maximum attainable density is model-dependent.
``UD" refers to the ultra-dense regime between TOV and perturbative densities where pQCD is applicable.

\begin{figure}
	\includegraphics[width=0.8\linewidth]{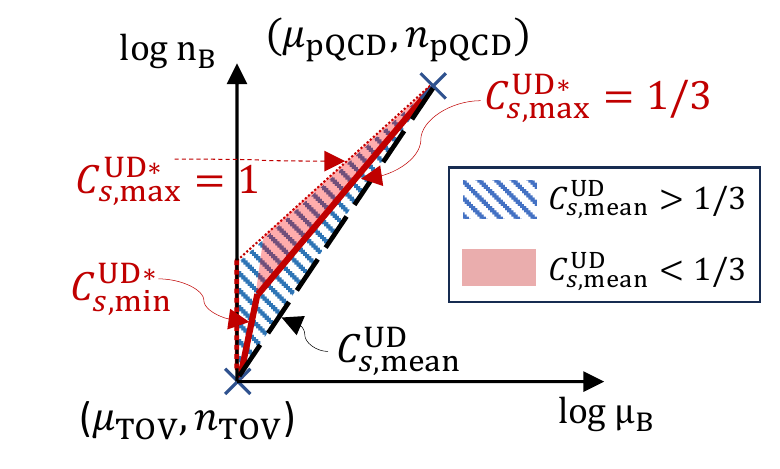}
	\caption{
Visualizing the sufficient conditions \cref{eq:SCA} relevant when $\cmeang>1/3$ 
    and \cref{eq:criterion} obtains when $\cmeang<1/3$. 
    They ensure a peak in $C_s$ above NS densities, and along with \cref{eq:cmaxns} guarantee two peaks in $C_s$.
    The upper boundary is due to causality.
	The dashed black line is the mean value EOS \cref{eq:meanEOS}, and the solid red curve depicts the maximally soft construction underlying \cref{eq:criterion}.
	}\label{fig:demo}
\end{figure}

I now derive a set of stronger conditions that, at the expense of becoming more conservative,  
are entirely expressible in terms of 
the EOS at NS and pQCD densities. 
To begin with, bounds on the magnitude of $\cming$ and $\cmaxg$ are reported in the preceding letter~\cite{Zhou:2024hdi}. A central ingredient is the mean value EOS, the constant sound speed model that directly connects TOV and pQCD points in the baryon number density $n_B$ versus baryon chemical potential $\mu_B$ plane.
It is shown as the dashed black line in \cref{fig:demo}.
Since slopes of $n_B(\mu_B)$ relations are related to the sound speed
\begin{equation}\label{eq:Cs}
\frac{\ud\log n_B}{\ud \log\mu_B}=C_s^{-1},
\end{equation}
the mean value EOS is found to be
\begin{equation}\label{eq:meanEOS}
C_s(\mu_B)=\cmeang,~\cmeang = \frac{\log\lrp{\mu_\pqcd/\mu_\tov}}{\log\lrp{n_\pqcd/n_\tov}},
\end{equation}
where $n_\tov=n_B(\mu_\tov)$ and $n_\pqcd=n_B(\mu_\pqcd)$ are  baryon number densities at TOV and pQCD points.
Ref~\cite{Zhou:2024hdi} showed that $\cmeang$ places stringent limits on the {\em global} minimum and maximum of $C_s$ in the ultra-dense phase denoted by $\cmingg$ and $\cmaxgg$:
\begin{equation}\label{eq:mvb}
\cmingg\leq\cmeang\leq\cmaxgg.
\end{equation}

To simplify the discussion, 
in the main text I assume the extrema are unique, i.e. $\cmingg=\cming$ and $\cmaxgg=\cmaxg$. 
A proof of the central results relaxing this assumption is presented in ~\cref{sec:proof}.
The mean value bound \cref{eq:mvb} can be used to rewrite \cref{eq:cmaxns_cminud} as 
\begin{equation} \label{eq:cmaxns}
\cmaxl > \cmeang.
\end{equation}
This is a stronger requirement for NS EOSs since $\cmeang\geq\cming$.

Formulating \cref{eq:ordering} 
is viable by 
considering the increment in pressure from $\mu_\tov$ to $\mu_\pqcd$ associated with the mean value EOS,  
$\Delta P_\mathrm{mean}\equiv\lrb{P(\mu_\pqcd)-P(\mu_\tov)}_\mathrm{mean EOS}$. 
It follows from the thermodynamic relation
\begin{equation}\label{eq:dp}
\Delta P\equiv P_\pqcd-P_\tov
=\int^{\mu_\pqcd}_{\mu_\tov} n_B(\mu_B)\ud \mu_B
\end{equation}
where for \cref{eq:meanEOS} 
$n_B(\mu_B)=n_\tov \lrp{\mu_B/\mu_\tov}^{{\cmeang}^{-1}}$
is obtained by integrating \cref{eq:Cs}. Denote $\gamma=1/\cmeang$, one finds
\begin{equation}\label{eq:dpmean}
	\Delta P_\mathrm{mean}=\frac{n_\pqcd \mu_\pqcd}{\gamma+1}\lrb{1-\lrp{\frac{\mu_\tov}{\mu_\pqcd}}^{\gamma+1}}.
\end{equation} 
A sufficient condition for \cref{eq:ordering} is 
\begin{equation}\label{eq:SCA}
\Delta P> \Delta P_\mathrm{mean}. 
\end{equation} 
It can be understood in terms of the geometric interpretation of \cref{eq:dp}, namely $\Delta P$ as the area in \cref{fig:demo}.
Continuations of NS EOSs satisfying \cref{eq:SCA} must have segments above the mean value EOS. 
If the sound speed of such high-density models were to decrease monotonically, i.e., the slope $1/C_s$ in \cref{fig:demo} were to monotonically increase, \cref{eq:SCA} would be violated because convex $\log n_B(\log\mu_B)$ functions cannot surpass the mean value EOS in \cref{fig:demo}.
By contradiction, \cref{eq:SCA} 
necessitates a minimum preceding a maximum.

\begin{table*}[!htbp]
\centering
\setlength{\tabcolsep}{1em} 
{\renewcommand{\arraystretch}{1.5}
\begin{tabular}{ l|c c c c c|c c c @{} }
\toprule
astro data	& $X=1$ & $1.5$ & $2$ 	& $3$ &  $4$  & $X\sim U[1,4]$ & $U[2,4]$ & $\mathcal{N}(2.5, 0.5)$ \\
\midrule	
$M_\tov\geq2.08^{+0.08}_{-0.08}M_\odot$ & 0 & $2.1^{+0.5}_{-0.3}$ & $4.2_{-0.3}^{+0.5}$ & $25_{-8}^{+5}$     & $>10^3$	 & $2.5_{-0.5}^{+0.6}$ & ${7.7}^{+2.2}_{-2.1}$ &  $4.9_{-0.7}^{+0.8}$ \\
+NICER  							& 0 & $2.2^{+0.5}_{-0.3}$     & $4.4_{-0.3}^{+0.5}$	& $26_{-9}^{+6}$	 & $>10^3$   & $2.6_{-0.5}^{+0.6}$ & ${7.9}^{+2.3}_{-2.1}$ &	$5.1_{-0.7}^{+0.9}$\\
\hline
+$M_\tov\leq2.23^{+0.05}_{-0.05}M_\odot$& 0 & $0.2^{+0.1}_{-0.1}$ & $0.9_{-0.2}^{+0.2}$	& $7.4_{-2.4}^{+1.8}$ & $>10^3$	 & $1.5_{-0.2}^{+0.3}$ & $3.0_{-0.9}^{+1.0}$ &	$1.9_{-0.3}^{+0.3}$\\
+$M_\tov\geq2.58^{+0.09}_{-0.09}M_\odot$ & 0 & $10^{+2}_{-2}$     & $32_{-14}^{+8}$     & $\gg10^3$          & $\gg10^6$ & $3.1_{-0.9}^{+1.7}$ & ${200}^{+60}_{-50}$ & $27_{-8}^{+11}$\\
\bottomrule
\end{tabular}
}
\caption{Bayes factors for the two-peak scenario obtained by successively incorporating the mass and radius measurements of PSR J0740+6620~\cite{NANOGrav:2019jur,Fonseca:2021wxt,Miller:2021qha,Riley:2021pdl,Salmi:2024aum,Dittmann:2024mbo}. From there, either the upper bound on the TOV limit $M_\tov$ from post-merger evolutions of GW170817~\cite{Margalit:2017dij,Shibata:2017xdx,Rezzolla:2017aly,Radice:2018xqa,Shibata:2019ctb} or the lower bound if GW190814~\cite{LIGOScientific:2020zkf} involves an NS is imposed. 
 Astrophysical uncertainties are approximated as Gaussian.
In the last three columns the pQCD renormalization scale $X$ is sampled from either uniform ($U$) or Gaussian ($\mathcal{N}$) priors.
Reported are the average and limiting values across 12 NS inner core models summarized in \cref{sec:eossum}.
}
\label{tab:bfactor}
\end{table*}

For a given NS model, the requirements \cref{eq:SCA,eq:cmax_asymp} can be simplified depending on the value of $\cmeang$.
When $\cmeang>1/3$, the mean value bound implies \cref{eq:cmax_asymp},
 so \cref{eq:SCA,eq:cmaxns} alone suffice in this scenario. 
I prove in \cref{sec:proof} 
that this deduction holds even if $\cming\neq\cmingg$ in the presence of multiple extrema.

The other possibility in which $\cmeang<1/3$ appears more likely because it is supported by the existence of two-solar-mass NSs 
assuming chiral effective field theory ($\ceft$)~\cite{Weinberg:1968de,Weinberg:1990rz,Weinberg:1991um,Weinberg:1992yk,Kaplan:1996xu,Kaplan:1998tg,Kaplan:1998we,Beane:2001bc} is valid up to $n_B=2n_0$~\cite{Zhou:2024hdi}.
Since massive pulsars also suggest $\cmaxl>1/3$~\cite{Bedaque:2014sqa,Tews:2018kmu,Drischler:2020fvz,Drischler:2021bup},
\cref{eq:cmaxns} is strongly favored in this scenario.
The remaining conditions \cref{eq:SCA,eq:cmax_asymp} can be further simplified with the help of the so-called maximally soft EOS~\cite{Rhoades:1974fn,Koranda:1996jm,Lattimer:2000nx,Drischler:2020fvz,Komoltsev:2021jzg,Zhou:2023zrm,Zhou:2024hdi}. It is shown as the red curve in \cref{fig:demo} and yields the largest $\Delta P$ at given $\cmaxgg$ and $\cmingg$.
This upper bound 
is given by~\cite{Zhou:2024hdi}
\begin{multline}
	\Delta P_\mathrm{max}(\cmingg,\cmaxgg)
	=n_\pqcd\mu_\pqcd \\
	\times\lrb{\frac{1}{\alpha+1}\lrp{1-x^{\frac{1+\alpha}{1-\delta} } y^{\frac{1+\alpha}{\beta-\alpha}} }	+\frac{1}{\beta+1}\lrp{x^{\frac{1+\alpha}{1-\delta} } y^{\frac{1+\alpha}{\beta-\alpha}}-\frac{x}{y}} }, \label{eq:dpmax}
\end{multline}
where $x=\frac{\mu_\tov}{\mu_\pqcd}$, $y=\frac{n_\pqcd}{n_\tov}$, $\alpha=1/\cmaxgg$, $\beta=1/\cmingg$, and $\delta=\alpha/\beta$.
EOSs obeying $\Delta P>\dpmax(\cmingg, \cmaxgg=1/3)$ must have a maximum in $C_s$ exceeding $1/3$ in the ultra-dense phase.
Note that $\dpmax>\dpmean$ since the maximally soft model lies above the mean value EOS in \cref{fig:demo}, 
\cref{eq:SCA,eq:cmax_asymp} 
are implied by
\begin{equation}\label{eq:criterion}
\Delta P > \Delta P_\mathrm{max}(\cmingg,\cmaxgg=1/3)
\end{equation}
when $\cmeang<1/3$. 
A proof is presented in \cref{sec:proof}.
Intuitively, \cref{eq:criterion} assures a second peak because EOSs obeying it are soft in the ultra-dense regime so must have segments resembling the maximally soft model characterized by a low trough preceding a high peak.

In summary, 
NS EOSs satisfying all of the following are guaranteed to have one peak in $C_s$ within and another above NS densities:
\begin{equation}\label{eq:summary}
    \begin{aligned}
        &\cmaxl>\cmeang,\\
        &\Delta P >
        \begin{cases}
            \dpmean, &\cmeang>1/3,\\
            \dpmax(\cmaxgg=1/3), &\cmeang<1/3.
        \end{cases}
    \end{aligned}
\end{equation}

\section{Bayesian evidence}\label{sec:bayesian}

To assess the evidence for \cref{eq:summary} I adopt a Bayesian approach where {\em ab-initio} nuclear theories are taken as prior knowledge and astrophysical observations are imposed as constraints.
The outer layers of NSs are constructed following the prescription in refs~\cite{Forbes:2019xaz,Zhou:2023zrm}. 
It centers on  
$\ceft$ calculations of the pure neutron matter up to N3LO ~\cite{Drischler:2017wtt,Drischler:2020yad} along with $2\sigma$ truncation errors 
and is informed by empirical constraints on the symmetric nuclear matter.
NS inner cores where the density exceeds $n_\ceft=2n_0$ are modeled by 
a class of agnostic parameterizations detailed in \cref{sec:parameos}.
As for the CQM EOS in the perturbative regime,
I take N2LO pQCD predictions~\cite{Freedman:1976ub,Freedman:1976xs}  at $\mu_B=2.4$ GeV supplemented by the soft contribution $\sim\mathcal{O}(\alpha_s^3\log^2\alpha_s)$ at N3LO~\cite{Gorda:2021kme,Gorda:2021znl}.

In each analysis, all NS EOSs are selected to be consistent with chosen pQCD calculations, i.e., satisfy~\cite{Komoltsev:2021jzg,Gorda:2022jvk,Somasundaram:2022ztm,Zhou:2023zrm}
\begin{multline}\label{eq:dp_prior}
    \dpmin(\cmingg=0,\cmaxgg=1)\\
    \leq\Delta P\leq \dpmax(\cmingg=0,\cmaxgg=1)
\end{multline} 
in the absence of superconducting  gaps~\cite{Zhou:2023zrm,Kurkela:2024xfh}.
Above, $\dpmin$ is based on the maximally stiff model and its expression is given in \cite{Zhou:2024hdi}.
\Cref{eq:dp_prior} is the result of causality and thermodynamic stability, and
ensures every sample is physical.
To address the phase space volume,
I impose equally probable distributions on the criterion 
$\Delta P-\Delta P_\mathrm{mean,max}$ 
that are flat on each side ($+$ and $-$), yielding neutral priors without preferences.
Other choices of priors are discussed in \cref{sec:assump}.

\Cref{tab:bfactor} shows Bayes evidence for scenarios where two peaks are required.
It is sensitive to the pQCD renormalization scale $X=\bar{\Lambda}/(\mu_B/3)$ and reveals diametric predilections in the extreme cases $X=1,4$.
Larger $X$'s lend stronger support because they predict higher $P_\pqcd$ thus higher $\Delta P$.
Marginalization over this uncertainty is achieved by sampling X from
a uniform distribution over the commonly assumed range $X\in[1,4]$ and more restrictive priors 
motivated by the fact that $P_\pqcd$ near $X=1$ is susceptible to pairing contributions~\cite{Zhou:2023zrm}, by resummed leading and next-to-leading soft contributions~\cite{Fernandez:2021jfr,Fernandez:2024ilg}, and by hints from lattice calculations at finite isospin~\cite{Abbott:2023coj,Abbott:2024vhj,Fujimoto:2024pcd}.
The results generally favor the two-peak scenario,
and are predominantly determined by the NS maximum mass $M_\tov$ where Bayes factors grow rapidly with increasing $M_\tov$.
Constraints on NS radii have very limited effects because current uncertainties are large, and NS radii may not robustly inform the EOS near maximum-mass except in extreme scenarios~\cite{Zhou:2023zrm}.
Gravitational wave constraints on tidal deformability $\Lambda_{1.4M_\odot}\lesssim600-800$~\cite{Abbott:2017aa,Abbott:2018exr,De:2018uhw,Radice:2017lry,Capano:2019eae} are implied by the N3LO $\ceft$ EOS up to $2n_0$~\cite{Drischler:2021bup,Zhou:2023zrm}.
 Results employing $\ceft$ to lower densities are reported in \cref{sec:assump}.

\begin{figure}
	\includegraphics[width=\linewidth]{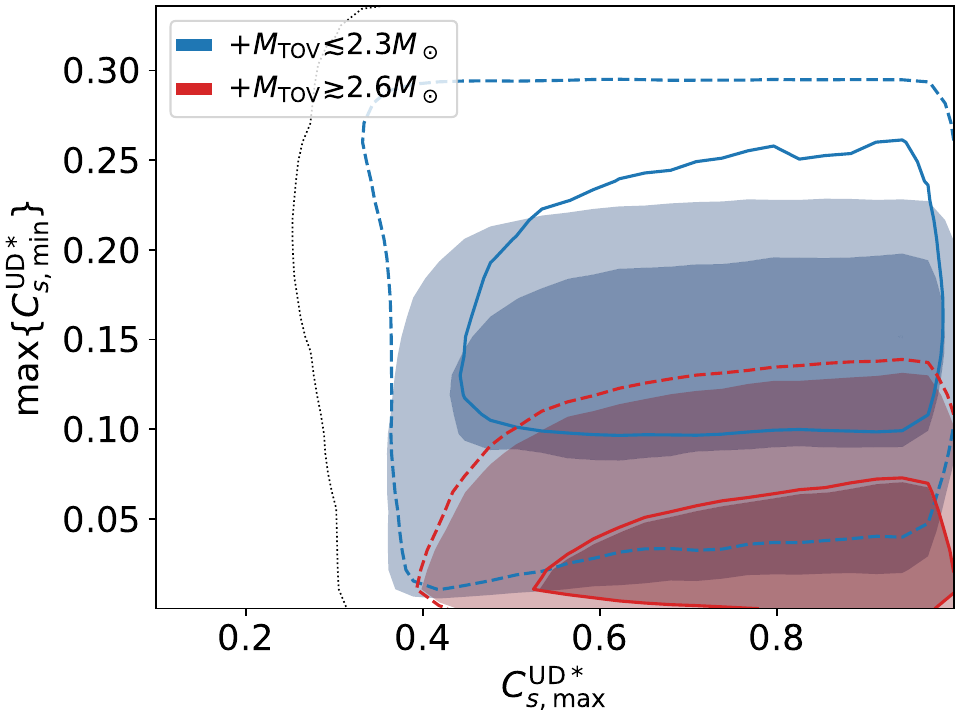}
	\caption{
		The $50\%$ and $90\%$ posterior credible intervals (CIs) on $\cmaxgg$ and $\maxcmingg$. The colored lines show the full posterior whereas the shaded regions depict CIs when two peaks are required. The dotted line shows the $90\%$ prior after imposing a likelihood similar to \cref{eq:dp_prior} but with $\cmaxgg$ randomly sampled. Its absence below $\cmaxgg\simeq0.2$ is the result of thermodynamic consistency implied in 
        \cref{eq:dp_prior}. All $\sim500$ million samples are utilized and $X\sim\mathcal{N}(2.5,0.5)$.
	}\label{fig:CIs}
\end{figure}

This framework also allows one to infer the height of the second peak and the preceding trough.
Here, in addition to the physical requirement \cref{eq:dp_prior}, I impose a similar constraint where $\cmaxgg$ is sampled randomly from a uniform distribution on $[0,1]$.
This injected likelihood excludes NS EOSs demanding $C_s^\mathrm{UD}>\cmaxgg$, so the resulting posterior on $\cmaxgg$ provides an estimate for the upper bound on the maximum sound speed 
\footnote{To be clear, $C_s^\mathrm{UD}$ may surpass this bound but is not required to.}.
An upper limit on $\cmingg$ is then obtained for each NS model by solving $\Delta P = \Delta P_\mathrm{max}(\maxcmingg,\cmaxgg)$
when $\Delta P>\dpmean$ and $\Delta P = \Delta P_\mathrm{min}(\maxcmingg,\cmaxgg)$ otherwise, see ref~\cite{Zhou:2024hdi}.

\Cref{fig:CIs} shows constraints on the magnitude of the peak and the trough.
While posteriors on $\cmaxgg$ only marginally deviate from the prior suggesting limited constraints on this upper bound,
 the minimum sound speed is informed.
For $M_\tov\gtrsim2.6M_\odot$, one finds 
$\maxcmingg=0.05^{+0.08}_{-0.04}$ at the $90\%$ level.
Considering the prior is $0.16^{+0.15}_{-0.13}$, this presents a strong case for first-order phase transitions above NS densities.
This evidence becomes stronger when the second peak is required (shaded).
Furthermore, it is interesting to note that among the subset of samples disfavoring low $\maxcming$, the preference for the two-peak scenario grows because $\dpmax(\cmingg, \cmaxgg=1/3)$  
decreases with increasing $\maxcming$. 
The strengthened evidence in the absence of phase transitions is discussed in \cref{sec:cminres}.
Finally, the onset density~\cite{Zhou:2024hdi} of the second peak is inferred to lie in the range $n_B\in[4.5_{-0.3}^{+0.5}, 24^{+3}_{-6}] n_0$ or $\mu_B\in[1.8^{+0.1}_{-0.2}, 1.9^{+0.2}_{-0.1}]$ GeV when $M_\tov\gtrsim2.6M_\odot$, and $n_B\in[5.8_{-0.9}^{+1.0}, 26^{+4}_{-9}]  n_0$ or $\mu_B\in[1.7^{+0.2}_{-0.2},2.0^{+0.3}_{-0.3}]$ GeV if $M_\tov\lesssim2.2M_\odot$.
Detailed examination of the magnitude and location of extrema of $C_s$ will be reported in a subsequent paper.

\section{Discussion}\label{sec:discussion} 

Using agnostic NS models and physical, neutral priors,  astrophysical data in particular the existence of massive pulsars is shown to favor (at least) two peaks in $C_s$ at supranuclear densities. From a Bayesian perspective, the evidence is at least substantial if the pQCD renormalization scale $X\gtrsim1.5-2$~\cite{Fernandez:2021jfr,Fernandez:2024ilg,Fujimoto:2024pcd}
or if phase transitions are absent~\cite{Schafer:1998ef,Schafer:1999pb,Schafer:1999fe} with $M_\tov\lesssim 2.2M_\odot$, and the evidence is very strong if $2.6M_\odot$ NSs exist.

The second peak above NS densities 
arises because of asymptotic freedom of QCD and indications that EOS in the NS core is stiff
\footnote{Earlier work (e.g. \cite{Gorda:2022jvk,Somasundaram:2022ztm,Fujimoto:2024pcd}) often finds \cref{eq:dp_prior} favoring soft NS cores. 
I clarified model-independently in \cite{Zhou:2023zrm,Zhou:2024hdi} that softening is only required in the ultra-dense phase. 
For a discussion of \cref{eq:dp_prior} in the context of Bayesian analyses see \cref{sec:cminres}. }. 
Since the weakly-coupled CQM is almost conformal~\footnote{One may not expect large sensitivities towards $X$ considering $C_s^\pqcd\simeq1/3$ regardless of its value. But the pressure $P(\mu_B)$ is an integrated quantity (\cref{eq:dp,eq:Cs}) and tiny deviations in $C_s$ originated at asymptotic densities cumulate to a considerable spread down to $\mu_B\lesssim3$ GeV.}
, a featureless and mostly constant $C_s=\cmeang\simeq1/3$ above NS densities is only feasible if the EOS is also conformal at the center of most massive NSs. 
Astrophysical observations suggest rapid stiffening of the EOS above $\sim2-5n_0$~(e.g.,~\cite{Gandolfi:2011xu,Tews:2018aa}),
yielding relatively low pressure at $\mu_B\simeq1.5-1.9$~\GeV.
Taken together with a high $P_\pqcd$, 
a minimum $\cming<1/3$ preceding a maximum $\cmaxg>1/3$
is thus required to bridge a large $\Delta P=P_\cqm-P_\tov$.
A related measure is the trace anomaly $P/\eden-1/3$ whose density evolution informs structures in $C_s$ ~\cite{Fujimoto:2022ohj}.
Note though successful matching of the $P(\eden)$ relation 
is not always 
consistent with thermodynamics as $\mu_B$ could be discontinuous. 

Non-perturbative effects in CQM are strongly indicated 
by the second peak near the weak-coupling regime because perturbative calculations generally reveal $C_s^\cqm<1/3$~\cite{Freedman:1976xs,Freedman:1976ub,Vuorinen:2003fs,Kurkela:2009gj,Gorda:2018gpy,Fernandez:2021jfr,Gorda:2021kme,Gorda:2021znl,Gorda:2023mkk,Fernandez:2021jfr}.
Color superconductivity is perhaps the most anticipated mechanism.
Since pairing contributions to the sound speed is
$$C_s^\Delta 
\approx 4\frac{\Delta^2}{\mu_B^2} -2\lrb{{\Delta'}^2+\frac{\Delta\Delta'}{\mu_B} } +\mathcal{O}(\Delta'^2,(\frac{\Delta}{\mu_B})^2,\Delta^{(n>1)}) 
$$ where $\Delta=\Delta(\mu_B)$ is the superconducting gap, a maximum exceeding $1/3$ on top of $C_s^\pqcd\simeq0.31-0.32$ suggests sizable gaps.
While estimates based on both leading-order pQCD and phenomenological models
agree with the postulated density dependence of $C_s^\cqm=C_s^\pqcd+C_s^\Delta$ since $\Delta\sim\exp(-1/g_s)$ 
falls off faster than any powers of $g_s$ in $C_s^\pqcd=1/3+\mathcal{O}(g_s^2)$ as $\mu_B\rightarrow\infty$,
only the latter predicts sufficiently large gaps $\Delta\gtrsim100~\MeV$ at $\mu_B\simeq1.6-2.4~\GeV$.
Therefore, if superconducting gap is the underlying mechanism, it likely receives sizable instanton enhancements~\cite{Rapp:1997zu,Alford:1997zt} or is underestimated by leading-order pQCD calculations in the not-so-weakly-coupled regime.
Alternatively, the second peak may be attributed to meson condensates~\cite{Kaplan:2001qk,Bedaque:2001je}.
These possibilities will be investigated in detail later.

Although Bayesian analyses (e.g.,~\cite{Legred:2021hdx,Fujimoto:2022ohj} and the present work) generally support
violations of the conformal limit~\cite{Cherman:2009tw,Hohler:2009tv,Bedaque:2014sqa,Hoyos:2016cob,Ecker:2017fyh,Hoyos:2021uff,Jokela:2020piw,Gursoy:2017wzz,Anabalon:2017eri} in NSs, 
$\cmaxl\leq1/3$ is yet to be ruled out by current astrophysical data.
This possibility is nevertheless tightly constrained~\cite{Drischler:2020fvz,Zhou:mi1,Zhou:mi2}, or from a Bayesian perspective highly fine-tuned
hence penalized by Occam's factors implicit in Bayesian inferences.
A fully model-independent assessment of the two-peak evidence eschewing parametrization of the NS EOS and statistical interpretations
will be reported later.

\section*{Acknowledgment}

I am grateful to Sanjay Reddy for carefully reading a draft of the manuscript.
The data analysis is performed on the Lawrencium cluster at the Lawrence Berkeley National Laboratory.
During the conception and completion of this work the author is supported 
by the Institute for Nuclear Theory Grant No. DE-FG02-00ER41132 from the Department of Energy,
and by NSF PFC 2020275 (Network for Neutrinos, Nuclear Astrophysics, and Symmetries (N3AS)).

\appendix
\newpage

\section{Proving \cref{eq:summary}}\label{sec:proof}
\begin{figure}[!htbp]
	\includegraphics[width=0.96\linewidth]{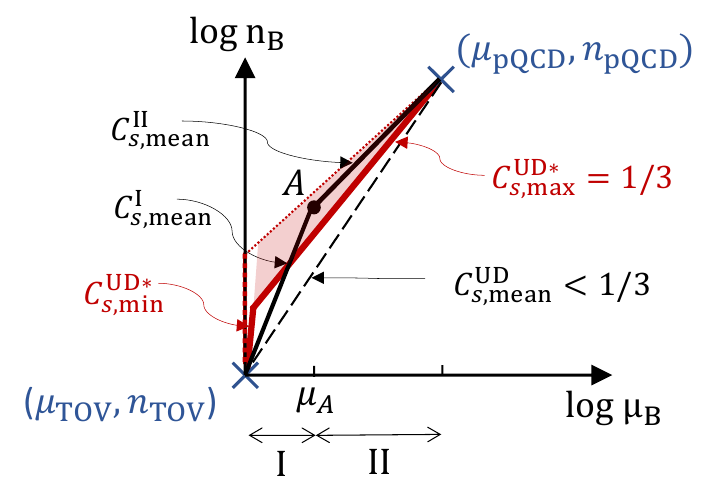}
	\caption{
	NS EOSs satisfying \cref{eq:criterion} (whose continuations in the range shown pass through the shaded region) are required to exhibit a second peak in $C_s$.
    Point A is arbitrary in the shaded region and divides the horizontal axis into region I where $\mu_B\in[\mu_\tov,\mu_A]$ and region II where $\mu_B\in[\mu_A,\mu_\pqcd]$.
    The central statement can be proven by applying the mean value bound \cref{eq:mvb} separately to regions I and II, see \cref{sec:proof}.
 	}\label{fig:proof}
\end{figure}

In this appendix I prove 
\cref{eq:SCA,eq:criterion} guarantee a peak in $C_s$ above NS densities, and that along with \cref{eq:cmaxns} they ensure two peaks at supranuclear densities.
I begin with the case when $\cmeang<1/3$.
The aim is to show \cref{eq:criterion} ensures that the peak bounded by the constant $\cmaxgg=1/3$ segment is preceded by a minimum less than $\cmeang$. 
The right hand side of \cref{eq:criterion} is given by the maximally soft EOS with $\cmaxgg=1/3$ depicted as the solid red curve in \cref{fig:proof}.
If \cref{eq:criterion} holds, the EOS must have segments lie above the solid red curve, i.e., pass through the shaded region.
I then pick an arbitrary point along such an EOS in the shaded region, for instance point A in \cref{fig:proof}, and label the density range preceding it as region I and the one subsequent to it as region II.
The mean value EOSs between point A and the TOV point and between point A and the pQCD point are shown as solid black lines in \cref{fig:proof}, and their associated mean values are labeled as $\cmean^\mathrm{I}$, $\cmean^\mathrm{II}$.
The mean value bound \cref{eq:mvb} informs the respective global minima and global maxima within regions I and II, requiring
\begin{equation}
\begin{aligned}\label{eq:mvb2}
    \cmin^\text{I}&\leq\cmean^\text{I} \leq\cmax^\text{I},\\
    \cmin^\text{II}&\leq\cmean^\text{II} \leq\cmax^\text{II}.
\end{aligned}
\end{equation}

Next, I show that $\cmean^\text{I}$ and $\cmean^\text{II}$ are themselves constrained.
These bounds can be obtained by comparing the slopes of these EOSs in \cref{fig:proof} to known values and recalling that the inverse of slopes are the speed of sound squared (\cref{eq:Cs}).
Since the mean value EOS in region I is steeper than the mean value EOS over the entire ultra-dense phase 
(underlying $\cmeanud$), and the mean value EOS in region II is shallower than the red segment with constant $C_s=\cmaxgg=1/3$,
one finds
\begin{align}
\cmean^\text{I} &< \cmeang,\\
\cmean^\text{II} &> 1/3.
\end{align}
Putting everything together I recover
\begin{align}\label{eq:cmean12}
    \cmin^\text{I} &< \cmeang<1/3,\\
    \cmax^\text{II}&>1/3,
\end{align}
i.e., \cref{eq:ordering,eq:cmax_asymp}. 
And since the minimum in region I is below $\cmeang$, \cref{eq:cmaxns} implies \cref{eq:cmaxns_cminud} regardless of the number of minima in the ultra-dense phase.

\begin{figure}[htbp]
	\includegraphics[width=0.96\linewidth]{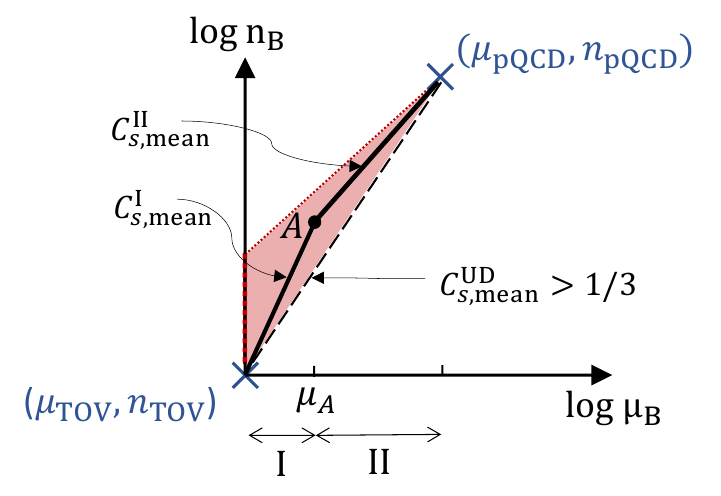}
	\caption{Proving \cref{eq:SCA} guarantees two peaks when $\cmeang>1/3$.
 The region indicated by \cref{eq:SCA} is shaded in red, from which I arbitrarily picked a point labeled ``A". Note that in \cref{fig:demo} this region is shown as the blue hatches.
 Here, the maximally soft EOS with $\cmaxgg=1/3$ is infeasible because $\cmaxgg\geq\cmeang>1/3$.
 	}\label{fig:proof2}
\end{figure}

Similar lines of reasoning can prove that \cref{eq:SCA} necessitates an extra peak when $\cmeang>1/3$.
This scenario is illustrated in \cref{fig:proof2}.
Here, by the mean value bound, the requirement \cref{eq:cmax_asymp} is automatically satisfied by EOSs obeying \cref{eq:SCA}, i.e., passing through the shaded region in \cref{fig:proof2}.
The same two-segment division of the $\mu_B$ axis by an arbitrary point A yields
\begin{align}
\cmean^\text{I} &< \cmeang,\\
\cmean^\text{II} &> \cmeang > 1/3.
\end{align}
Upon substituting in \cref{eq:mvb2} one again recovers
\begin{align}
    \cmin^\text{I} &< \cmeang,\\
    \cmax^\text{II}&>1/3.
\end{align}
Even though this scenario where $\cmeang>1/3$ is disfavored by the existence of two-solar-mass NSs, it is still included because (i) in the Bayesian framework eventually excluded samples still require a physical likelihood to be evaluated; and (ii) it is relevant if $\ceft$ breaks down earlier, for instance due to a low-density quarkyonic transition~\cite{McLerran:2018hbz}.
Analyses that follow $\ceft$ to lower densities are presented in \cref{sec:assump}.

\section{Agnostic models of NS cores}\label{sec:parameos}

This appendix presents a family of sound speed based parameterizations used to describe the poorly-understood inner cores of cold NSs.
Parameterizing the sound speed, the slope of the EOS, instead of the EOS itself is convenient because physical constraints can be easily imposed. 
$C_s$ is bounded from below by stability (equivalent to the convexity requirement) as $C_s>0$ and above by causality as $C_s<1$.
I initially developed this ensemble in~\cite{mckeen:2018xwc} and later incorporated developments in ref ~\cite{Landry:2018prl,Drischler:2020yad}.
Some subsets have also been reported elsewhere in e.g.~\cite{Tews:2018kmu,Annala:2019puf}.
Note that these parameterizations are only used to describe the beta-equilibrium matter in neutron star interiors where each EOS is truncated at the central density of the heaviest neutron stars i.e., the TOV point.
%

Starting from a given low-density nuclear theory calculation, the task of determining the EOS at higher densities for specified $C_s$ is an initial value problem. 
The initial values are given by the EOS at the starting point (in our case, $\ceft$-based low-density EOS at $n_\ceft$), and the density ``evolution" of the EOS is governed by zero-temperature thermodynamics.
Depending on the thermodynamic variable against which $C_s$ is parameterized, the differential equations differ in form.
Additional freedom also arises when representing $C_s$ as a function of densities leading to an extended family of models.
These choices influence the structure of resulting EOSs and the probability distributions of the ensemble in statistical inferences.
These aspects are discussed below in details.

\subsection{independent variable}\label{sec:eosvar}

In principal, $C_s$ may be parameterized in terms of any thermodynamic variables at zero temperature, i.e.
pressure $P$, energy density $\eden$, baryon number density $n$, and baryon chemical potential $\mu=(P+\eden)/n$. 
Zero-temperature thermodynamics ensures that knowledge of the relation between any pair would uniquely determine the rest (except when $C_s=0$, see discussions below).
These possibilities and the corresponding differential equations are

\begin{itemize}
\item $C_s(n)$:

\begin{equation}\label{eq:Csn}
\begin{aligned}
\diff{\eden}{n}&=\frac{P+\eden}{n},\\
\diff{P}{n}&=\frac{P+\eden}{n}C_s;\\
\end{aligned}
\end{equation}

\item $C_s(\eden)$:

\begin{equation}\label{eq:Cse}
\begin{aligned}
\diff{P}{\eden}&=C_s,\\
\diff{n}{\eden}&=\frac{n}{P+\eden};\\
\end{aligned}
\end{equation}

\item $C_s(P)$:

\begin{equation}\label{eq:Csp}
\begin{aligned}
\diff{\eden}{P}&=C_s^{-1},\\
\diff{n}{P}&=\frac{n}{P+\eden}C_s^{-1};\\
\end{aligned}
\end{equation}

\item $C_s(\mu)$:
\begin{equation}\label{eq:Csmu}
\begin{aligned}
\diff{P}{\mu}&=n,\\
\diff{\eden}{\mu}&=n C_s^{-1},\\
\diff{n}{\mu}&=\frac{n^2}{P+\eden}C_s^{-1}.\\
\end{aligned}
\end{equation}
\end{itemize}
Although formally equivalent, these options 
have important implications for EOS parameterizations in practice. In particular, segments of low $C_s$ manifest as large jumps in $n$ and $\mathcal{E}$ but tiny increments in $P$ and $\mu$, imprinting disparate inter-density correlations in the EOS.
This point will be expanded further in \cref{sec:interp}.
   
A related and more obvious caveat concerns when $C_s$ is low or vanishes identically.
Singularities of this type can be tamed by adding a safeguard $C_s\geq\eps$ where $\eps$ is some positive value greater than the machine precision. 
A more general approach restricting $\cmin$ and $\cmax$ (that is actually implemented) is discussed below in \cref{sec:mapping}.

Since $P$ and $\eden$ can vary up to 2 orders of magnitude in the core of neutron stars, it is also convenient to parameterize $C_s$ against the logarithm of these thermodynamic potentials leading to
\begin{itemize}
\item $C_s(\log\eden)$
\begin{equation}
\begin{aligned}
\diff{P}{\log\eden}&=\eden C_s,\\
\diff{n}{\log\eden}&=\frac{\eden n}{P+\eden};\\
\end{aligned}
\end{equation}

\item $C_s(\log P)$~\cite{Landry:2018prl}
\begin{equation}
\begin{aligned}
\diff{\eden}{\log P}&=P C_s^{-1},\\
\diff{n}{\log P}&=\frac{P n}{P+\eden}C_s^{-1};\\
\end{aligned}
\end{equation}
\end{itemize}

As a technical side note, when splines are used in conjunction with the option $C_s(\eden)$, closed form integrations exist, providing a test for numerical accuracy. 
I adopt an embedded 8th-order Rugge-Kutta routine~\cite{10.5555/153158} and have verified that my implementation with a local error tolerance $\eps\sim10^{-9}$ performs better than $10^{-5}$ globally.

Once $x$ is chosen, one proceeds to parameterize $C_s(x)$ above $n_\ceft$ either by introducing explicit parameters that directly control $C_s(x)$ (e.g.~\cite{Tews:2018iwm}) or via implicit hyper-parameters that indirectly influences $C_s(x)$  (e.g.~\cite{Landry:2018prl}).
Below I shall collectively refer to $x$ as ``densities'', which can be any of the thermodynamic variables discussed in this section.
Parameterizing $C_s(x)$ is equivalent to picking  curves from the infinite-dimensional functional space spanned by all physical possibilities.
In practice, this is achieved via discretizing the density axis and choosing a interpolation that passes through specified knots $\lrc{(x_i,C_{s,i}\equiv C_s(x_i))}$.
The rest of this appendix will focus on these aspects.

\subsection{Mappings of $C_s$ to unbounded intervals} \label{sec:mapping}

Before delving into the details, a general remark on the functional form of $C_s(x)$ that enables wider flexibility in our models. 
Most interpolation schemes do not respect the physical requirements $0\leq C_s\leq 1$ even if the specified knots $\lrc{(x_i,C_{s,i})}$ are physical.
To address this, one can perform a change of variable $y=f(C_s)$ so that $y\in (-\infty,\infty)$ is unbounded, and interpolating $\lrc{x_i, y(x_i)}$ before inverting the mapping to obtain the interpolated $C_s(x)$.
The mappings employed in this work are listed in \cref{tab:map} along with their inverses.

\begin{table}[htbp]
\setlength{\tabcolsep}{1em} 
{\renewcommand{\arraystretch}{1.85}
\centering
\begin{tabular}{ l|c|c @{} }
\toprule
	  	& mapping $f: C_s\rightarrow y$ 	& inverse $f^{-1}: y\rightarrow C_s$ \\
\midrule	
Erf  &   $y=\erf^{-1}(2C_s-1)$ 		& $C_s=[\erf(y)+1]/2$   \\
``Lorentz" &	$y=\dfrac{C_s - 1/2}{\sqrt{C_s(1-C_s)}}$	& $C_s=\dfrac{1}{2}\left[\dfrac{y}{\sqrt{1+y^2}}+1\right]$ 	\\
Tanh  &   $y=\log\dfrac{C_s}{1-C_s}$ 		& $C_s=[\tanh(y)+1]/2$   \\
Tan &    $y=\tan\left[\pi(C_s-\frac{1}{2})\right]$     &   $C_s=\dfrac{\tan^{-1}(y)}{\pi}+\dfrac{1}{2}$          \\
\bottomrule
\end{tabular}
}
\caption{A few mappings between $C_s\in[0,1]$ and $y\in(-\infty, \infty)$.
In addition to influencing the functional form of represented EOSs, these mappings implicitly determine the shape of priors on $C_s$.
For instance, see \cref{fig:csmap} when a standard Gaussian distribution is assumed for $y\sim\mathcal{N}(0,1)$.
}
\label{tab:map}
\end{table}
The change of variable also enables a straightforward way of incorporating restrictions on the minimum and maximum of $C_s$.
A simple linear transformation is handy and maps $\widetilde{C_s}\in[0,1]$, the image of $f$, to $C_s=g^{-1}(\widetilde{C_s})\in[\cmin,\cmax]$.
The linear mapping 
\begin{align}
g(C_s)&=\frac{C_s-\cmin}{\cmax-\cmin}, \\
g^{-1}(\widetilde{C_s})&=\widetilde{C_s}\cmax + (1-\widetilde{C_s})\cmin   
\end{align}
is then composed with $f$ to transform between $y$ and $C_s$: $\widetilde{f}=f\circ g$.
The present work only assumes $\cmax=1$, i.e. causality, and $\cmin=10^{-6}$ for numerical stability. Stronger bounds on $\cmax$ have been proposed in the literature~\cite{Hippert:2024hum}.
Note that the quantity $\cmaxl$ in the main text is found for each EOS sample by actually searching for the maximum, which is different from $\cmax$ imposed in this manner as the later is only a bound not necessarily reached in resulting EOSs.

\subsection{Interpolation Methods}\label{sec:interp}

In this work I interpolate $\{(x_i, y_i)\}$ with piecewise polynomials and Gaussian processes~\cite{Landry:2018prl,Melendez:2019izc,Drischler:2020fvz}: 
\begin{itemize}
\item {\bf Splines (linear, quadratic, cubic)} are perhaps the simplest choices.
My approach differs from those available in the market in that I do not impose the number of pieces, whose physical meaning is obscure.
Instead, I directly work with the length scale $l\equiv\Delta x$ of the underlying thermodynamic variable $x$. 
$l(x)$ is generally density-dependent.

The length scales are indirectly controlled by hyper-parameters $\lmin$ and $\lmax$ such that $\lmin\leq l(x)\leq\lmax$.
Starting from the low-density $\ceft$ EOS,
I construct $y(x)$ by randomly sampling increments in density $\Delta x$ in the range [$\lmin$, $\lmax$] until the highest attainable $x_\mathrm{max}$ in NSs
are reached~\cite{Drischler:2020yad,Zhou:2023zrm,Zhou:2024hdi}.
The ordinates $y$ are randomly sampled indirectly via a flat prior on $C_s\in[0,1]$, though one can also directly impose priors on $y=f(C_s)$, see \cref{sec:eosprior}.

The hyper-parameters $\lmin$ and $\lmax$ are themselves randomly selected from $\mathcal{O}(10^5)$ priors each, either uniformly or from log-uniform distributions.
The resulting normalized length scales $\tilde{l}\equiv l/(x_\mathrm{max}-x_\ceft)$ covers a wide range from $1/2000$ to $2$, where $x_\ceft$ refers to the value of $x$ at $n_B=n_\ceft$.
Each choice of $\lmin$ and $\lmax$ are used to generate about a thousand EOSs, yielding $\mathcal{O}(10^8)$ samples in total.

\item {\bf Gaussian Process (GP)}
Gaussian processes assume that the interpolant $y(x)$ follows a multivariate normal distribution in the infinite dimensional function space
$$
y(x)\sim \mathcal{N} (\bar{y}, K)
$$
where $\bar{y}$ is the mean and $K=K(x,x')$ is the covariance also known as the kernel.
Each GP kernel in this work is a sum of $N_k$ Gaussian kernels:
\begin{equation}\label{eq:GK}
K(x, x')=\sum^{N_k}_{a=1}\sigma(l_a) \exp[-(x-x')^2/(2l_a^2)].
\end{equation}
An implicit assumption in GP is density-independent correlation lengths and strengths implied by the translational-invariant kernels $K(x,x')=K(|x-x'|)$.
Ref ~\cite{Drischler:2020yad} used GP to model $\ceft$ predictions of the energy per particle where the authors carefully examined the inter-density correlations to ensure a translational invariant kernel over the Fermi momentum is faithful. 
Density-independent correlations in $C_s$ throughout neutron star densities are yet to be justified.
This issue is especially notable when $C_s$ is low.
As mentioned earlier in \cref{sec:eosvar}, first-order phase transitions introduce tiny length scales in $P$ and $\mu$ whereas in $\eden$ and $n$ they manifest as large jumps.
As such, translational invariant correlations limit the possible behaviors of generated EOSs~\footnote{To be clear, in the continuum and infinite sample size limits, reasonable implementations of GP models and other parameterizations are expected to cover the entire physical phase space. The rate of convergence and the shape of probability distributions however are sensitive to these choices~\cite{Zhou:mi1}.}.
A simple remedy is to include all possible $C_s(x)$ discussed in \cref{sec:eosvar}, so that the mixture contains a wider variety of behaviors of $C_s$.

The strategy for choosing the length scales $l_a$ is similar to that in the previous section except here $l_a$ is independent of $x$.
I fix the grid spacing $\Delta x$ to be the lesser of $\min\{l_a\}/3$ and $(x_\mathrm{max}-x_\mathrm{\ceft})/500$, so that the resolution is adequate for short correlation lengths.
The strength of each Gaussian kernel $\sigma(l)$ is either taken to be completely random or follow a power law $\sigma(l)\propto l^\beta$ where the exponent $\beta$ is sampled from a flat distribution on $[-1,1]$. 
A negative (positive) exponent stresses shorter (longer) correlation lengths, and $\beta=0$ reflects a flat spectrum. 
I normalize the strengths by randomly picking $\Sigma=\sum_a \sigma(l_a)$ from a uniform distribution on $[\Sigma_\mathrm{min}, \Sigma_\mathrm{max}]$.
Larger $\Sigma$ leads to more volatile features whereas lower $\Sigma$ tends to produce flatter sound speed.
$\Sigma_\mathrm{min}$ and $\Sigma_\mathrm{max}$ are chosen to ensure a wide range of variances in $C_s$  are realized.
The mean $\bar{y}$ is sampled randomly from a standard normal distribution $\bar{y}\sim\mathcal{N}(0,1)$.
Finally, a white noise (not shown in \cref{eq:GK}) of strength $\sigma=10^{-4}$ is added to the kernel  for numerical stability.

I generated $\mathcal{O}(10^5)$ kernels of the form \cref{eq:GK}, each a sum of $N_k=1-100$ Gaussian kernels, and have accumulated over $\mathcal{O}(10^8)$ EOS samples.
\end{itemize}

When numerically integrating the ODEs in \cref{sec:eosvar} the step sizes are restricted to be at most $\lmin/3$ so that sharp features in $C_s$ are captured.
The 8th order formula discussed earlier is found to be very accurate and efficient.

\subsection{prior distributions}\label{sec:eosprior}

\begin{figure}
\includegraphics[width=0.98\linewidth]{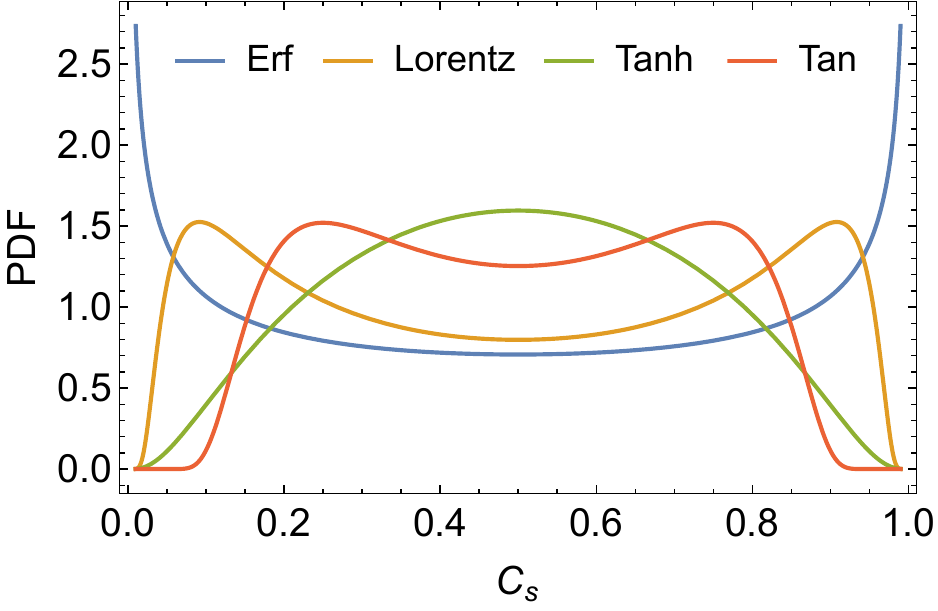}
\caption{
Probability distribution function on the squared sound speed $C_s$ for mappings $f: C_s\rightarrow y$ listed in \cref{tab:map} assuming a standard normal distribution on the mapped variable $y\sim\mathcal{N}(0,1)$. Choices of $f$ implicitly determine the shape of prior on $C_s$ at fixed densities. 
The prior for $C_s(x)$ as a function of densities 
are further influenced by choices listed in \cref{sec:eosvar,sec:interp}.
}
\label{fig:csmap}
\end{figure}

Choices of priors are perhaps the most significant factor influencing posteriors in Bayesian analyses.
The abscissae of the knots are controlled by the length scale $l\equiv\Delta x$, and its prior is discussed in the previous section.
Priors on $C_s$ can either be imposed directly or indirectly through priors on $y$.
\Cref{fig:csmap} shows the probability density function (pdf) of $C_s$ assuming a standard normal distribution on $y\sim\mathcal{N}(0,1)$.
Note that the standard deviation of the prior on $y$ is degenerate with the scaling of $y$ in the mappings, which is set to $1$ in \cref{tab:map}. 
The choice of mappings implicitly determines the shape of pdf of $C_s$.
I have included all four options in the ensemble, and have also
repeated the above steps but impose priors on $c_s=\sqrt{C_s}$ instead of $C_s$.
Compared to $C_s$-based models, directly working with $\sqrt{C_s}$ favors lower sound speed in the prior, and sometimes notable differences in the posteriors are observed between the choices of $C_s$ and $\sqrt{C_s}$~\cite{Zhou:mi1}.

The commonly adopted EOS prior is a uniform distribution on the observable being inferred.
This work concerns the criteria \cref{eq:summary}, and I discuss choices of priors in \cref{sec:assump}.
Priors in conventional settings such as inferring the EOS or NS observables will be discussed in an ensuing paper.

\subsection{Summary of models}\label{sec:eossum}

A significant number of models arise from taking combinations of the choices presented in the previous sections.
However, this proliferation of models greatly increases the computational cost if Bayesian analyses are to be performed separately on every subset.
To reduce the number of Bayesian runs several subsets are merged, leaving 12 collections of models in total. 
They are labeled by the 6 choices of independent variable ($P$, $\eden$, $\mu$, $n$, $\log P$, and $\log\eden$, see \cref{sec:eosvar}) and 2 options of the interpolation schemes (piecewise polynomials and Gaussian processes, see \cref{sec:interp}).
Each of the 12 ensembles contains $10^7-10^8$ samples that support at least $1.4M_\odot$ NSs,  consistent with theories of core-collapse supernovae~\cite{Baade_1934b,Lattimer:2004pg,Burrows:2012ew}.
Bayesian analyses are then carried out independently on every subsets.
The average, minimum, and maximum of the resulting Bayes factors are reported in~\cref{tab:bfactor} and the supplemental material.
Note that neither the spline-based nor the Gaussian-process-based models contain explicit parameters, and are only indirectly controlled by the hyper-parameters concerning the length scale $l\equiv\Delta x$ and the magnitude of $C_s(x)$.
So in the sense commonly implied in the literature, both subsets of models are ``non-parametric".
A robust measure of implicit assumptions underlying these models is reported in a following paper \cite{Zhou:mi1}.

\section{Effects of priors, $\cmaxgg$, $\cmingg$,  $n_\ceft$, and $\mu_\pqcd$}
\label{sec:cminres}
\label{sec:assump}

This appendix examines assumptions underlie the main results in addition to those discussed in \cref{sec:parameos}.
I begin by discussing choices of prior distributions.
In the main text, neutral priors without preferences for the criterion $\Delta P-\Delta P_\mathrm{mean,max}$ are chosen.
They are obtained by imposing separate flat distributions over regions where $\Delta P-\Delta P_\mathrm{mean,max}>0$ and where $\Delta P-\Delta P_\mathrm{mean,max}<0$. 
The two segments are normalized such that their integrated probability equals.
The effect of \cref{eq:cmaxns} is minor,
and I have verified that neutral priors on $\Delta P-\Delta P_\mathrm{mean,max}$ alone lead to unbiased priors for the two-peak scenario.

Another common choice in the literature is a simple flat prior over possible ranges of $\Delta P-\Delta P_\mathrm{mean,max}$.
It is deemed the least informative and generally preferred.
Bayes factors based on this choice are shown in \cref{tab:BsNE}.
Here, the priors display clear preferences.
The two-peak scenario is generally disfavored by these priors unless $X\gtrsim 3$.
Although the statistical significance of the Bayesian evidence decreases mildly in these disadvantageous settings, the two-peak scenario is still favored.

\begin{table*}[!htbp]
\centering
\setlength{\tabcolsep}{1em} 
{\renewcommand{\arraystretch}{1.5}
\begin{tabular}{ l|c c c | c c c c c @{} }
	\toprule
	astro data                   & $X=1.5$ &   $2$  &  $3$   & $X\sim U[1,4]$ & $U[1.5,4]$ & $U[2,3]$ & $U[2,4]$ & $\mathcal{N}(2.5, 0.5)$ \\ \midrule
    prior 						 &   ${0.2}^{+0.1}_{-0.1}$  & ${0.7}^{+0.2}_{-0.3}$ & ${1.7}^{+0.5}_{-0.6}$ & ${0.8}^{+0.2}_{-0.2}$ & ${1.3}^{+0.3}_{-0.4}$ & ${1.2}^{+0.4}_{-0.5}$ & ${1.8}^{+0.5}_{-0.6}$ &  ${1.2}^{+0.2}_{-0.3}$       \\
	+J0740+6620                  &   ${0.4}^{+0.1}_{-0.1}$  & ${2.6}^{+0.8}_{-1.0}$ & ${46}^{+31}_{-24}$    & ${1.8}^{+0.2}_{-0.1}$ & ${4.8}^{+0.9}_{-1.1}$ & ${6.5}^{+4.0}_{-3.0}$ & ${13}^{+7}_{-6}$     &   ${5.6}^{+1.0}_{-1.1}$      \\ 
    \hline
	+$M_\tov\lesssim2.2M_\odot$ &  $\lesssim0.1$                       & ${0.5}^{+0.2}_{-0.2}$ & ${13}^{+9}_{-7}$       & ${1.1}^{+0.3}_{-0.2}$ & ${2.6}^{+0.6}_{-0.5}$ & ${1.9}^{+1.1}_{-0.9}$ & ${4.7}^{+2.1}_{-1.8}$ &  ${2.1}^{+0.5}_{-0.5}$      \\ 
	+$M_\tov\gtrsim2.6M_\odot$  &  ${2.0}^{+0.6}_{-0.7}$   & ${26}^{+15}_{-14}$    &  $\gg10^3$             & ${2.1}^{+0.5}_{-0.3}$ & ${24}^{+3}_{-5}$      & ${160}^{+120}_{-100}$ & ${310}^{+180}_{-170}$ &  ${29}^{+5}_{-3}$       \\ 
    \bottomrule
\end{tabular}
}
\caption{Bayes evidence assuming simple flat priors on the criterion \cref{eq:summary}. Here, the priors are no longer neutral and favors the two-peak scenario when $X\gtrsim2.4$. Analyses with $X=1,4$ always yield contradicting yet extreme evidence and are not shown.
}
\label{tab:BsNE}

\vspace{3em}
{\renewcommand{\arraystretch}{1.5}
\begin{tabular}{ l|c c c | c c c c c @{} }
	\toprule
	astro data                   & $X=1.5$ &   $2$  &  $3$   & $X\sim U[1,4]$ & $U[1.5,4]$ & $U[2,3]$ & $U[2,4]$ & $\mathcal{N}(2.5, 0.5)$ \\ \midrule
    prior 						 & ${0.2}^{+0.1}_{-0.1}$ & ${0.6}^{+0.2}_{-0.3}$ & ${1.9}^{+0.5}_{-0.7}$ & ${0.7}^{+0.2}_{-0.2}$ & ${1.1}^{+0.3}_{-0.4}$ & ${1.1}^{+0.4}_{-0.4}$ & ${1.6}^{+0.5}_{-0.6}$ & ${0.9}^{+0.2}_{-0.3}$\\
	+J0740+6620                  & ${0.3}^{+0.1}_{-0.1}$ & ${2.0}^{+0.7}_{-0.7}$ & ${36}^{+22}_{-18}$    & ${1.6}^{+0.2}_{-0.1}$ & ${4.0}^{+0.8}_{-0.9}$ & ${5.6}^{+3.7}_{-2.7}$ & ${11.3}^{+6.4}_{-4.9}$ & ${4.3}^{+0.6}_{-0.6}$ \\ 
    \hline
	+$M_\tov\lesssim2.2M_\odot$ & $\lesssim0.1$                  & ${0.4}^{+0.2}_{-0.2}$  & ${12}^{+7}_{-6}$    & ${1.1}^{+0.2}_{-0.2}$ & ${2.0}^{+0.5}_{-0.5}$ & ${1.9}^{+1.1}_{-0.9}$ & ${4.4}^{+2.2}_{-1.9}$ & ${1.7}^{+0.4}_{-0.4}$  \\ 
	+$M_\tov\gtrsim2.6M_\odot$  & ${1.6}^{+0.4}_{-0.5}$  & ${19}^{+10}_{-10}$     & $\gg10^3$           & ${2.0}^{+0.4}_{-0.3}$ & ${14}^{+12}_{-3}$     & ${100}^{+60}_{-50}$  & ${190}^{+110}_{-90}$  & ${12}^{+5}_{-3}$ \\ 
    \bottomrule
\end{tabular}
}
\caption{Similar to \cref{tab:BsNE} but restricts $\cmaxgg\leq2/3$ in the prior (see \cref{eq:dp_prior23}).
The evidence is not significantly affected by the prior unfavorable for high $C_s$.
}
\label{tab:BsC23}

\end{table*}

It is noteworthy that the criteria for the two-peak scenario \cref{eq:summary} are not equivalent to a single requirement on the maximum sound speed $\cmaxgg>1/3$. 
For instance, a monotonically decreasing $C_s^\mathrm{UD}$ with $\cmaxgg>1/3$ does not lead to a second peak at high densities.
In fact, monotonically decreasing $C_s$ between TOV and pQCD points are strongly disfavored, see \cref{tab:BsND} in \cref{sec:csdemo}.
The upshot is that flat priors on $\cmaxgg$ are not the least informative ones.
Nevertheless, it is illuminating to explicitly exclude high sound speed by imposing upper bounds on $\cmaxgg$ in the prior.
For instance, let us demand $\cmaxgg\leq2/3$ by replacing the requirement on the prior \cref{eq:dp_prior} with the following:
\begin{multline}\label{eq:dp_prior23}
    \dpmin(\cmingg,\cmaxgg=2/3)\\
    \leq\Delta P\leq \dpmax(\cmingg,\cmaxgg=2/3).
\end{multline}
The expression for $\dpmin$ 
is related to \cref{eq:dpmax} via swapping $\alpha$ and $\beta$ and is given in ref \cite{Zhou:2024hdi}.
The resulting Bayes factors are shown in \cref{tab:BsC23}.
Narrowing the phase-space volume in the prior reduces the evidence, but the main findings remain unaffected.

A side note on the treatment of \cref{eq:dp_prior}.
In Bayesian analyses in the literature, e.g. refs \cite{Gorda:2022jvk,Somasundaram:2022ztm,Zhou:2023zrm,Fujimoto:2024pcd}, \cref{eq:dp_prior} is often treated as a likelihood. 
Since \cref{eq:dp_prior} simply comes from thermodynamic stability and causality, {\em a priori} one expects it to hold.
Therefore, I consider it a requirement on the prior, the same way causality and stability are built into most parameterizations of NS EOSs.
Furthermore, because fully model-independent studies of the pressure-energy density relation~\cite{Komoltsev:2021jzg} 
and in the mass-radius plane~\cite{Zhou:2023zrm} are available, the effects of \cref{eq:dp_prior} are robustly determined as the boundaries that delimit physical regions in the phase spaces.
Treating \cref{eq:dp_prior} as a prior implicitly incorporates these clean and simple results.
As such, the choice made here appears to be more sensible, at least for the present purpose, both from a physical viewpoint and a Bayesian perspective.

When taken as a likelihood, \cref{eq:dp_prior} is found to favor soft EOSs in NS cores~\cite{Gorda:2022jvk,Somasundaram:2022ztm,Fujimoto:2024pcd} since acausal extensions in the ultra-dense phase are more common among stiff NS cores. 
When treated as a prior, almost any arbitrarily stiff core can be compatible with \cref{eq:dp_prior}~\cite{Zhou:2023zrm}.
To sum up, stiffening of the EOS in the density range $\sim2-5n_0$ is a physical consequence of two-solar-mass pulsars~\cite{LattimerPrakash:2010,Gandolfi:2011xu}, and only additional constraints on $M_\tov$ can robustly inform the NS EOS at higher densities.\\

\begin{table*}
\centering
\setlength{\tabcolsep}{1em} 
{\renewcommand{\arraystretch}{1.5}
\begin{tabular}{ l|c c c | c c c c c @{} }
	\toprule
	astro data                   & $X=1.5$ &   $2$  &  $3$   & $X\sim U[1,4]$ & $U[1.5,4]$ & $U[2,3]$ & $U[2,4]$ & $\mathcal{N}(2.5, 0.5)$ \\ \midrule
	J0740+6620                &   ${2.1}^{+0.3}_{-0.3}$  & ${3.9}^{+0.4}_{-0.2}$ & ${20}^{+7}_{-6}$ & $2.7^{+0.6}_{-0.6}$& ${4.9}^{+1.0}_{-1.5}$ & ${6.4}^{+1.3}_{-1.0}$ &  ${9.6}^{+2.1}_{-1.6}$ &  ${4.8}^{+0.6}_{-0.5}$ \\ 
    \hline
	+$M_\tov\lesssim2.2M_\odot$ &    ${0.2}^{+0.1}_{-0.1}$ & ${0.7}^{+0.2}_{-0.2}$ & ${6.4}^{+1.8}_{-1.9}$  & ${1.7}^{+0.2}_{-0.2}$ & ${2.1}^{+0.5}_{-0.6}$  & ${2.0}^{+0.4}_{-0.3}$ & ${3.6}^{+0.7}_{-0.6}$  &  ${1.9}^{+0.1}_{-0.1}$  \\ 
	+$M_\tov\gtrsim2.6M_\odot$  &    ${12}^{+0.9}_{-0.8}$ & ${52}^{+4}_{-5}$      & $>10^3$                & ${3.4}^{+1.6}_{-1.2}$ & ${21}^{+9}_{-8}$  & ${160}^{+70}_{-40}$ &  ${220}^{+50}_{-40}$       &  ${19}^{+7}_{-5}$  \\ 
    \bottomrule
\end{tabular}
}
\caption{Similar to \cref{tab:bfactor} but expands the correlated $\ceft$ uncertainty band from $2\sigma$ to $4\sigma$ when constructing NSs. 
Uniform distributions within the $\ceft$ uncertainty band are assumed. 
The Bayes factors here are obtained from a (much) smaller pool of samples, and are largely similar to \cref{tab:bfactor}.
}
\label{tab:Bs4sigma}

\vspace{3em}
{\renewcommand{\arraystretch}{1.5}
\begin{tabular}{ l|c c c | c c c c c @{} }
	\toprule
	astro data                   & $X=1.5$ &   $2$  &  $3$   & $X\sim U[1,4]$ & $U[1.5,4]$ & $U[2,3]$ & $U[2,4]$ & $\mathcal{N}(2.5, 0.5)$ \\ \midrule
    J0740+6620 					 &   ${1.1}^{+0.2}_{-0.2}$  & ${2.1}^{+0.2}_{-0.3}$ & ${12.8}^{+2.0}_{-2.9}$ & ${2.0}^{+0.2}_{-0.2}$ & ${2.8}^{+0.2}_{-0.2}$ & ${3.4}^{+0.9}_{-0.7}$ & ${4.4}^{+0.4}_{-0.4}$  & ${2.8}^{+0.2}_{-0.3}$  \\
	+GW170817                    &   ${1.0}^{+0.2}_{-0.2}$  & ${1.9}^{+0.3}_{-0.4}$ & ${12.8}^{+1.7}_{-2.8}$ & ${1.9}^{+0.2}_{-0.2}$& ${2.5}^{+0.1}_{-0.2}$ & ${3.1}^{+0.7}_{-0.7}$ &  ${4.0}^{+0.3}_{-0.4}$ &  ${2.6}^{+0.3}_{-0.3}$ \\ 
    \hline
	+$M_\tov\lesssim2.2M_\odot$ &    ${0.2}^{+0.1}_{-0.1}$ & ${0.3}^{+0.1}_{-0.1}$ & ${3.5}^{+0.7}_{-0.9}$  & ${0.9}^{+0.1}_{-0.1}$ & ${1.0}^{+0.1}_{-0.2}$  & ${0.8}^{+0.2}_{-0.2}$ & ${1.4}^{+0.6}_{-0.4}$  &  ${0.8}^{+0.1}_{-0.1}$  \\ 
	+$M_\tov\gtrsim2.6M_\odot$  &    ${2.9}^{+0.8}_{-0.5}$ & ${14}^{+3}_{-4}$      & $>10^3$                & ${3.7}^{+1.0}_{-0.8}$ & ${9.3}^{+2.3}_{-2.1}$  & ${50}^{+32}_{-19}$ &  ${60}^{+22}_{-23}$       &  ${13}^{+3}_{-2}$  \\ 
    \bottomrule
\end{tabular}
}
\caption{Similar to \cref{tab:bfactor} but only follows $\ceft$ up to $1.5 
n_0$, above which density agnostic parameterizations ~\cref{sec:parameos} are used to describe NS inner cores.
The number of generated samples is roughly the same as when $n_\ceft=2n_0$ is assumed.
Even if $\ceft$ breaks down earlier, $M_\tov\leq2.2M_\odot$ would still yield substantial evidence if first-order phase transitions are absent, see \cref{fig:BsNeft}.
Constraints from GW170817 are approximated as a hard cut off $\Lambda_{1.4M_\odot}\leq600$. 
This simplification renders the bound stronger than implied by the actual data, and is more unfavorable to the two-peak scenario.
Yet the reduction to the evidence is mild, demonstrating that with agnostic models the EOS at the TOV point largely decouples from that at lower densities relevant for most NSs~\cite{Zhou:2023zrm,Zhou:2024hdi}.
}
\label{tab:Bsn15}

\end{table*}

The phase space for the two-peak scenario is also controlled by the lower bound of sound speed squared $\cmingg$.
So far, 
only $\cmingg=0$ required by thermodynamic stability is imposed.
Low sound speed generally signals first-order phase transitions where low- and high-density phases in mechanical equilibrium are separated by a finite latent heat so that $C_s=\ud P/\ud \mathcal{E}\simeq 0$.
Based on similar symmetry breaking patterns and low-lying excitations in nucleon superfluid and in quark color superconductor, it has been conjectured that the quark-hadron transition at finite density is a smooth crossover~\cite{Schafer:1998ef,Schafer:1999pb,Schafer:1999fe}
 \footnote{though topological phase transitions are possible~\cite{Cherman:2018jir,Hirono:2018fjr,Cherman:2020hbe,Dumitrescu:2023hbe}.
 The ultra-dense regime would still be free of phase transitions if they occur inside NSs.}, motivating more restrictive $\cmingg$.
As discussed in the main text, when $\cmingg>0$ the area $\dpmax(\cmingg,\cmaxgg=1/3)$ decreases rendering the criterion \cref{eq:criterion} less demanding.
The Bayes factors as functions of $\cmingg$ are shown in \cref{fig:BsNeft,fig:BsMus}.
With $\cmingg\gtrsim0.1-0.15$, even the unfavorable data $M_\tov\lesssim2.2M_\odot$ would yield substantial support for two peaks. 
When $M_\tov\gtrsim2.6M_\odot$, most EOSs are incompatible with $\cmingg\gtrsim0.1-0.2$~\cite{Zhou:2024hdi}, rendering the statistics less reliable with stronger assumptions about $\cmingg$. \\

Other parameters of this framework are $n_\ceft$, the highest density $\ceft$ is adopted, and $\mu_\pqcd$, the lowest chemical potential pQCD remains reliable.
The convergence of $\ceft$ up to $n_\ceft=2n_0$ is demonstrated by statistical analyses exploiting the structure of EFT~\cite{Drischler:2020yad}, although its uncertainty may have been underestimated~\cite{Cirigliano:2024ocg}.
The evidence is only marginally affected if the EFT error band is broadened from $2\sigma$ to $4\sigma$, see~\cref{tab:Bs4sigma}.
Here and throughout this work, flat priors within the $\ceft$ uncertainty band are imposed by sampling uniformly the coefficients of basis functions in our $\ceft$-based low-density model, see eq 2 in \cite{Zhou:2023zrm}.
Additionally, $\ceft$ may not be suitable if the ground state is no longer nucleonic at $\sim2n_0$, such as the case of low-density quarkyonic transitions~\cite{McLerran:2018hbz}.
To examine these possibilities, I repeat the Bayesian analyses but only adopt $\ceft$ up to $n_\ceft=n_0, 1.5n_0$.
The updated Bayes evidence is shown in \cref{tab:Bsn15} and \cref{fig:BsNeft}.
If the NS maximum mass is indeed low near $2.2M_\odot$, low-density nuclear theory inputs between $n_0$ and $2n_0$ would be helpful to secure clear evidence for two peaks in $C_s$.

On the high-density end, 
$\mu_\pqcd=2.4-2.6$ GeV is commonly adopted in the literature, although the large sensitivity on $X$ raises questions about this assumption.
Refs ~\cite{Fernandez:2021jfr,Fernandez:2024ilg} resummed to all order the leading and next-to-leading soft modes, and found reduced dependence on $X$ favoring $X\gtrsim1.5$.
Ref~\cite{Fujimoto:2024pcd} inferred $X$ from comparing the N2LO pQCD EOS with lattice calculations at finite isospin~\cite{Abbott:2023coj,Abbott:2024vhj}.
By including the LO pQCD prediction for the superconducting gap and exploiting the similar structure of pQCD at finite isospin and at finite $\mu_B$, 
$X$ is found to be large $X\in[2.4,4]$ and the superconducting gap at finite $\mu_B$ is found to be negligible with $\Delta\lesssim\MeV$ near $\mu_B=2.4~\GeV$.
This bound on $X$ translates to a tightly constrained $P_\pqcd\in[2.8,3.0]$ GeV/fm$^3$ at $\mu_B=2.4~\GeV$, which roughly corresponds to $X\gtrsim2.3$ when soft modes at N3LO are included (the pQCD EOS adopted in the main text).
According to \cref{tab:bfactor}, such high values of $X$ lead to a notable preference for the two-peak scenario even with $M_\tov\lesssim2.2M_\odot$.
On the other hand, the suppressed superconducting gap seems to suggest that other mechanisms are behind the postulated second peak in $C_s$. 
If the superconducting gap is underestimated by LO pQCD, the inferred $X$ thus $P_\pqcd$ is reduced when matching to lattice data at finite isopsin, but the full pressure $P_\cqm=P_\pqcd+P_\Delta$ receives a pairing contribution $P_\Delta=\mu_B^2\Delta^2/(3\pi^2)$ that is important if $\Delta\gtrsim100$ MeV and $X\lesssim1.5$.
The bottom line is that pairing gaps much larger than inferred in ref \cite{Fujimoto:2024pcd} can still yield a large $\Delta P$ essential for the two-peak evidence, even though inferred $X$ would be reduced.
These large gaps are near the high end of current estimates, and are compatible with astrophysics~\cite{Zhou:2023zrm,Kurkela:2024xfh}.
In short, while there are indications that pQCD is reliable down to $\mu_B=2.4$ GeV, higher-order calculations and detailed statistical analyses will be essential to confirm the applicable range of asymptotic expansions in $\alpha_s$ and to pinpoint the size of superconducting gaps.

\Cref{fig:BsMus} explores a pessimistic scenario and shows Bayes evidence if the matching to pQCD can only be performed at higher baryon chemical potentials $\mu_B=2.8, 3.2$ GeV.
The reduction of Bayes factors is considerable if $M_\tov\lesssim2.2M_\odot$, in which case only stronger assumptions $\cmingg\gtrsim0.15-0.2$ can lead to substantial evidence.
Once again, utilizing {\em ab-initio} calculations to the fullest extent is helpful to extract features in $C_s$ at ultra-high densities.

\begin{figure}
	\includegraphics[width=\linewidth]{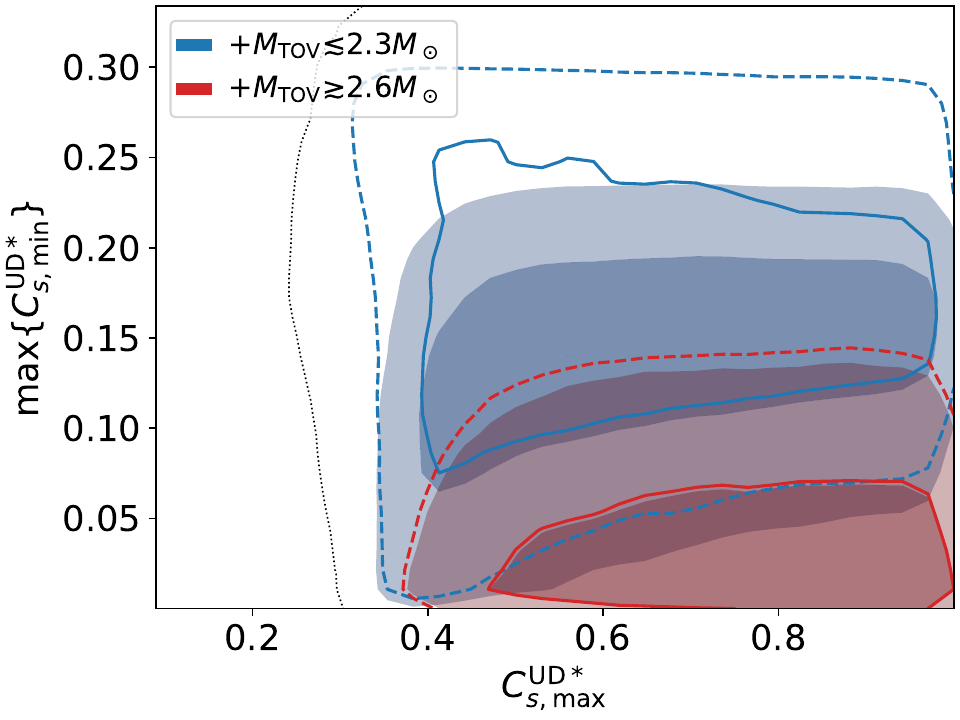}
	\caption{Similar to \cref{fig:CIs} but assumes a flat prior on $\sqrt{\cmaxgg}$ instead of $\cmaxgg$ so that on average $\cmaxgg$ $\simeq0.25$ in the prior. Posteriors here are almost indistinguishable from those in \cref{fig:CIs}.
	}\label{fig:CIs_cmax}
\end{figure}

Lastly, I demonstrate the robustness of the inferred bounds on the magnitude of the peak and trough.
\Cref{fig:CIs_cmax} shows the posterior CIs if a flat prior on $\sqrt{\cmaxg}$ is imposed.
Now, the mean of the prior is $\cmaxgg\approx0.25$ favoring low sound speed.
The posteriors are only marginally shifted, confirming that the preference for high $\cmaxgg$ is a physical consequence.
More detailed investigations into the height of peaks and troughs in $C_s$ will be reported later.

\clearpage

\begin{widetext}
\onecolumngrid

\begin{figure}[!htbp]
	\includegraphics[width=0.99\linewidth]{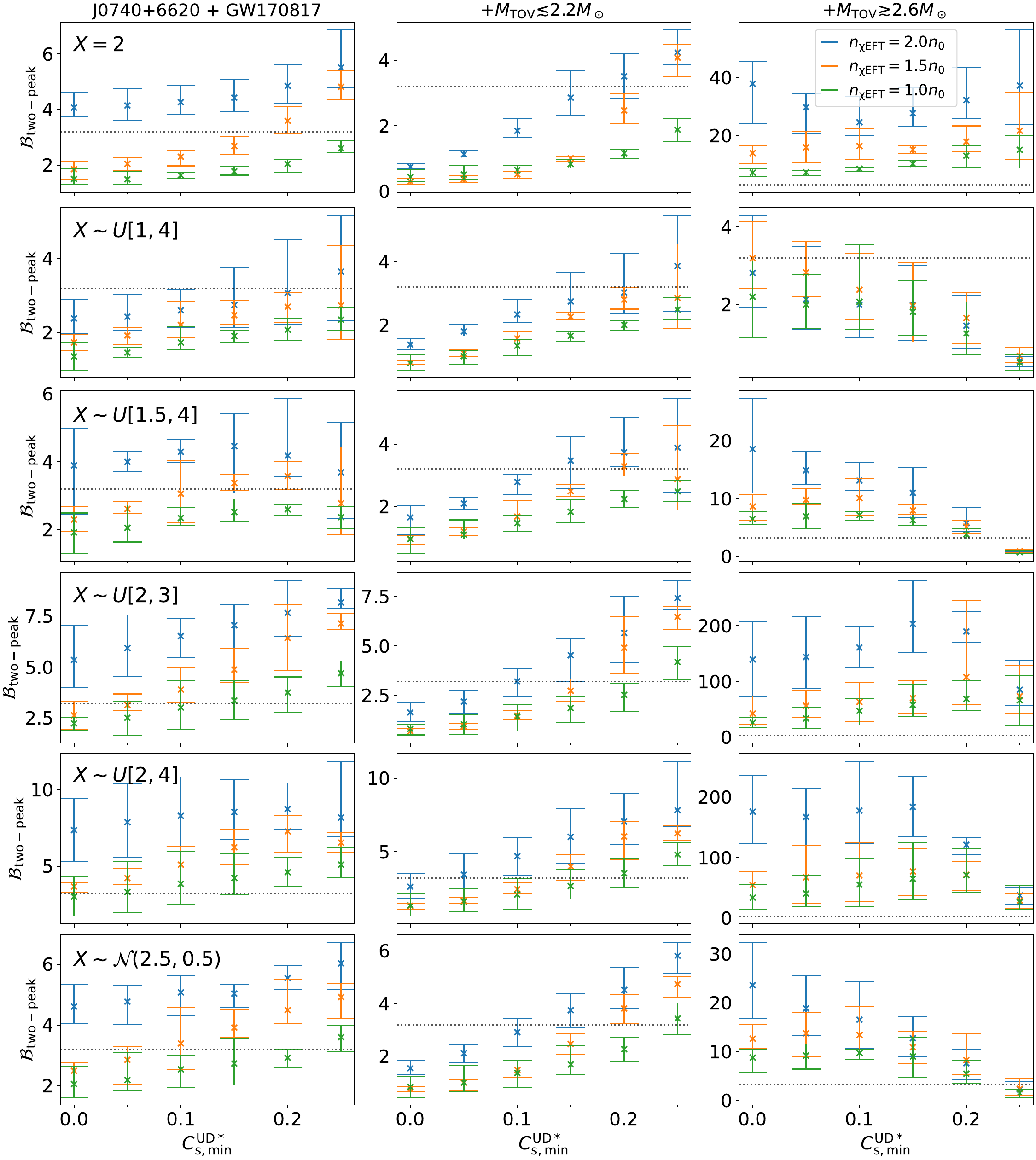}
	\caption{Bayes factors based on neutral priors as functions of $\cmingg$. 
    Each row assumes a different value of or distribution on $X$.
    By the geometric interpretation of $\Delta P$, higher $\cmingg$ requires higher $\cmaxgg$ to make up for the same area $\Delta P$.
    The evidence for two peaks thus becomes stronger. An exception is for $M_\tov\gtrsim2.6M_\odot$, when thermodynamics demands $\cmingg\lesssim0.1-0.2$~\cite{Zhou:2024hdi}.
    In this case, there are significantly fewer samples, so the evidence becomes more uncertain and may decrease with increasing $\cmingg$.
    In the first column, both the mass and radius measurements of PSR J0740+6620 are imposed; the second and third columns are identical to the last two rows in \cref{tab:bfactor}. Error bars reflect systematics associated with NS inner core models.
	}\label{fig:BsNeft}
\end{figure}

\begin{figure}[!htbp]
	\includegraphics[width=0.99\linewidth]{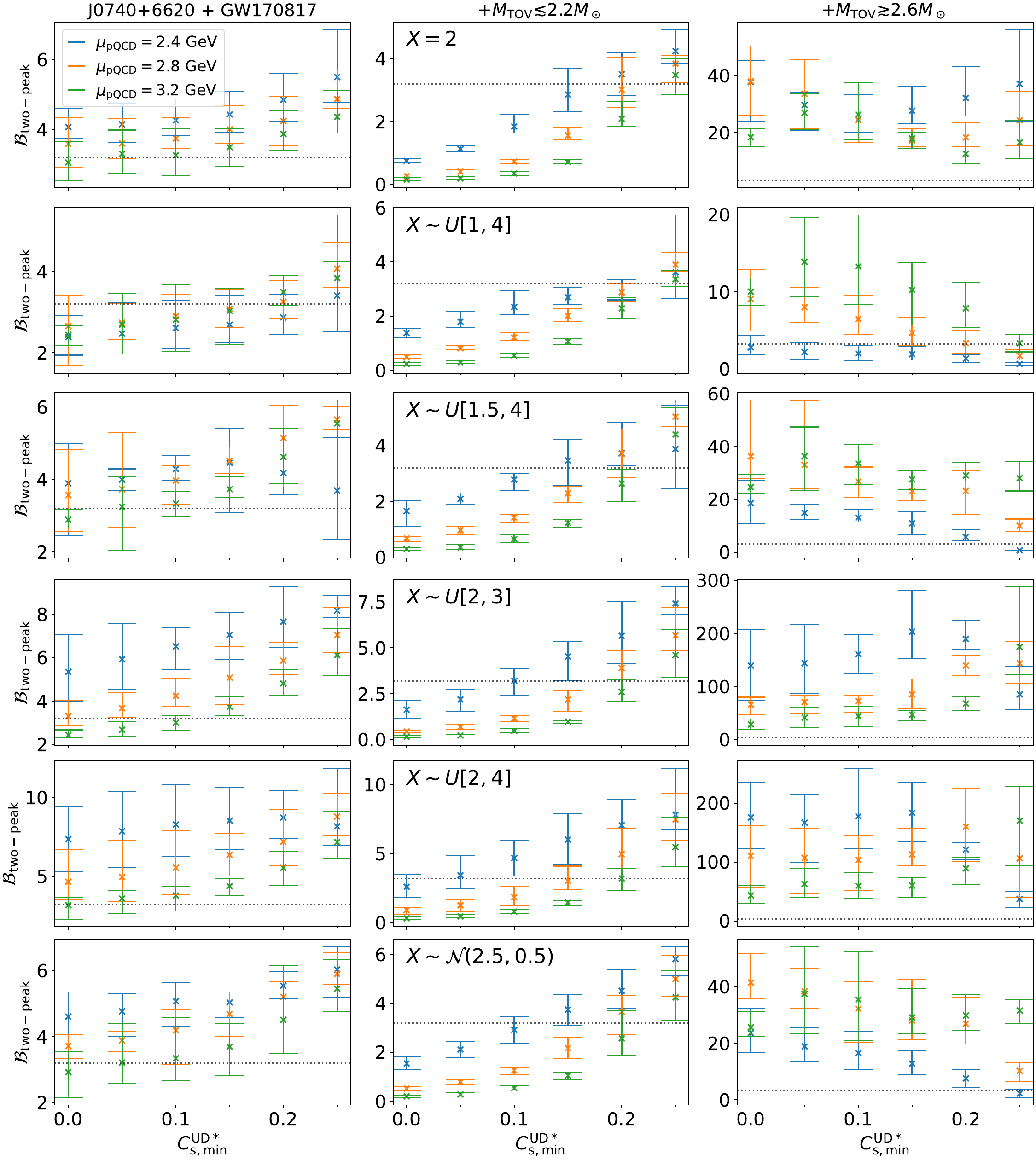}
	\caption{Similar to \cref{fig:BsMus} but demonstrating the effects of $\mu_\pqcd$. Neutral priors and $n_\ceft=2n_0$ are assumed.
	}\label{fig:BsMus}
\end{figure}

\twocolumngrid
\end{widetext}

\clearpage


\newpage

\section{Illustrations of favored and disfavored scenarios}\label{sec:csdemo}

Here, I illustrate behaviors of $C_s$ at high densities that are favored and disfavored.
The examples listed below are only demonstrations.
Their functional form and numerical values are not to be construed as posterior distributions or bounds.
For instance, the value of $n_\tov$ is model-dependent and in reality would not be identical for the examples shown below.
The only requirement is the existence of a peak in $C_s$ exceeding $1/3$ following a trough above $n_\tov$.
More refined constraints on $C_s(n_B)$ will be reported in a subsequent paper.

Note that \cref{eq:summary} does not explicitly require $\cmaxl$ to be greater than $1/3$, though astrophysics strongly suggests violation of conformal limit inside NSs. 
Moreover, while $C_s(n_B)$ is allowed to have a finite number of discontinuities, a fact implicit in the central results ~\cref{eq:summary} and in NS models in ~\cref{sec:parameos} (via very short length scales $\tilde{l}$),
for aesthetic reasons I only show smooth curves here.
Finally, at low densities the possibility of a peak in $C_s$ in the $\ceft$-based outer core or in the crust is not counted towards the number of peaks in this work.

\begin{figure}[htbp]
	\includegraphics[width=0.99\linewidth]{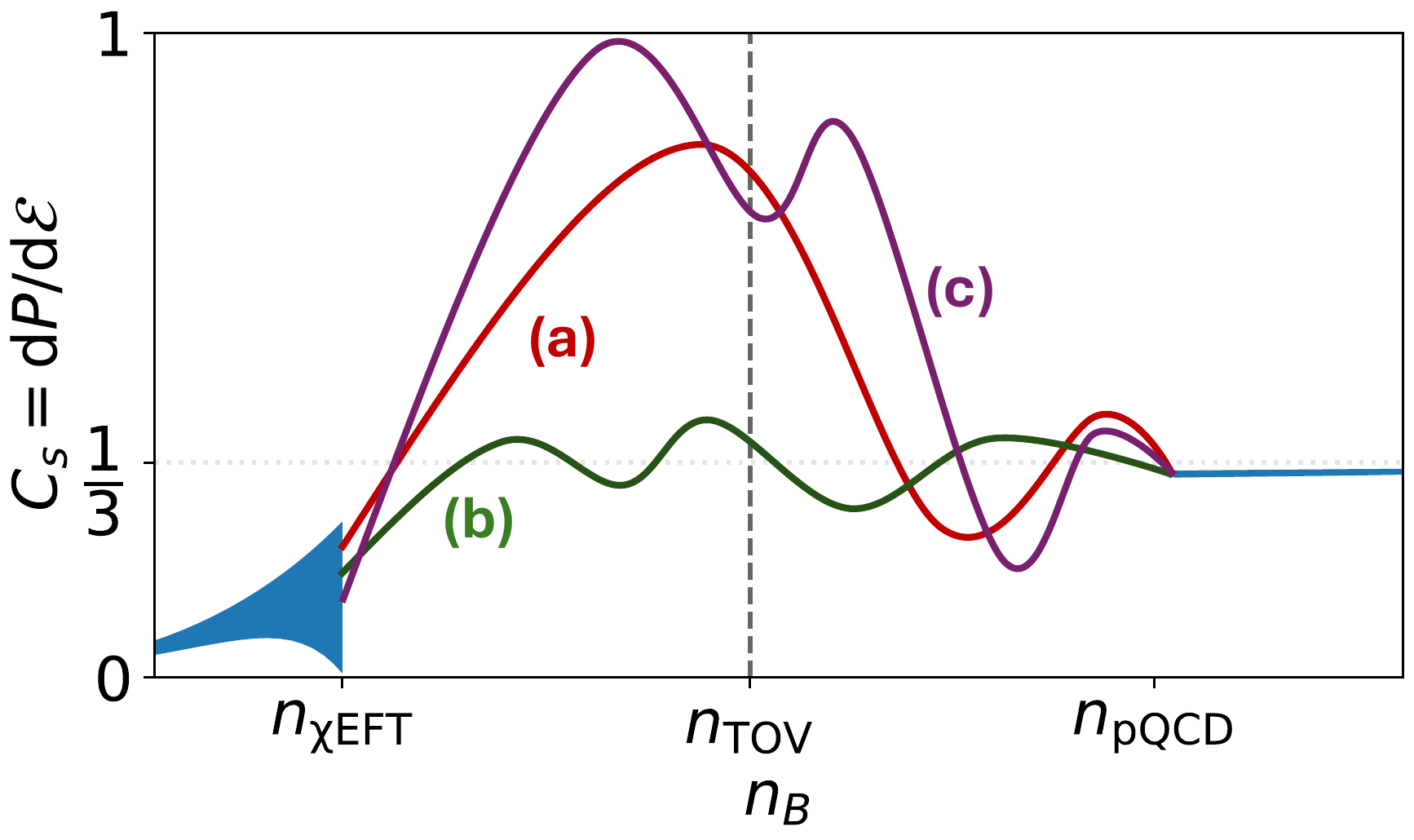}
	\caption{ Behaviors of $C_s$ compatible with the sufficient conditions~\cref{eq:summary}.
    (a): one peak within and another peak outside NS densities;
    (b): more than one peak inside NS densities, one peak in the ultra-dense phase;
    (c): one peak inside NSs, more than one peak above NS densities;
    and (d): more than one peak within and beyond NS densities (not shown).
    Note that as written \cref{eq:summary} does not require peaks inside NSs to surpass $1/3$, though astrophysics strongly suggests $\cmaxl>1/3$.
        }\label{fig:goodex}
\end{figure}

\begin{figure}[htbp]
	\includegraphics[width=0.99\linewidth]{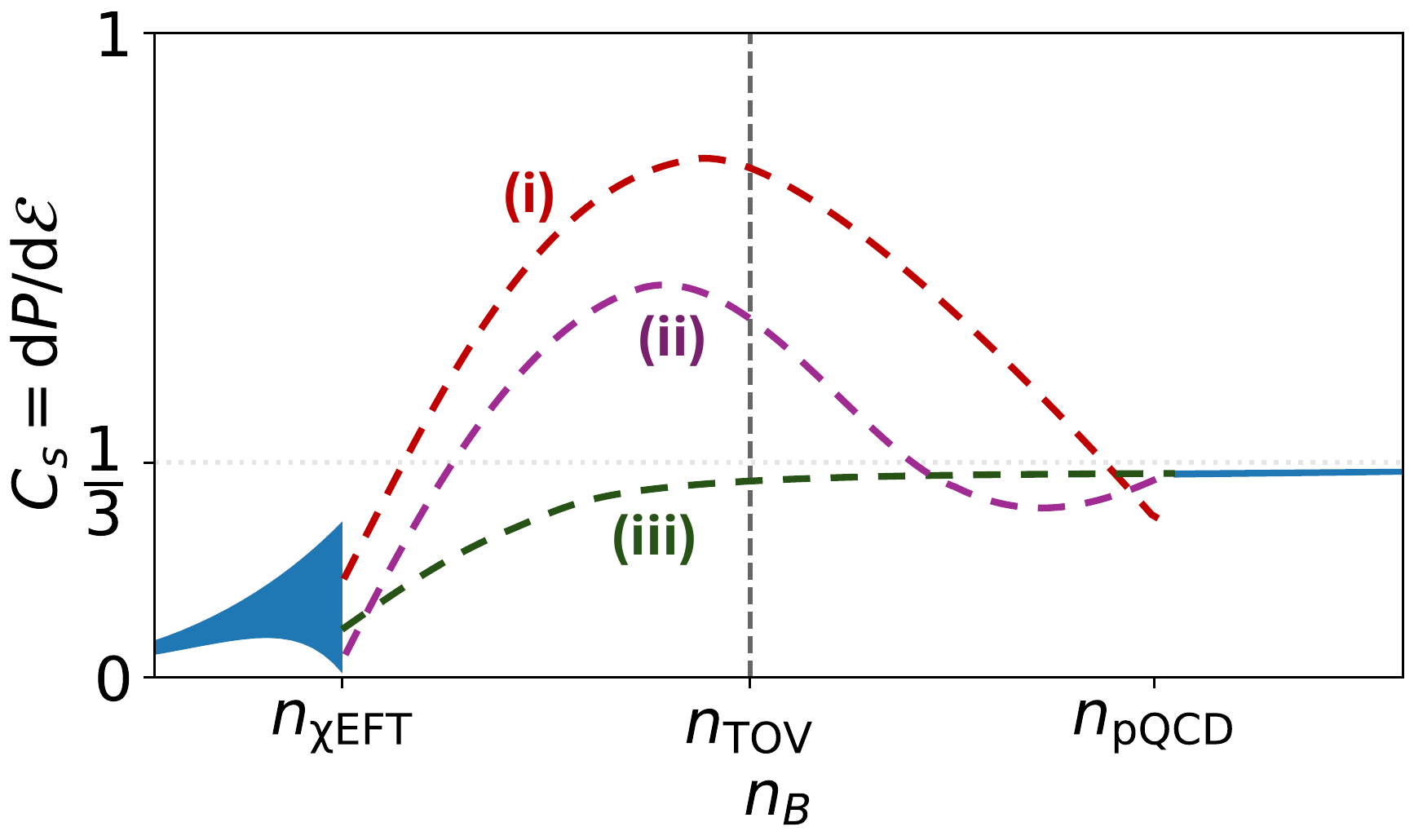}
	\caption{Examples of disfavored $C_s(n_B)$.
    (i): a monotonically decreasing $C_s^\mathrm{UD}$ above NS densities. This is severely constrained and almost ruled out, see \cref{tab:BsND};
    (ii): absence of a peak above $1/3$ following a trough.
    One peak inside NSs followed by a trough has been considered the minimal and standard picture of $C_s$ in cold QCD, and the present work shows that this scenario is likely incompatible with astrophysics and pQCD;
    (iii): a monotonic $C_s$ across cold dense QCD. This possibility is completely ruled out unless $X\simeq1$. A fully model independent evidence for this statement will be reported in an ensuing work.
	}\label{fig:goodex}
\end{figure}


\begin{table}[!hbt]
\centering
\setlength{\tabcolsep}{1em} 
{\renewcommand{\arraystretch}{1.5}
\begin{tabular}{ l|c c c | c c c c c @{} }
	\toprule
	astro data                   & $X=1.5$ &   $2$  &  $3$   & $X\sim U[1,4]$ & $U[1.5,4]$ & $U[2,3]$ & $U[2,4]$ & $\mathcal{N}(2.5, 0.5)$ \\ \midrule
	+J0740+6620                  & ${7.1}^{+0.8}_{-1.3}$ & $70^{+59}_{-45}$ & $\gg10^3$ & $10^{+2}_{-2}$        &  $70^{+25}_{-23}$ & $>10^3$ & $>10^3$  & $70^{+14}_{-12}$\\ 
    \hline
	+$M_\tov\lesssim2.2~M_\odot$ & ${0.7}^{+0.3}_{-0.5}$ & $15^{+11}_{-7}$  & $\gg10^3$ & ${2.4}^{+0.5}_{-0.3}$ &  $13^{+4}_{-5}$  & ${200}^{+180}_{-120}$ & ${280}^{+240}_{-150}$ & $13^{+3}_{-2}$\\ 
	+$M_\tov\gtrsim2.6~M_\odot$  & $>10^3$               & $\gg10^3$        & $\gg10^3$ & $160^{+33}_{-26}$     &  $>10^3$         & $\gg10^3$ & $\gg10^3$ &  $>10^3$ \\ 
    \bottomrule
\end{tabular}
}
\caption{Bayes factors for a non-decreasing $C_s$ between NS and pQCD densities. \Cref{eq:SCA} $\Delta P>\dpmean$ is the sole requirement for this scenario.
}
\label{tab:BsND}
\end{table}


\clearpage
\newpage

\input{npeaks.bbl}

\end{document}

%% file: npeaks.bbl
%

%% file: npeaks1.bbl
\begin{thebibliography}{119}%
\makeatletter
\providecommand \@ifxundefined [1]{%
 \@ifx{#1\undefined}
}%
\providecommand \@ifnum [1]{%
 \ifnum #1\expandafter \@firstoftwo
 \else \expandafter \@secondoftwo
 \fi
}%
\providecommand \@ifx [1]{%
 \ifx #1\expandafter \@firstoftwo
 \else \expandafter \@secondoftwo
 \fi
}%
\providecommand \natexlab [1]{#1}%
\providecommand \enquote  [1]{``#1''}%
\providecommand \bibnamefont  [1]{#1}%
\providecommand \bibfnamefont [1]{#1}%
\providecommand \citenamefont [1]{#1}%
\providecommand \href@noop [0]{\@secondoftwo}%
\providecommand \href [0]{\begingroup \@sanitize@url \@href}%
\providecommand \@href[1]{\@@startlink{#1}\@@href}%
\providecommand \@@href[1]{\endgroup#1\@@endlink}%
\providecommand \@sanitize@url [0]{\catcode `\\12\catcode `\$12\catcode
  `\&12\catcode `\#12\catcode `\^12\catcode `\_12\catcode `\%12\relax}%
\providecommand \@@startlink[1]{}%
\providecommand \@@endlink[0]{}%
\providecommand \url  [0]{\begingroup\@sanitize@url \@url }%
\providecommand \@url [1]{\endgroup\@href {#1}{\urlprefix }}%
\providecommand \urlprefix  [0]{URL }%
\providecommand \Eprint [0]{\href }%
\providecommand \doibase [0]{https://doi.org/}%
\providecommand \selectlanguage [0]{\@gobble}%
\providecommand \bibinfo  [0]{\@secondoftwo}%
\providecommand \bibfield  [0]{\@secondoftwo}%
\providecommand \translation [1]{[#1]}%
\providecommand \BibitemOpen [0]{}%
\providecommand \bibitemStop [0]{}%
\providecommand \bibitemNoStop [0]{.\EOS\space}%
\providecommand \EOS [0]{\spacefactor3000\relax}%
\providecommand \BibitemShut  [1]{\csname bibitem#1\endcsname}%
\let\auto@bib@innerbib\@empty
\bibitem [{\citenamefont {Troyer}\ and\ \citenamefont
  {Wiese}(2005)}]{Troyer:2004ge}%
  \BibitemOpen
  \bibfield  {author} {\bibinfo {author} {\bibfnamefont {M.}~\bibnamefont
  {Troyer}}\ and\ \bibinfo {author} {\bibfnamefont {U.-J.}\ \bibnamefont
  {Wiese}},\ }\bibfield  {title} {\bibinfo {title} {{Computational complexity
  and fundamental limitations to fermionic quantum Monte Carlo simulations}},\
  }\href {https://doi.org/10.1103/PhysRevLett.94.170201} {\bibfield  {journal}
  {\bibinfo  {journal} {Phys. Rev. Lett.}\ }\textbf {\bibinfo {volume} {94}},\
  \bibinfo {pages} {170201} (\bibinfo {year} {2005})},\ \Eprint
  {https://arxiv.org/abs/cond-mat/0408370} {arXiv:cond-mat/0408370}
  \BibitemShut {NoStop}%
\bibitem [{\citenamefont {de~Forcrand}(2009)}]{deForcrand:2009zkb}%
  \BibitemOpen
  \bibfield  {author} {\bibinfo {author} {\bibfnamefont {P.}~\bibnamefont
  {de~Forcrand}},\ }\bibfield  {title} {\bibinfo {title} {{Simulating QCD at
  finite density}},\ }\href {https://doi.org/10.22323/1.091.0010} {\bibfield
  {journal} {\bibinfo  {journal} {PoS}\ }\textbf {\bibinfo {volume}
  {LAT2009}},\ \bibinfo {pages} {010} (\bibinfo {year} {2009})},\ \Eprint
  {https://arxiv.org/abs/1005.0539} {arXiv:1005.0539 [hep-lat]} \BibitemShut
  {NoStop}%
\bibitem [{\citenamefont {Kaplan}(2009)}]{Kaplan:2009yg}%
  \BibitemOpen
  \bibfield  {author} {\bibinfo {author} {\bibfnamefont {D.~B.}\ \bibnamefont
  {Kaplan}},\ }\bibfield  {title} {\bibinfo {title} {{Chiral Symmetry and
  Lattice Fermions}},\ }in\ \href@noop {} {\emph {\bibinfo {booktitle} {{Les
  Houches Summer School: Session 93: Modern perspectives in lattice QCD:
  Quantum field theory and high performance computing}}}}\ (\bibinfo {year}
  {2009})\ \Eprint {https://arxiv.org/abs/0912.2560} {arXiv:0912.2560
  [hep-lat]} \BibitemShut {NoStop}%
\bibitem [{\citenamefont {{Lattimer}}\ and\ \citenamefont
  {{Prakash}}(2010)}]{LattimerPrakash:2010}%
  \BibitemOpen
  \bibfield  {author} {\bibinfo {author} {\bibfnamefont {J.~M.}\ \bibnamefont
  {{Lattimer}}}\ and\ \bibinfo {author} {\bibfnamefont {M.}~\bibnamefont
  {{Prakash}}},\ }\bibfield  {title} {\bibinfo {title} {{What a Two Solar Mass
  Neutron Star Really Means}},\ }\href@noop {} {\bibfield  {journal} {\bibinfo
  {journal} {ArXiv e-prints}\ } (\bibinfo {year} {2010})},\ \Eprint
  {https://arxiv.org/abs/1012.3208} {arXiv:1012.3208 [astro-ph.SR]}
  \BibitemShut {NoStop}%
\bibitem [{\citenamefont {Watts}\ \emph {et~al.}(2016)\citenamefont {Watts}
  \emph {et~al.}}]{Watts:2016uzu}%
  \BibitemOpen
  \bibfield  {author} {\bibinfo {author} {\bibfnamefont {A.~L.}\ \bibnamefont
  {Watts}} \emph {et~al.},\ }\bibfield  {title} {\bibinfo {title} {{Colloquium
  : Measuring the neutron star equation of state using x-ray timing}},\ }\href
  {https://doi.org/10.1103/RevModPhys.88.021001} {\bibfield  {journal}
  {\bibinfo  {journal} {Rev. Mod. Phys.}\ }\textbf {\bibinfo {volume} {88}},\
  \bibinfo {pages} {021001} (\bibinfo {year} {2016})},\ \Eprint
  {https://arxiv.org/abs/1602.01081} {arXiv:1602.01081 [astro-ph.HE]}
  \BibitemShut {NoStop}%
\bibitem [{\citenamefont {et~al. LIGO Scientific~Collaboration}\ and\
  \citenamefont {Collaboration}(2017)}]{Abbott:2017aa}%
  \BibitemOpen
  \bibfield  {author} {\bibinfo {author} {\bibfnamefont {B.~P.~A.}\
  \bibnamefont {et~al. LIGO Scientific~Collaboration}}\ and\ \bibinfo {author}
  {\bibfnamefont {V.}~\bibnamefont {Collaboration}},\ }\bibfield  {title}
  {\bibinfo {title} {Gw170817: Observation of gravitational waves from a binary
  neutron star inspiral},\ }\bibfield  {journal} {\bibinfo  {journal} {Physical
  Review Letters}\ }\textbf {\bibinfo {volume} {119}},\ \href
  {https://doi.org/10.1103/PhysRevLett.119.161101}
  {10.1103/PhysRevLett.119.161101} (\bibinfo {year} {2017})\BibitemShut
  {NoStop}%
\bibitem [{\citenamefont {Abbott}\ \emph {et~al.}(2018)\citenamefont {Abbott}
  \emph {et~al.}}]{Abbott:2018exr}%
  \BibitemOpen
  \bibfield  {author} {\bibinfo {author} {\bibfnamefont {B.~P.}\ \bibnamefont
  {Abbott}} \emph {et~al.} (\bibinfo {collaboration} {LIGO Scientific,
  Virgo}),\ }\bibfield  {title} {\bibinfo {title} {{GW170817: Measurements of
  neutron star radii and equation of state}},\ }\href
  {https://doi.org/10.1103/PhysRevLett.121.161101} {\bibfield  {journal}
  {\bibinfo  {journal} {Phys. Rev. Lett.}\ }\textbf {\bibinfo {volume} {121}},\
  \bibinfo {pages} {161101} (\bibinfo {year} {2018})},\ \Eprint
  {https://arxiv.org/abs/1805.11581} {arXiv:1805.11581 [gr-qc]} \BibitemShut
  {NoStop}%
\bibitem [{\citenamefont {Tews}\ \emph
  {et~al.}(2018{\natexlab{a}})\citenamefont {Tews}, \citenamefont {Margueron},\
  and\ \citenamefont {Reddy}}]{Tews:2018aa}%
  \BibitemOpen
  \bibfield  {author} {\bibinfo {author} {\bibfnamefont {I.}~\bibnamefont
  {Tews}}, \bibinfo {author} {\bibfnamefont {J.}~\bibnamefont {Margueron}},\
  and\ \bibinfo {author} {\bibfnamefont {S.}~\bibnamefont {Reddy}},\ }\bibfield
   {title} {\bibinfo {title} {A critical examination of constraints on the
  equation of state of dense matter obtained from gw170817},\ }\href
  {https://arxiv.org/pdf/1804.02783} {\bibfield  {journal} {\bibinfo  {journal}
  {Phys. Rev. C}\ }\textbf {\bibinfo {volume} {98}},\ \bibinfo {pages} {045804}
  (\bibinfo {year} {2018}{\natexlab{a}})},\ \Eprint
  {https://arxiv.org/abs/1804.02783} {1804.02783} \BibitemShut {NoStop}%
\bibitem [{\citenamefont {De}\ \emph {et~al.}(2018)\citenamefont {De},
  \citenamefont {Finstad}, \citenamefont {Lattimer}, \citenamefont {Brown},
  \citenamefont {Berger},\ and\ \citenamefont {Biwer}}]{De:2018uhw}%
  \BibitemOpen
  \bibfield  {author} {\bibinfo {author} {\bibfnamefont {S.}~\bibnamefont
  {De}}, \bibinfo {author} {\bibfnamefont {D.}~\bibnamefont {Finstad}},
  \bibinfo {author} {\bibfnamefont {J.~M.}\ \bibnamefont {Lattimer}}, \bibinfo
  {author} {\bibfnamefont {D.~A.}\ \bibnamefont {Brown}}, \bibinfo {author}
  {\bibfnamefont {E.}~\bibnamefont {Berger}},\ and\ \bibinfo {author}
  {\bibfnamefont {C.~M.}\ \bibnamefont {Biwer}},\ }\bibfield  {title} {\bibinfo
  {title} {Constraining the nuclear equation of state with {GW170817}},\ }\href
  {https://doi.org/10.1103/PhysRevLett.121.091102} {\bibfield  {journal}
  {\bibinfo  {journal} {Phys. Rev. Lett.}\ }\textbf {\bibinfo {volume} {121}},\
  \bibinfo {pages} {091102} (\bibinfo {year} {2018})},\ \Eprint
  {https://arxiv.org/abs/1804.08583} {arXiv:1804.08583 [astro-ph.HE]}
  \BibitemShut {NoStop}%
\bibitem [{\citenamefont {Radice}\ \emph
  {et~al.}(2018{\natexlab{a}})\citenamefont {Radice}, \citenamefont {Perego},
  \citenamefont {Zappa},\ and\ \citenamefont {Bernuzzi}}]{Radice:2017lry}%
  \BibitemOpen
  \bibfield  {author} {\bibinfo {author} {\bibfnamefont {D.}~\bibnamefont
  {Radice}}, \bibinfo {author} {\bibfnamefont {A.}~\bibnamefont {Perego}},
  \bibinfo {author} {\bibfnamefont {F.}~\bibnamefont {Zappa}},\ and\ \bibinfo
  {author} {\bibfnamefont {S.}~\bibnamefont {Bernuzzi}},\ }\bibfield  {title}
  {\bibinfo {title} {{GW170817: Joint Constraint on the Neutron Star Equation
  of State from Multimessenger Observations}},\ }\href
  {https://doi.org/10.3847/2041-8213/aaa402} {\bibfield  {journal} {\bibinfo
  {journal} {Astrophys. J.}\ }\textbf {\bibinfo {volume} {852}},\ \bibinfo
  {pages} {L29} (\bibinfo {year} {2018}{\natexlab{a}})},\ \Eprint
  {https://arxiv.org/abs/1711.03647} {arXiv:1711.03647 [astro-ph.HE]}
  \BibitemShut {NoStop}%
\bibitem [{\citenamefont {Capano}\ \emph {et~al.}(2020)\citenamefont {Capano},
  \citenamefont {Tews}, \citenamefont {Brown}, \citenamefont {Margalit},
  \citenamefont {De}, \citenamefont {Kumar}, \citenamefont {Brown},
  \citenamefont {Krishnan},\ and\ \citenamefont {Reddy}}]{Capano:2019eae}%
  \BibitemOpen
  \bibfield  {author} {\bibinfo {author} {\bibfnamefont {C.~D.}\ \bibnamefont
  {Capano}}, \bibinfo {author} {\bibfnamefont {I.}~\bibnamefont {Tews}},
  \bibinfo {author} {\bibfnamefont {S.~M.}\ \bibnamefont {Brown}}, \bibinfo
  {author} {\bibfnamefont {B.}~\bibnamefont {Margalit}}, \bibinfo {author}
  {\bibfnamefont {S.}~\bibnamefont {De}}, \bibinfo {author} {\bibfnamefont
  {S.}~\bibnamefont {Kumar}}, \bibinfo {author} {\bibfnamefont {D.~A.}\
  \bibnamefont {Brown}}, \bibinfo {author} {\bibfnamefont {B.}~\bibnamefont
  {Krishnan}},\ and\ \bibinfo {author} {\bibfnamefont {S.}~\bibnamefont
  {Reddy}},\ }\bibfield  {title} {\bibinfo {title} {{Stringent constraints on
  neutron-star radii from multimessenger observations and nuclear theory}},\
  }\href {https://doi.org/10.1038/s41550-020-1014-6} {\bibfield  {journal}
  {\bibinfo  {journal} {Nature Astron.}\ }\textbf {\bibinfo {volume} {4}},\
  \bibinfo {pages} {625} (\bibinfo {year} {2020})},\ \Eprint
  {https://arxiv.org/abs/1908.10352} {arXiv:1908.10352 [astro-ph.HE]}
  \BibitemShut {NoStop}%
\bibitem [{\citenamefont {Miller}\ \emph {et~al.}(2019)\citenamefont {Miller}
  \emph {et~al.}}]{Miller:2019cac}%
  \BibitemOpen
  \bibfield  {author} {\bibinfo {author} {\bibfnamefont {M.~C.}\ \bibnamefont
  {Miller}} \emph {et~al.},\ }\bibfield  {title} {\bibinfo {title} {{PSR
  J0030+0451 Mass and Radius from $NICER$ Data and Implications for the
  Properties of Neutron Star Matter}},\ }\href
  {https://doi.org/10.3847/2041-8213/ab50c5} {\bibfield  {journal} {\bibinfo
  {journal} {Astrophys. J. Lett.}\ }\textbf {\bibinfo {volume} {887}},\
  \bibinfo {pages} {L24} (\bibinfo {year} {2019})},\ \Eprint
  {https://arxiv.org/abs/1912.05705} {arXiv:1912.05705 [astro-ph.HE]}
  \BibitemShut {NoStop}%
\bibitem [{\citenamefont {Legred}\ \emph {et~al.}(2021)\citenamefont {Legred},
  \citenamefont {Chatziioannou}, \citenamefont {Essick}, \citenamefont {Han},\
  and\ \citenamefont {Landry}}]{Legred:2021hdx}%
  \BibitemOpen
  \bibfield  {author} {\bibinfo {author} {\bibfnamefont {I.}~\bibnamefont
  {Legred}}, \bibinfo {author} {\bibfnamefont {K.}~\bibnamefont
  {Chatziioannou}}, \bibinfo {author} {\bibfnamefont {R.}~\bibnamefont
  {Essick}}, \bibinfo {author} {\bibfnamefont {S.}~\bibnamefont {Han}},\ and\
  \bibinfo {author} {\bibfnamefont {P.}~\bibnamefont {Landry}},\ }\bibfield
  {title} {\bibinfo {title} {{Impact of the PSR J0740+6620 radius constraint on
  the properties of high-density matter}},\ }\href
  {https://doi.org/10.1103/PhysRevD.104.063003} {\bibfield  {journal} {\bibinfo
   {journal} {Phys. Rev. D}\ }\textbf {\bibinfo {volume} {104}},\ \bibinfo
  {pages} {063003} (\bibinfo {year} {2021})},\ \Eprint
  {https://arxiv.org/abs/2106.05313} {arXiv:2106.05313 [astro-ph.HE]}
  \BibitemShut {NoStop}%
\bibitem [{\citenamefont {Riley}\ \emph {et~al.}(2019)\citenamefont {Riley}
  \emph {et~al.}}]{Riley:2019yda}%
  \BibitemOpen
  \bibfield  {author} {\bibinfo {author} {\bibfnamefont {T.~E.}\ \bibnamefont
  {Riley}} \emph {et~al.},\ }\bibfield  {title} {\bibinfo {title} {{A $NICER$
  View of PSR J0030+0451: Millisecond Pulsar Parameter Estimation}},\ }\href
  {https://doi.org/10.3847/2041-8213/ab481c} {\bibfield  {journal} {\bibinfo
  {journal} {Astrophys. J.}\ }\textbf {\bibinfo {volume} {887}},\ \bibinfo
  {pages} {L21} (\bibinfo {year} {2019})},\ \Eprint
  {https://arxiv.org/abs/1912.05702} {arXiv:1912.05702 [astro-ph.HE]}
  \BibitemShut {NoStop}%
\bibitem [{\citenamefont {Miller}\ \emph {et~al.}(2021)\citenamefont {Miller}
  \emph {et~al.}}]{Miller:2021qha}%
  \BibitemOpen
  \bibfield  {author} {\bibinfo {author} {\bibfnamefont {M.~C.}\ \bibnamefont
  {Miller}} \emph {et~al.},\ }\bibfield  {title} {\bibinfo {title} {{The Radius
  of PSR J0740+6620 from NICER and XMM-Newton Data}},\ }\href@noop {} {\
  (\bibinfo {year} {2021})},\ \Eprint {https://arxiv.org/abs/2105.06979}
  {arXiv:2105.06979 [astro-ph.HE]} \BibitemShut {NoStop}%
\bibitem [{\citenamefont {Riley}\ \emph {et~al.}(2021)\citenamefont {Riley}
  \emph {et~al.}}]{Riley:2021pdl}%
  \BibitemOpen
  \bibfield  {author} {\bibinfo {author} {\bibfnamefont {T.~E.}\ \bibnamefont
  {Riley}} \emph {et~al.},\ }\bibfield  {title} {\bibinfo {title} {{A NICER
  View of the Massive Pulsar PSR J0740+6620 Informed by Radio Timing and
  XMM-Newton Spectroscopy}},\ }\href {https://doi.org/10.3847/2041-8213/ac0a81}
  {\bibfield  {journal} {\bibinfo  {journal} {Astrophys. J. Lett.}\ }\textbf
  {\bibinfo {volume} {918}},\ \bibinfo {pages} {L27} (\bibinfo {year}
  {2021})},\ \Eprint {https://arxiv.org/abs/2105.06980} {arXiv:2105.06980
  [astro-ph.HE]} \BibitemShut {NoStop}%
\bibitem [{\citenamefont {Salmi}\ \emph {et~al.}(2024)\citenamefont {Salmi}
  \emph {et~al.}}]{Salmi:2024aum}%
  \BibitemOpen
  \bibfield  {author} {\bibinfo {author} {\bibfnamefont {T.}~\bibnamefont
  {Salmi}} \emph {et~al.},\ }\bibfield  {title} {\bibinfo {title} {{The Radius
  of the High-mass Pulsar PSR J0740+6620 with 3.6 yr of NICER Data}},\ }\href
  {https://doi.org/10.3847/1538-4357/ad5f1f} {\bibfield  {journal} {\bibinfo
  {journal} {Astrophys. J.}\ }\textbf {\bibinfo {volume} {974}},\ \bibinfo
  {pages} {294} (\bibinfo {year} {2024})},\ \Eprint
  {https://arxiv.org/abs/2406.14466} {arXiv:2406.14466 [astro-ph.HE]}
  \BibitemShut {NoStop}%
\bibitem [{\citenamefont {Dittmann}\ \emph {et~al.}(2024)\citenamefont
  {Dittmann} \emph {et~al.}}]{Dittmann:2024mbo}%
  \BibitemOpen
  \bibfield  {author} {\bibinfo {author} {\bibfnamefont {A.~J.}\ \bibnamefont
  {Dittmann}} \emph {et~al.},\ }\bibfield  {title} {\bibinfo {title} {{A More
  Precise Measurement of the Radius of PSR J0740+6620 Using Updated NICER
  Data}},\ }\href {https://doi.org/10.3847/1538-4357/ad5f1e} {\bibfield
  {journal} {\bibinfo  {journal} {Astrophys. J.}\ }\textbf {\bibinfo {volume}
  {974}},\ \bibinfo {pages} {295} (\bibinfo {year} {2024})},\ \Eprint
  {https://arxiv.org/abs/2406.14467} {arXiv:2406.14467 [astro-ph.HE]}
  \BibitemShut {NoStop}%
\bibitem [{\citenamefont {Alford}\ \emph {et~al.}(1998)\citenamefont {Alford},
  \citenamefont {Rajagopal},\ and\ \citenamefont {Wilczek}}]{Alford:1997zt}%
  \BibitemOpen
  \bibfield  {author} {\bibinfo {author} {\bibfnamefont {M.~G.}\ \bibnamefont
  {Alford}}, \bibinfo {author} {\bibfnamefont {K.}~\bibnamefont {Rajagopal}},\
  and\ \bibinfo {author} {\bibfnamefont {F.}~\bibnamefont {Wilczek}},\
  }\bibfield  {title} {\bibinfo {title} {{QCD at finite baryon density: Nucleon
  droplets and color superconductivity}},\ }\href
  {https://doi.org/10.1016/S0370-2693(98)00051-3} {\bibfield  {journal}
  {\bibinfo  {journal} {Phys. Lett. B}\ }\textbf {\bibinfo {volume} {422}},\
  \bibinfo {pages} {247} (\bibinfo {year} {1998})},\ \Eprint
  {https://arxiv.org/abs/hep-ph/9711395} {arXiv:hep-ph/9711395} \BibitemShut
  {NoStop}%
\bibitem [{\citenamefont {Berges}\ and\ \citenamefont
  {Rajagopal}(1999)}]{Berges:1998rc}%
  \BibitemOpen
  \bibfield  {author} {\bibinfo {author} {\bibfnamefont {J.}~\bibnamefont
  {Berges}}\ and\ \bibinfo {author} {\bibfnamefont {K.}~\bibnamefont
  {Rajagopal}},\ }\bibfield  {title} {\bibinfo {title} {{Color
  superconductivity and chiral symmetry restoration at nonzero baryon density
  and temperature}},\ }\href {https://doi.org/10.1016/S0550-3213(98)00620-8}
  {\bibfield  {journal} {\bibinfo  {journal} {Nucl. Phys. B}\ }\textbf
  {\bibinfo {volume} {538}},\ \bibinfo {pages} {215} (\bibinfo {year}
  {1999})},\ \Eprint {https://arxiv.org/abs/hep-ph/9804233}
  {arXiv:hep-ph/9804233} \BibitemShut {NoStop}%
\bibitem [{\citenamefont {Carter}\ and\ \citenamefont
  {Diakonov}(1999)}]{Carter:1998ji}%
  \BibitemOpen
  \bibfield  {author} {\bibinfo {author} {\bibfnamefont {G.~W.}\ \bibnamefont
  {Carter}}\ and\ \bibinfo {author} {\bibfnamefont {D.}~\bibnamefont
  {Diakonov}},\ }\bibfield  {title} {\bibinfo {title} {{Light quarks in the
  instanton vacuum at finite baryon density}},\ }\href
  {https://doi.org/10.1103/PhysRevD.60.016004} {\bibfield  {journal} {\bibinfo
  {journal} {Phys. Rev. D}\ }\textbf {\bibinfo {volume} {60}},\ \bibinfo
  {pages} {016004} (\bibinfo {year} {1999})},\ \Eprint
  {https://arxiv.org/abs/hep-ph/9812445} {arXiv:hep-ph/9812445} \BibitemShut
  {NoStop}%
\bibitem [{\citenamefont {Pisarski}\ and\ \citenamefont
  {Rischke}(2000)}]{Pisarski:1999bf}%
  \BibitemOpen
  \bibfield  {author} {\bibinfo {author} {\bibfnamefont {R.~D.}\ \bibnamefont
  {Pisarski}}\ and\ \bibinfo {author} {\bibfnamefont {D.~H.}\ \bibnamefont
  {Rischke}},\ }\bibfield  {title} {\bibinfo {title} {{Gaps and critical
  temperature for color superconductivity}},\ }\href
  {https://doi.org/10.1103/PhysRevD.61.051501} {\bibfield  {journal} {\bibinfo
  {journal} {Phys. Rev. D}\ }\textbf {\bibinfo {volume} {61}},\ \bibinfo
  {pages} {051501} (\bibinfo {year} {2000})},\ \Eprint
  {https://arxiv.org/abs/nucl-th/9907041} {arXiv:nucl-th/9907041} \BibitemShut
  {NoStop}%
\bibitem [{\citenamefont {Son}(1999)}]{Son:1998uk}%
  \BibitemOpen
  \bibfield  {author} {\bibinfo {author} {\bibfnamefont {D.~T.}\ \bibnamefont
  {Son}},\ }\bibfield  {title} {\bibinfo {title} {{Superconductivity by long
  range color magnetic interaction in high density quark matter}},\ }\href
  {https://doi.org/10.1103/PhysRevD.59.094019} {\bibfield  {journal} {\bibinfo
  {journal} {Phys. Rev. D}\ }\textbf {\bibinfo {volume} {59}},\ \bibinfo
  {pages} {094019} (\bibinfo {year} {1999})},\ \Eprint
  {https://arxiv.org/abs/hep-ph/9812287} {arXiv:hep-ph/9812287} \BibitemShut
  {NoStop}%
\bibitem [{\citenamefont {Alford}\ \emph {et~al.}(1999)\citenamefont {Alford},
  \citenamefont {Rajagopal},\ and\ \citenamefont {Wilczek}}]{Alford:1998mk}%
  \BibitemOpen
  \bibfield  {author} {\bibinfo {author} {\bibfnamefont {M.~G.}\ \bibnamefont
  {Alford}}, \bibinfo {author} {\bibfnamefont {K.}~\bibnamefont {Rajagopal}},\
  and\ \bibinfo {author} {\bibfnamefont {F.}~\bibnamefont {Wilczek}},\
  }\bibfield  {title} {\bibinfo {title} {{Color flavor locking and chiral
  symmetry breaking in high density QCD}},\ }\href
  {https://doi.org/10.1016/S0550-3213(98)00668-3} {\bibfield  {journal}
  {\bibinfo  {journal} {Nucl. Phys. B}\ }\textbf {\bibinfo {volume} {537}},\
  \bibinfo {pages} {443} (\bibinfo {year} {1999})},\ \Eprint
  {https://arxiv.org/abs/hep-ph/9804403} {arXiv:hep-ph/9804403} \BibitemShut
  {NoStop}%
\bibitem [{\citenamefont {Son}\ and\ \citenamefont
  {Stephanov}(2000)}]{Son:1999cm}%
  \BibitemOpen
  \bibfield  {author} {\bibinfo {author} {\bibfnamefont {D.~T.}\ \bibnamefont
  {Son}}\ and\ \bibinfo {author} {\bibfnamefont {M.~A.}\ \bibnamefont
  {Stephanov}},\ }\bibfield  {title} {\bibinfo {title} {{Inverse meson mass
  ordering in color flavor locking phase of high density QCD}},\ }\href
  {https://doi.org/10.1103/PhysRevD.61.074012} {\bibfield  {journal} {\bibinfo
  {journal} {Phys. Rev. D}\ }\textbf {\bibinfo {volume} {61}},\ \bibinfo
  {pages} {074012} (\bibinfo {year} {2000})},\ \Eprint
  {https://arxiv.org/abs/hep-ph/9910491} {arXiv:hep-ph/9910491} \BibitemShut
  {NoStop}%
\bibitem [{\citenamefont {Casalbuoni}\ and\ \citenamefont
  {Gatto}(1999)}]{Casalbuoni:1999wu}%
  \BibitemOpen
  \bibfield  {author} {\bibinfo {author} {\bibfnamefont {R.}~\bibnamefont
  {Casalbuoni}}\ and\ \bibinfo {author} {\bibfnamefont {R.}~\bibnamefont
  {Gatto}},\ }\bibfield  {title} {\bibinfo {title} {{Effective theory for color
  flavor locking in high density QCD}},\ }\href
  {https://doi.org/10.1016/S0370-2693(99)01032-1} {\bibfield  {journal}
  {\bibinfo  {journal} {Phys. Lett. B}\ }\textbf {\bibinfo {volume} {464}},\
  \bibinfo {pages} {111} (\bibinfo {year} {1999})},\ \Eprint
  {https://arxiv.org/abs/hep-ph/9908227} {arXiv:hep-ph/9908227} \BibitemShut
  {NoStop}%
\bibitem [{\citenamefont {Hong}\ \emph {et~al.}(2000)\citenamefont {Hong},
  \citenamefont {Lee},\ and\ \citenamefont {Min}}]{Hong:1999ei}%
  \BibitemOpen
  \bibfield  {author} {\bibinfo {author} {\bibfnamefont {D.~K.}\ \bibnamefont
  {Hong}}, \bibinfo {author} {\bibfnamefont {T.}~\bibnamefont {Lee}},\ and\
  \bibinfo {author} {\bibfnamefont {D.-P.}\ \bibnamefont {Min}},\ }\bibfield
  {title} {\bibinfo {title} {{Meson mass at large baryon chemical potential in
  dense QCD}},\ }\href {https://doi.org/10.1016/S0370-2693(00)00188-X}
  {\bibfield  {journal} {\bibinfo  {journal} {Phys. Lett. B}\ }\textbf
  {\bibinfo {volume} {477}},\ \bibinfo {pages} {137} (\bibinfo {year}
  {2000})},\ \Eprint {https://arxiv.org/abs/hep-ph/9912531}
  {arXiv:hep-ph/9912531} \BibitemShut {NoStop}%
\bibitem [{\citenamefont {Sch\"afer}\ and\ \citenamefont
  {Wilczek}(1999{\natexlab{a}})}]{Schafer:1999jg}%
  \BibitemOpen
  \bibfield  {author} {\bibinfo {author} {\bibfnamefont {T.}~\bibnamefont
  {Sch\"afer}}\ and\ \bibinfo {author} {\bibfnamefont {F.}~\bibnamefont
  {Wilczek}},\ }\bibfield  {title} {\bibinfo {title} {{Superconductivity from
  perturbative one gluon exchange in high density quark matter}},\ }\href
  {https://doi.org/10.1103/PhysRevD.60.114033} {\bibfield  {journal} {\bibinfo
  {journal} {Phys. Rev. D}\ }\textbf {\bibinfo {volume} {60}},\ \bibinfo
  {pages} {114033} (\bibinfo {year} {1999}{\natexlab{a}})},\ \Eprint
  {https://arxiv.org/abs/hep-ph/9906512} {arXiv:hep-ph/9906512} \BibitemShut
  {NoStop}%
\bibitem [{\citenamefont {Rajagopal}\ and\ \citenamefont
  {Wilczek}(2000)}]{Rajagopal:2000wf}%
  \BibitemOpen
  \bibfield  {author} {\bibinfo {author} {\bibfnamefont {K.}~\bibnamefont
  {Rajagopal}}\ and\ \bibinfo {author} {\bibfnamefont {F.}~\bibnamefont
  {Wilczek}},\ }\bibinfo {title} {{The Condensed matter physics of QCD}},\ in\
  \href {https://doi.org/10.1142/9789812810458_0043} {\emph {\bibinfo
  {booktitle} {{At the frontier of particle physics. Handbook of QCD. Vol.
  1-3}}}},\ \bibinfo {editor} {edited by\ \bibinfo {editor} {\bibfnamefont
  {M.}~\bibnamefont {Shifman}}\ and\ \bibinfo {editor} {\bibfnamefont
  {B.}~\bibnamefont {Ioffe}}}\ (\bibinfo {year} {2000})\ pp.\ \bibinfo {pages}
  {2061--2151},\ \Eprint {https://arxiv.org/abs/hep-ph/0011333}
  {arXiv:hep-ph/0011333} \BibitemShut {NoStop}%
\bibitem [{\citenamefont {Alford}\ \emph {et~al.}(2008)\citenamefont {Alford},
  \citenamefont {Schmitt}, \citenamefont {Rajagopal},\ and\ \citenamefont
  {Sch\"afer}}]{Alford:2007xm}%
  \BibitemOpen
  \bibfield  {author} {\bibinfo {author} {\bibfnamefont {M.~G.}\ \bibnamefont
  {Alford}}, \bibinfo {author} {\bibfnamefont {A.}~\bibnamefont {Schmitt}},
  \bibinfo {author} {\bibfnamefont {K.}~\bibnamefont {Rajagopal}},\ and\
  \bibinfo {author} {\bibfnamefont {T.}~\bibnamefont {Sch\"afer}},\ }\bibfield
  {title} {\bibinfo {title} {{Color superconductivity in dense quark matter}},\
  }\href {https://doi.org/10.1103/RevModPhys.80.1455} {\bibfield  {journal}
  {\bibinfo  {journal} {Rev. Mod. Phys.}\ }\textbf {\bibinfo {volume} {80}},\
  \bibinfo {pages} {1455} (\bibinfo {year} {2008})},\ \Eprint
  {https://arxiv.org/abs/0709.4635} {arXiv:0709.4635 [hep-ph]} \BibitemShut
  {NoStop}%
\bibitem [{\citenamefont {Braun}\ and\ \citenamefont
  {Schallmo}(2022)}]{Braun:2021uua}%
  \BibitemOpen
  \bibfield  {author} {\bibinfo {author} {\bibfnamefont {J.}~\bibnamefont
  {Braun}}\ and\ \bibinfo {author} {\bibfnamefont {B.}~\bibnamefont
  {Schallmo}},\ }\bibfield  {title} {\bibinfo {title} {{From quarks and gluons
  to color superconductivity at supranuclear densities}},\ }\href
  {https://doi.org/10.1103/PhysRevD.105.036003} {\bibfield  {journal} {\bibinfo
   {journal} {Phys. Rev. D}\ }\textbf {\bibinfo {volume} {105}},\ \bibinfo
  {pages} {036003} (\bibinfo {year} {2022})},\ \Eprint
  {https://arxiv.org/abs/2106.04198} {arXiv:2106.04198 [hep-ph]} \BibitemShut
  {NoStop}%
\bibitem [{\citenamefont {Braun}\ \emph {et~al.}(2024)\citenamefont {Braun},
  \citenamefont {Gei\ss{}el},\ and\ \citenamefont {Schallmo}}]{Braun:2022jme}%
  \BibitemOpen
  \bibfield  {author} {\bibinfo {author} {\bibfnamefont {J.}~\bibnamefont
  {Braun}}, \bibinfo {author} {\bibfnamefont {A.}~\bibnamefont {Gei\ss{}el}},\
  and\ \bibinfo {author} {\bibfnamefont {B.}~\bibnamefont {Schallmo}},\
  }\bibfield  {title} {\bibinfo {title} {{Speed of sound in dense
  strong-interaction matter}},\ }\href
  {https://doi.org/10.21468/SciPostPhysCore.7.2.015} {\bibfield  {journal}
  {\bibinfo  {journal} {SciPost Phys. Core}\ }\textbf {\bibinfo {volume} {7}},\
  \bibinfo {pages} {015} (\bibinfo {year} {2024})},\ \Eprint
  {https://arxiv.org/abs/2206.06328} {arXiv:2206.06328 [nucl-th]} \BibitemShut
  {NoStop}%
\bibitem [{\citenamefont {Gei\ss{}el}\ \emph {et~al.}(2024)\citenamefont
  {Gei\ss{}el}, \citenamefont {Gorda},\ and\ \citenamefont
  {Braun}}]{Geissel:2024nmx}%
  \BibitemOpen
  \bibfield  {author} {\bibinfo {author} {\bibfnamefont {A.}~\bibnamefont
  {Gei\ss{}el}}, \bibinfo {author} {\bibfnamefont {T.}~\bibnamefont {Gorda}},\
  and\ \bibinfo {author} {\bibfnamefont {J.}~\bibnamefont {Braun}},\ }\bibfield
   {title} {\bibinfo {title} {{Pressure and speed of sound in two-flavor
  color-superconducting quark matter at next-to-leading order}},\ }\href
  {https://doi.org/10.1103/PhysRevD.110.014034} {\bibfield  {journal} {\bibinfo
   {journal} {Phys. Rev. D}\ }\textbf {\bibinfo {volume} {110}},\ \bibinfo
  {pages} {014034} (\bibinfo {year} {2024})},\ \Eprint
  {https://arxiv.org/abs/2403.18010} {arXiv:2403.18010 [hep-ph]} \BibitemShut
  {NoStop}%
\bibitem [{\citenamefont {Abbott}\ \emph {et~al.}(2024)\citenamefont {Abbott},
  \citenamefont {Detmold}, \citenamefont {Illa}, \citenamefont {Parre\~no},
  \citenamefont {Perry}, \citenamefont {Romero-L\'opez}, \citenamefont
  {Shanahan},\ and\ \citenamefont {Wagman}}]{Abbott:2024vhj}%
  \BibitemOpen
  \bibfield  {author} {\bibinfo {author} {\bibfnamefont {R.}~\bibnamefont
  {Abbott}}, \bibinfo {author} {\bibfnamefont {W.}~\bibnamefont {Detmold}},
  \bibinfo {author} {\bibfnamefont {M.}~\bibnamefont {Illa}}, \bibinfo {author}
  {\bibfnamefont {A.}~\bibnamefont {Parre\~no}}, \bibinfo {author}
  {\bibfnamefont {R.~J.}\ \bibnamefont {Perry}}, \bibinfo {author}
  {\bibfnamefont {F.}~\bibnamefont {Romero-L\'opez}}, \bibinfo {author}
  {\bibfnamefont {P.~E.}\ \bibnamefont {Shanahan}},\ and\ \bibinfo {author}
  {\bibfnamefont {M.~L.}\ \bibnamefont {Wagman}},\ }\bibfield  {title}
  {\bibinfo {title} {{QCD constraints on isospin-dense matter and the nuclear
  equation of state}},\ }\href@noop {} {\  (\bibinfo {year} {2024})},\ \Eprint
  {https://arxiv.org/abs/2406.09273} {arXiv:2406.09273 [hep-lat]} \BibitemShut
  {NoStop}%
\bibitem [{\citenamefont {Fujimoto}(2024)}]{Fujimoto:2024pcd}%
  \BibitemOpen
  \bibfield  {author} {\bibinfo {author} {\bibfnamefont {Y.}~\bibnamefont
  {Fujimoto}},\ }\bibfield  {title} {\bibinfo {title} {{Interplay between the
  weak-coupling results and the lattice data in dense QCD}},\ }\href@noop {} {\
   (\bibinfo {year} {2024})},\ \Eprint {https://arxiv.org/abs/2408.12514}
  {arXiv:2408.12514 [hep-ph]} \BibitemShut {NoStop}%
\bibitem [{\citenamefont {Tolman}(1939)}]{Tolman:1939jz}%
  \BibitemOpen
  \bibfield  {author} {\bibinfo {author} {\bibfnamefont {R.~C.}\ \bibnamefont
  {Tolman}},\ }\bibfield  {title} {\bibinfo {title} {{Static solutions of
  Einstein's field equations for spheres of fluid}},\ }\href
  {https://doi.org/10.1103/PhysRev.55.364} {\bibfield  {journal} {\bibinfo
  {journal} {Phys. Rev.}\ }\textbf {\bibinfo {volume} {55}},\ \bibinfo {pages}
  {364} (\bibinfo {year} {1939})}\BibitemShut {NoStop}%
\bibitem [{\citenamefont {Oppenheimer}\ and\ \citenamefont
  {Volkoff}(1939)}]{Oppenheimer:1939ne}%
  \BibitemOpen
  \bibfield  {author} {\bibinfo {author} {\bibfnamefont {J.~R.}\ \bibnamefont
  {Oppenheimer}}\ and\ \bibinfo {author} {\bibfnamefont {G.~M.}\ \bibnamefont
  {Volkoff}},\ }\bibfield  {title} {\bibinfo {title} {{On Massive neutron
  cores}},\ }\href {https://doi.org/10.1103/PhysRev.55.374} {\bibfield
  {journal} {\bibinfo  {journal} {Phys. Rev.}\ }\textbf {\bibinfo {volume}
  {55}},\ \bibinfo {pages} {374} (\bibinfo {year} {1939})}\BibitemShut
  {NoStop}%
\bibitem [{\citenamefont {Lattimer}\ and\ \citenamefont
  {Prakash}(2001)}]{Lattimer:2000nx}%
  \BibitemOpen
  \bibfield  {author} {\bibinfo {author} {\bibfnamefont {J.~M.}\ \bibnamefont
  {Lattimer}}\ and\ \bibinfo {author} {\bibfnamefont {M.}~\bibnamefont
  {Prakash}},\ }\bibfield  {title} {\bibinfo {title} {{Neutron star structure
  and the equation of state}},\ }\href {https://doi.org/10.1086/319702}
  {\bibfield  {journal} {\bibinfo  {journal} {Astrophys. J.}\ }\textbf
  {\bibinfo {volume} {550}},\ \bibinfo {pages} {426} (\bibinfo {year}
  {2001})},\ \Eprint {https://arxiv.org/abs/astro-ph/0002232}
  {arXiv:astro-ph/0002232} \BibitemShut {NoStop}%
\bibitem [{\citenamefont {McLerran}\ and\ \citenamefont
  {Pisarski}(2007)}]{McLerran:2007qj}%
  \BibitemOpen
  \bibfield  {author} {\bibinfo {author} {\bibfnamefont {L.}~\bibnamefont
  {McLerran}}\ and\ \bibinfo {author} {\bibfnamefont {R.~D.}\ \bibnamefont
  {Pisarski}},\ }\bibfield  {title} {\bibinfo {title} {{Phases of cold, dense
  quarks at large N(c)}},\ }\href
  {https://doi.org/10.1016/j.nuclphysa.2007.08.013} {\bibfield  {journal}
  {\bibinfo  {journal} {Nucl. Phys. A}\ }\textbf {\bibinfo {volume} {796}},\
  \bibinfo {pages} {83} (\bibinfo {year} {2007})},\ \Eprint
  {https://arxiv.org/abs/0706.2191} {arXiv:0706.2191 [hep-ph]} \BibitemShut
  {NoStop}%
\bibitem [{\citenamefont {Gandolfi}\ \emph {et~al.}(2012)\citenamefont
  {Gandolfi}, \citenamefont {Carlson},\ and\ \citenamefont
  {Reddy}}]{Gandolfi:2011xu}%
  \BibitemOpen
  \bibfield  {author} {\bibinfo {author} {\bibfnamefont {S.}~\bibnamefont
  {Gandolfi}}, \bibinfo {author} {\bibfnamefont {J.}~\bibnamefont {Carlson}},\
  and\ \bibinfo {author} {\bibfnamefont {S.}~\bibnamefont {Reddy}},\ }\bibfield
   {title} {\bibinfo {title} {{The maximum mass and radius of neutron stars and
  the nuclear symmetry energy}},\ }\href
  {https://doi.org/10.1103/PhysRevC.85.032801} {\bibfield  {journal} {\bibinfo
  {journal} {Phys. Rev.}\ }\textbf {\bibinfo {volume} {C85}},\ \bibinfo {pages}
  {032801} (\bibinfo {year} {2012})},\ \Eprint
  {https://arxiv.org/abs/1101.1921} {arXiv:1101.1921 [nucl-th]} \BibitemShut
  {NoStop}%
\bibitem [{\citenamefont {McLerran}\ and\ \citenamefont
  {Reddy}(2019)}]{McLerran:2018hbz}%
  \BibitemOpen
  \bibfield  {author} {\bibinfo {author} {\bibfnamefont {L.}~\bibnamefont
  {McLerran}}\ and\ \bibinfo {author} {\bibfnamefont {S.}~\bibnamefont
  {Reddy}},\ }\bibfield  {title} {\bibinfo {title} {{Quarkyonic Matter and
  Neutron Stars}},\ }\href {https://doi.org/10.1103/PhysRevLett.122.122701}
  {\bibfield  {journal} {\bibinfo  {journal} {Phys. Rev. Lett.}\ }\textbf
  {\bibinfo {volume} {122}},\ \bibinfo {pages} {122701} (\bibinfo {year}
  {2019})},\ \Eprint {https://arxiv.org/abs/1811.12503} {arXiv:1811.12503
  [nucl-th]} \BibitemShut {NoStop}%
\bibitem [{\citenamefont {Demorest}\ \emph {et~al.}(2010)\citenamefont
  {Demorest}, \citenamefont {Pennucci}, \citenamefont {Ransom}, \citenamefont
  {Roberts},\ and\ \citenamefont {Hessels}}]{Demorest:2010bx}%
  \BibitemOpen
  \bibfield  {author} {\bibinfo {author} {\bibfnamefont {P.}~\bibnamefont
  {Demorest}}, \bibinfo {author} {\bibfnamefont {T.}~\bibnamefont {Pennucci}},
  \bibinfo {author} {\bibfnamefont {S.}~\bibnamefont {Ransom}}, \bibinfo
  {author} {\bibfnamefont {M.}~\bibnamefont {Roberts}},\ and\ \bibinfo {author}
  {\bibfnamefont {J.}~\bibnamefont {Hessels}},\ }\bibfield  {title} {\bibinfo
  {title} {{Shapiro Delay Measurement of A Two Solar Mass Neutron Star}},\
  }\href {https://doi.org/10.1038/nature09466} {\bibfield  {journal} {\bibinfo
  {journal} {Nature}\ }\textbf {\bibinfo {volume} {467}},\ \bibinfo {pages}
  {1081} (\bibinfo {year} {2010})},\ \Eprint {https://arxiv.org/abs/1010.5788}
  {arXiv:1010.5788 [astro-ph.HE]} \BibitemShut {NoStop}%
\bibitem [{\citenamefont {Antoniadis}\ \emph {et~al.}(2013)\citenamefont
  {Antoniadis} \emph {et~al.}}]{Antoniadis:2013pzd}%
  \BibitemOpen
  \bibfield  {author} {\bibinfo {author} {\bibfnamefont {J.}~\bibnamefont
  {Antoniadis}} \emph {et~al.},\ }\bibfield  {title} {\bibinfo {title} {{A
  Massive Pulsar in a Compact Relativistic Binary}},\ }\href
  {https://doi.org/10.1126/science.1233232} {\bibfield  {journal} {\bibinfo
  {journal} {Science}\ }\textbf {\bibinfo {volume} {340}},\ \bibinfo {pages}
  {6131} (\bibinfo {year} {2013})},\ \Eprint {https://arxiv.org/abs/1304.6875}
  {arXiv:1304.6875 [astro-ph.HE]} \BibitemShut {NoStop}%
\bibitem [{\citenamefont {Romani}\ \emph {et~al.}(2021)\citenamefont {Romani},
  \citenamefont {Kandel}, \citenamefont {Filippenko}, \citenamefont {Brink},\
  and\ \citenamefont {Zheng}}]{Romani:2021xmb}%
  \BibitemOpen
  \bibfield  {author} {\bibinfo {author} {\bibfnamefont {R.~W.}\ \bibnamefont
  {Romani}}, \bibinfo {author} {\bibfnamefont {D.}~\bibnamefont {Kandel}},
  \bibinfo {author} {\bibfnamefont {A.~V.}\ \bibnamefont {Filippenko}},
  \bibinfo {author} {\bibfnamefont {T.~G.}\ \bibnamefont {Brink}},\ and\
  \bibinfo {author} {\bibfnamefont {W.}~\bibnamefont {Zheng}},\ }\bibfield
  {title} {\bibinfo {title} {{PSR J1810+1744: Companion Darkening and a Precise
  High Neutron Star Mass}},\ }\href {https://doi.org/10.3847/2041-8213/abe2b4}
  {\bibfield  {journal} {\bibinfo  {journal} {Astrophys. J. Lett.}\ }\textbf
  {\bibinfo {volume} {908}},\ \bibinfo {pages} {L46} (\bibinfo {year}
  {2021})},\ \Eprint {https://arxiv.org/abs/2101.09822} {arXiv:2101.09822
  [astro-ph.HE]} \BibitemShut {NoStop}%
\bibitem [{\citenamefont {Cromartie}\ \emph {et~al.}(2019)\citenamefont
  {Cromartie} \emph {et~al.}}]{NANOGrav:2019jur}%
  \BibitemOpen
  \bibfield  {author} {\bibinfo {author} {\bibfnamefont {H.~T.}\ \bibnamefont
  {Cromartie}} \emph {et~al.} (\bibinfo {collaboration} {NANOGrav}),\
  }\bibfield  {title} {\bibinfo {title} {{Relativistic Shapiro delay
  measurements of an extremely massive millisecond pulsar}},\ }\href
  {https://doi.org/10.1038/s41550-019-0880-2} {\bibfield  {journal} {\bibinfo
  {journal} {Nature Astron.}\ }\textbf {\bibinfo {volume} {4}},\ \bibinfo
  {pages} {72} (\bibinfo {year} {2019})},\ \Eprint
  {https://arxiv.org/abs/1904.06759} {arXiv:1904.06759 [astro-ph.HE]}
  \BibitemShut {NoStop}%
\bibitem [{\citenamefont {Fonseca}\ \emph {et~al.}(2021)\citenamefont {Fonseca}
  \emph {et~al.}}]{Fonseca:2021wxt}%
  \BibitemOpen
  \bibfield  {author} {\bibinfo {author} {\bibfnamefont {E.}~\bibnamefont
  {Fonseca}} \emph {et~al.},\ }\bibfield  {title} {\bibinfo {title} {{Refined
  Mass and Geometric Measurements of the High-mass PSR J0740+6620}},\ }\href
  {https://doi.org/10.3847/2041-8213/ac03b8} {\bibfield  {journal} {\bibinfo
  {journal} {Astrophys. J. Lett.}\ }\textbf {\bibinfo {volume} {915}},\
  \bibinfo {pages} {L12} (\bibinfo {year} {2021})},\ \Eprint
  {https://arxiv.org/abs/2104.00880} {arXiv:2104.00880 [astro-ph.HE]}
  \BibitemShut {NoStop}%
\bibitem [{\citenamefont {Bedaque}\ and\ \citenamefont
  {Steiner}(2015)}]{Bedaque:2014sqa}%
  \BibitemOpen
  \bibfield  {author} {\bibinfo {author} {\bibfnamefont {P.}~\bibnamefont
  {Bedaque}}\ and\ \bibinfo {author} {\bibfnamefont {A.~W.}\ \bibnamefont
  {Steiner}},\ }\bibfield  {title} {\bibinfo {title} {{Sound velocity bound and
  neutron stars}},\ }\href {https://doi.org/10.1103/PhysRevLett.114.031103}
  {\bibfield  {journal} {\bibinfo  {journal} {Phys. Rev. Lett.}\ }\textbf
  {\bibinfo {volume} {114}},\ \bibinfo {pages} {031103} (\bibinfo {year}
  {2015})},\ \Eprint {https://arxiv.org/abs/1408.5116} {arXiv:1408.5116
  [nucl-th]} \BibitemShut {NoStop}%
\bibitem [{\citenamefont {Tews}\ \emph
  {et~al.}(2018{\natexlab{b}})\citenamefont {Tews}, \citenamefont {Carlson},
  \citenamefont {Gandolfi},\ and\ \citenamefont {Reddy}}]{Tews:2018kmu}%
  \BibitemOpen
  \bibfield  {author} {\bibinfo {author} {\bibfnamefont {I.}~\bibnamefont
  {Tews}}, \bibinfo {author} {\bibfnamefont {J.}~\bibnamefont {Carlson}},
  \bibinfo {author} {\bibfnamefont {S.}~\bibnamefont {Gandolfi}},\ and\
  \bibinfo {author} {\bibfnamefont {S.}~\bibnamefont {Reddy}},\ }\bibfield
  {title} {\bibinfo {title} {{Constraining the speed of sound inside neutron
  stars with chiral effective field theory interactions and observations}},\
  }\href@noop {} {\  (\bibinfo {year} {2018}{\natexlab{b}})},\ \Eprint
  {https://arxiv.org/abs/1801.01923} {arXiv:1801.01923 [nucl-th]} \BibitemShut
  {NoStop}%
\bibitem [{\citenamefont {Drischler}\ \emph {et~al.}(2021)\citenamefont
  {Drischler}, \citenamefont {Han}, \citenamefont {Lattimer}, \citenamefont
  {Prakash}, \citenamefont {Reddy},\ and\ \citenamefont
  {Zhao}}]{Drischler:2020fvz}%
  \BibitemOpen
  \bibfield  {author} {\bibinfo {author} {\bibfnamefont {C.}~\bibnamefont
  {Drischler}}, \bibinfo {author} {\bibfnamefont {S.}~\bibnamefont {Han}},
  \bibinfo {author} {\bibfnamefont {J.~M.}\ \bibnamefont {Lattimer}}, \bibinfo
  {author} {\bibfnamefont {M.}~\bibnamefont {Prakash}}, \bibinfo {author}
  {\bibfnamefont {S.}~\bibnamefont {Reddy}},\ and\ \bibinfo {author}
  {\bibfnamefont {T.}~\bibnamefont {Zhao}},\ }\bibfield  {title} {\bibinfo
  {title} {{Limiting masses and radii of neutron stars and their
  implications}},\ }\href {https://doi.org/10.1103/PhysRevC.103.045808}
  {\bibfield  {journal} {\bibinfo  {journal} {Phys. Rev. C}\ }\textbf {\bibinfo
  {volume} {103}},\ \bibinfo {pages} {045808} (\bibinfo {year} {2021})},\
  \Eprint {https://arxiv.org/abs/2009.06441} {arXiv:2009.06441 [nucl-th]}
  \BibitemShut {NoStop}%
\bibitem [{\citenamefont {Drischler}\ \emph {et~al.}(2022)\citenamefont
  {Drischler}, \citenamefont {Han},\ and\ \citenamefont
  {Reddy}}]{Drischler:2021bup}%
  \BibitemOpen
  \bibfield  {author} {\bibinfo {author} {\bibfnamefont {C.}~\bibnamefont
  {Drischler}}, \bibinfo {author} {\bibfnamefont {S.}~\bibnamefont {Han}},\
  and\ \bibinfo {author} {\bibfnamefont {S.}~\bibnamefont {Reddy}},\ }\bibfield
   {title} {\bibinfo {title} {{Large and massive neutron stars: Implications
  for the sound speed within QCD of dense matter}},\ }\href
  {https://doi.org/10.1103/PhysRevC.105.035808} {\bibfield  {journal} {\bibinfo
   {journal} {Phys. Rev. C}\ }\textbf {\bibinfo {volume} {105}},\ \bibinfo
  {pages} {035808} (\bibinfo {year} {2022})},\ \Eprint
  {https://arxiv.org/abs/2110.14896} {arXiv:2110.14896 [nucl-th]} \BibitemShut
  {NoStop}%
\bibitem [{\citenamefont {Freedman}\ and\ \citenamefont
  {McLerran}(1977{\natexlab{a}})}]{Freedman:1976xs}%
  \BibitemOpen
  \bibfield  {author} {\bibinfo {author} {\bibfnamefont {B.~A.}\ \bibnamefont
  {Freedman}}\ and\ \bibinfo {author} {\bibfnamefont {L.~D.}\ \bibnamefont
  {McLerran}},\ }\bibfield  {title} {\bibinfo {title} {{Fermions and Gauge
  Vector Mesons at Finite Temperature and Density. 1. Formal Techniques}},\
  }\href {https://doi.org/10.1103/PhysRevD.16.1130} {\bibfield  {journal}
  {\bibinfo  {journal} {Phys. Rev. D}\ }\textbf {\bibinfo {volume} {16}},\
  \bibinfo {pages} {1130} (\bibinfo {year} {1977}{\natexlab{a}})}\BibitemShut
  {NoStop}%
\bibitem [{\citenamefont {Freedman}\ and\ \citenamefont
  {McLerran}(1977{\natexlab{b}})}]{Freedman:1976ub}%
  \BibitemOpen
  \bibfield  {author} {\bibinfo {author} {\bibfnamefont {B.~A.}\ \bibnamefont
  {Freedman}}\ and\ \bibinfo {author} {\bibfnamefont {L.~D.}\ \bibnamefont
  {McLerran}},\ }\bibfield  {title} {\bibinfo {title} {{Fermions and Gauge
  Vector Mesons at Finite Temperature and Density. 3. The Ground State Energy
  of a Relativistic Quark Gas}},\ }\href
  {https://doi.org/10.1103/PhysRevD.16.1169} {\bibfield  {journal} {\bibinfo
  {journal} {Phys. Rev. D}\ }\textbf {\bibinfo {volume} {16}},\ \bibinfo
  {pages} {1169} (\bibinfo {year} {1977}{\natexlab{b}})}\BibitemShut {NoStop}%
\bibitem [{\citenamefont {Vuorinen}(2003)}]{Vuorinen:2003fs}%
  \BibitemOpen
  \bibfield  {author} {\bibinfo {author} {\bibfnamefont {A.}~\bibnamefont
  {Vuorinen}},\ }\bibfield  {title} {\bibinfo {title} {{The Pressure of QCD at
  finite temperatures and chemical potentials}},\ }\href
  {https://doi.org/10.1103/PhysRevD.68.054017} {\bibfield  {journal} {\bibinfo
  {journal} {Phys. Rev. D}\ }\textbf {\bibinfo {volume} {68}},\ \bibinfo
  {pages} {054017} (\bibinfo {year} {2003})},\ \Eprint
  {https://arxiv.org/abs/hep-ph/0305183} {arXiv:hep-ph/0305183} \BibitemShut
  {NoStop}%
\bibitem [{\citenamefont {Kurkela}\ \emph {et~al.}(2010)\citenamefont
  {Kurkela}, \citenamefont {Romatschke},\ and\ \citenamefont
  {Vuorinen}}]{Kurkela:2009gj}%
  \BibitemOpen
  \bibfield  {author} {\bibinfo {author} {\bibfnamefont {A.}~\bibnamefont
  {Kurkela}}, \bibinfo {author} {\bibfnamefont {P.}~\bibnamefont
  {Romatschke}},\ and\ \bibinfo {author} {\bibfnamefont {A.}~\bibnamefont
  {Vuorinen}},\ }\bibfield  {title} {\bibinfo {title} {{Cold Quark Matter}},\
  }\href {https://doi.org/10.1103/PhysRevD.81.105021} {\bibfield  {journal}
  {\bibinfo  {journal} {Phys. Rev. D}\ }\textbf {\bibinfo {volume} {81}},\
  \bibinfo {pages} {105021} (\bibinfo {year} {2010})},\ \Eprint
  {https://arxiv.org/abs/0912.1856} {arXiv:0912.1856 [hep-ph]} \BibitemShut
  {NoStop}%
\bibitem [{\citenamefont {Gorda}\ \emph {et~al.}(2018)\citenamefont {Gorda},
  \citenamefont {Kurkela}, \citenamefont {Romatschke}, \citenamefont
  {S\"appi},\ and\ \citenamefont {Vuorinen}}]{Gorda:2018gpy}%
  \BibitemOpen
  \bibfield  {author} {\bibinfo {author} {\bibfnamefont {T.}~\bibnamefont
  {Gorda}}, \bibinfo {author} {\bibfnamefont {A.}~\bibnamefont {Kurkela}},
  \bibinfo {author} {\bibfnamefont {P.}~\bibnamefont {Romatschke}}, \bibinfo
  {author} {\bibfnamefont {S.}~\bibnamefont {S\"appi}},\ and\ \bibinfo {author}
  {\bibfnamefont {A.}~\bibnamefont {Vuorinen}},\ }\bibfield  {title} {\bibinfo
  {title} {{Next-to-Next-to-Next-to-Leading Order Pressure of Cold Quark
  Matter: Leading Logarithm}},\ }\href
  {https://doi.org/10.1103/PhysRevLett.121.202701} {\bibfield  {journal}
  {\bibinfo  {journal} {Phys. Rev. Lett.}\ }\textbf {\bibinfo {volume} {121}},\
  \bibinfo {pages} {202701} (\bibinfo {year} {2018})},\ \Eprint
  {https://arxiv.org/abs/1807.04120} {arXiv:1807.04120 [hep-ph]} \BibitemShut
  {NoStop}%
\bibitem [{\citenamefont {Fernandez}\ and\ \citenamefont
  {Kneur}(2022)}]{Fernandez:2021jfr}%
  \BibitemOpen
  \bibfield  {author} {\bibinfo {author} {\bibfnamefont {L.}~\bibnamefont
  {Fernandez}}\ and\ \bibinfo {author} {\bibfnamefont {J.-L.}\ \bibnamefont
  {Kneur}},\ }\bibfield  {title} {\bibinfo {title} {{All Order Resummed Leading
  and Next-to-Leading Soft Modes of Dense QCD Pressure}},\ }\href
  {https://doi.org/10.1103/PhysRevLett.129.212001} {\bibfield  {journal}
  {\bibinfo  {journal} {Phys. Rev. Lett.}\ }\textbf {\bibinfo {volume} {129}},\
  \bibinfo {pages} {212001} (\bibinfo {year} {2022})},\ \Eprint
  {https://arxiv.org/abs/2109.02410} {arXiv:2109.02410 [hep-ph]} \BibitemShut
  {NoStop}%
\bibitem [{\citenamefont {Gorda}\ \emph
  {et~al.}(2021{\natexlab{a}})\citenamefont {Gorda}, \citenamefont {Kurkela},
  \citenamefont {Paatelainen}, \citenamefont {S\"appi},\ and\ \citenamefont
  {Vuorinen}}]{Gorda:2021kme}%
  \BibitemOpen
  \bibfield  {author} {\bibinfo {author} {\bibfnamefont {T.}~\bibnamefont
  {Gorda}}, \bibinfo {author} {\bibfnamefont {A.}~\bibnamefont {Kurkela}},
  \bibinfo {author} {\bibfnamefont {R.}~\bibnamefont {Paatelainen}}, \bibinfo
  {author} {\bibfnamefont {S.}~\bibnamefont {S\"appi}},\ and\ \bibinfo {author}
  {\bibfnamefont {A.}~\bibnamefont {Vuorinen}},\ }\bibfield  {title} {\bibinfo
  {title} {{Cold quark matter at N3LO: Soft contributions}},\ }\href
  {https://doi.org/10.1103/PhysRevD.104.074015} {\bibfield  {journal} {\bibinfo
   {journal} {Phys. Rev. D}\ }\textbf {\bibinfo {volume} {104}},\ \bibinfo
  {pages} {074015} (\bibinfo {year} {2021}{\natexlab{a}})},\ \Eprint
  {https://arxiv.org/abs/2103.07427} {arXiv:2103.07427 [hep-ph]} \BibitemShut
  {NoStop}%
\bibitem [{\citenamefont {Gorda}\ \emph
  {et~al.}(2021{\natexlab{b}})\citenamefont {Gorda}, \citenamefont {Kurkela},
  \citenamefont {Paatelainen}, \citenamefont {S\"appi},\ and\ \citenamefont
  {Vuorinen}}]{Gorda:2021znl}%
  \BibitemOpen
  \bibfield  {author} {\bibinfo {author} {\bibfnamefont {T.}~\bibnamefont
  {Gorda}}, \bibinfo {author} {\bibfnamefont {A.}~\bibnamefont {Kurkela}},
  \bibinfo {author} {\bibfnamefont {R.}~\bibnamefont {Paatelainen}}, \bibinfo
  {author} {\bibfnamefont {S.}~\bibnamefont {S\"appi}},\ and\ \bibinfo {author}
  {\bibfnamefont {A.}~\bibnamefont {Vuorinen}},\ }\bibfield  {title} {\bibinfo
  {title} {{Soft Interactions in Cold Quark Matter}},\ }\href
  {https://doi.org/10.1103/PhysRevLett.127.162003} {\bibfield  {journal}
  {\bibinfo  {journal} {Phys. Rev. Lett.}\ }\textbf {\bibinfo {volume} {127}},\
  \bibinfo {pages} {162003} (\bibinfo {year} {2021}{\natexlab{b}})},\ \Eprint
  {https://arxiv.org/abs/2103.05658} {arXiv:2103.05658 [hep-ph]} \BibitemShut
  {NoStop}%
\bibitem [{\citenamefont {Gorda}\ \emph
  {et~al.}(2023{\natexlab{a}})\citenamefont {Gorda}, \citenamefont
  {Paatelainen}, \citenamefont {S\"appi},\ and\ \citenamefont
  {Sepp\"anen}}]{Gorda:2023mkk}%
  \BibitemOpen
  \bibfield  {author} {\bibinfo {author} {\bibfnamefont {T.}~\bibnamefont
  {Gorda}}, \bibinfo {author} {\bibfnamefont {R.}~\bibnamefont {Paatelainen}},
  \bibinfo {author} {\bibfnamefont {S.}~\bibnamefont {S\"appi}},\ and\ \bibinfo
  {author} {\bibfnamefont {K.}~\bibnamefont {Sepp\"anen}},\ }\bibfield  {title}
  {\bibinfo {title} {{Equation of State of Cold Quark Matter to $O(\alpha_s^3
  \ln \alpha_s)$}},\ }\href {https://doi.org/10.1103/PhysRevLett.131.181902}
  {\bibfield  {journal} {\bibinfo  {journal} {Phys. Rev. Lett.}\ }\textbf
  {\bibinfo {volume} {131}},\ \bibinfo {pages} {181902} (\bibinfo {year}
  {2023}{\natexlab{a}})},\ \Eprint {https://arxiv.org/abs/2307.08734}
  {arXiv:2307.08734 [hep-ph]} \BibitemShut {NoStop}%
\bibitem [{\citenamefont {Fernandez}\ and\ \citenamefont
  {Kneur}(2024)}]{Fernandez:2024ilg}%
  \BibitemOpen
  \bibfield  {author} {\bibinfo {author} {\bibfnamefont {L.}~\bibnamefont
  {Fernandez}}\ and\ \bibinfo {author} {\bibfnamefont {J.-L.}\ \bibnamefont
  {Kneur}},\ }\bibfield  {title} {\bibinfo {title} {{Cold Quark Matter:
  Renormalization Group Improvement at next-to-next-to leading order}},\
  }\href@noop {} {\  (\bibinfo {year} {2024})},\ \Eprint
  {https://arxiv.org/abs/2408.16674} {arXiv:2408.16674 [hep-ph]} \BibitemShut
  {NoStop}%
\bibitem [{\citenamefont {Annala}\ \emph {et~al.}(2018)\citenamefont {Annala},
  \citenamefont {Gorda}, \citenamefont {Kurkela},\ and\ \citenamefont
  {Vuorinen}}]{Annala:2017llu}%
  \BibitemOpen
  \bibfield  {author} {\bibinfo {author} {\bibfnamefont {E.}~\bibnamefont
  {Annala}}, \bibinfo {author} {\bibfnamefont {T.}~\bibnamefont {Gorda}},
  \bibinfo {author} {\bibfnamefont {A.}~\bibnamefont {Kurkela}},\ and\ \bibinfo
  {author} {\bibfnamefont {A.}~\bibnamefont {Vuorinen}},\ }\bibfield  {title}
  {\bibinfo {title} {{Gravitational-wave constraints on the neutron-star-matter
  Equation of State}},\ }\href {https://doi.org/10.1103/PhysRevLett.120.172703}
  {\bibfield  {journal} {\bibinfo  {journal} {Phys. Rev. Lett.}\ }\textbf
  {\bibinfo {volume} {120}},\ \bibinfo {pages} {172703} (\bibinfo {year}
  {2018})},\ \Eprint {https://arxiv.org/abs/1711.02644} {arXiv:1711.02644
  [astro-ph.HE]} \BibitemShut {NoStop}%
\bibitem [{\citenamefont {Komoltsev}\ and\ \citenamefont
  {Kurkela}(2022)}]{Komoltsev:2021jzg}%
  \BibitemOpen
  \bibfield  {author} {\bibinfo {author} {\bibfnamefont {O.}~\bibnamefont
  {Komoltsev}}\ and\ \bibinfo {author} {\bibfnamefont {A.}~\bibnamefont
  {Kurkela}},\ }\bibfield  {title} {\bibinfo {title} {{How Perturbative QCD
  Constrains the Equation of State at Neutron-Star Densities}},\ }\href
  {https://doi.org/10.1103/PhysRevLett.128.202701} {\bibfield  {journal}
  {\bibinfo  {journal} {Phys. Rev. Lett.}\ }\textbf {\bibinfo {volume} {128}},\
  \bibinfo {pages} {202701} (\bibinfo {year} {2022})},\ \Eprint
  {https://arxiv.org/abs/2111.05350} {arXiv:2111.05350 [nucl-th]} \BibitemShut
  {NoStop}%
\bibitem [{\citenamefont {Gorda}\ \emph
  {et~al.}(2023{\natexlab{b}})\citenamefont {Gorda}, \citenamefont
  {Komoltsev},\ and\ \citenamefont {Kurkela}}]{Gorda:2022jvk}%
  \BibitemOpen
  \bibfield  {author} {\bibinfo {author} {\bibfnamefont {T.}~\bibnamefont
  {Gorda}}, \bibinfo {author} {\bibfnamefont {O.}~\bibnamefont {Komoltsev}},\
  and\ \bibinfo {author} {\bibfnamefont {A.}~\bibnamefont {Kurkela}},\
  }\bibfield  {title} {\bibinfo {title} {{Ab-initio QCD Calculations Impact the
  Inference of the Neutron-star-matter Equation of State}},\ }\href
  {https://doi.org/10.3847/1538-4357/acce3a} {\bibfield  {journal} {\bibinfo
  {journal} {Astrophys. J.}\ }\textbf {\bibinfo {volume} {950}},\ \bibinfo
  {pages} {107} (\bibinfo {year} {2023}{\natexlab{b}})},\ \Eprint
  {https://arxiv.org/abs/2204.11877} {arXiv:2204.11877 [nucl-th]} \BibitemShut
  {NoStop}%
\bibitem [{\citenamefont {Somasundaram}\ \emph {et~al.}(2023)\citenamefont
  {Somasundaram}, \citenamefont {Tews},\ and\ \citenamefont
  {Margueron}}]{Somasundaram:2022ztm}%
  \BibitemOpen
  \bibfield  {author} {\bibinfo {author} {\bibfnamefont {R.}~\bibnamefont
  {Somasundaram}}, \bibinfo {author} {\bibfnamefont {I.}~\bibnamefont {Tews}},\
  and\ \bibinfo {author} {\bibfnamefont {J.}~\bibnamefont {Margueron}},\
  }\bibfield  {title} {\bibinfo {title} {{Perturbative QCD and the neutron star
  equation~of state}},\ }\href {https://doi.org/10.1103/PhysRevC.107.L052801}
  {\bibfield  {journal} {\bibinfo  {journal} {Phys. Rev. C}\ }\textbf {\bibinfo
  {volume} {107}},\ \bibinfo {pages} {L052801} (\bibinfo {year} {2023})},\
  \Eprint {https://arxiv.org/abs/2204.14039} {arXiv:2204.14039 [nucl-th]}
  \BibitemShut {NoStop}%
\bibitem [{\citenamefont {Zhou}(2023)}]{Zhou:2023zrm}%
  \BibitemOpen
  \bibfield  {author} {\bibinfo {author} {\bibfnamefont {D.}~\bibnamefont
  {Zhou}},\ }\bibfield  {title} {\bibinfo {title} {{Reexamining constraints on
  neutron star properties from perturbative QCD}},\ }\href@noop {} {\
  (\bibinfo {year} {2023})},\ \Eprint {https://arxiv.org/abs/2307.11125}
  {arXiv:2307.11125 [astro-ph.HE]} \BibitemShut {NoStop}%
\bibitem [{\citenamefont {Kurkela}\ \emph {et~al.}(2024)\citenamefont
  {Kurkela}, \citenamefont {Rajagopal},\ and\ \citenamefont
  {Steinhorst}}]{Kurkela:2024xfh}%
  \BibitemOpen
  \bibfield  {author} {\bibinfo {author} {\bibfnamefont {A.}~\bibnamefont
  {Kurkela}}, \bibinfo {author} {\bibfnamefont {K.}~\bibnamefont {Rajagopal}},\
  and\ \bibinfo {author} {\bibfnamefont {R.}~\bibnamefont {Steinhorst}},\
  }\bibfield  {title} {\bibinfo {title} {{Astrophysical Equation-of-State
  Constraints on the Color-Superconducting Gap}},\ }\href@noop {} {\  (\bibinfo
  {year} {2024})},\ \Eprint {https://arxiv.org/abs/2401.16253}
  {arXiv:2401.16253 [astro-ph.HE]} \BibitemShut {NoStop}%
\bibitem [{\citenamefont {Zhou}(2024{\natexlab{a}})}]{Zhou:2024hdi}%
  \BibitemOpen
  \bibfield  {author} {\bibinfo {author} {\bibfnamefont {D.}~\bibnamefont
  {Zhou}},\ }\bibfield  {title} {\bibinfo {title} {{Bounds on the minimum sound
  speed above neutron star densities}},\ }\href@noop {} {\  (\bibinfo {year}
  {2024}{\natexlab{a}})},\ \Eprint {https://arxiv.org/abs/2408.16738}
  {arXiv:2408.16738 [nucl-th]} \BibitemShut {NoStop}%
\bibitem [{\citenamefont {Margalit}\ and\ \citenamefont
  {Metzger}(2017)}]{Margalit:2017dij}%
  \BibitemOpen
  \bibfield  {author} {\bibinfo {author} {\bibfnamefont {B.}~\bibnamefont
  {Margalit}}\ and\ \bibinfo {author} {\bibfnamefont {B.~D.}\ \bibnamefont
  {Metzger}},\ }\bibfield  {title} {\bibinfo {title} {{Constraining the Maximum
  Mass of Neutron Stars From Multi-Messenger Observations of GW170817}},\
  }\href {https://doi.org/10.3847/2041-8213/aa991c} {\bibfield  {journal}
  {\bibinfo  {journal} {Astrophys. J.}\ }\textbf {\bibinfo {volume} {850}},\
  \bibinfo {pages} {L19} (\bibinfo {year} {2017})},\ \Eprint
  {https://arxiv.org/abs/1710.05938} {arXiv:1710.05938 [astro-ph.HE]}
  \BibitemShut {NoStop}%
\bibitem [{\citenamefont {Shibata}\ \emph {et~al.}(2017)\citenamefont
  {Shibata}, \citenamefont {Fujibayashi}, \citenamefont {Hotokezaka},
  \citenamefont {Kiuchi}, \citenamefont {Kyutoku}, \citenamefont {Sekiguchi},\
  and\ \citenamefont {Tanaka}}]{Shibata:2017xdx}%
  \BibitemOpen
  \bibfield  {author} {\bibinfo {author} {\bibfnamefont {M.}~\bibnamefont
  {Shibata}}, \bibinfo {author} {\bibfnamefont {S.}~\bibnamefont
  {Fujibayashi}}, \bibinfo {author} {\bibfnamefont {K.}~\bibnamefont
  {Hotokezaka}}, \bibinfo {author} {\bibfnamefont {K.}~\bibnamefont {Kiuchi}},
  \bibinfo {author} {\bibfnamefont {K.}~\bibnamefont {Kyutoku}}, \bibinfo
  {author} {\bibfnamefont {Y.}~\bibnamefont {Sekiguchi}},\ and\ \bibinfo
  {author} {\bibfnamefont {M.}~\bibnamefont {Tanaka}},\ }\bibfield  {title}
  {\bibinfo {title} {{Modeling GW170817 based on numerical relativity and its
  implications}},\ }\href {https://doi.org/10.1103/PhysRevD.96.123012}
  {\bibfield  {journal} {\bibinfo  {journal} {Phys. Rev.}\ }\textbf {\bibinfo
  {volume} {D96}},\ \bibinfo {pages} {123012} (\bibinfo {year} {2017})},\
  \Eprint {https://arxiv.org/abs/1710.07579} {arXiv:1710.07579 [astro-ph.HE]}
  \BibitemShut {NoStop}%
\bibitem [{\citenamefont {Rezzolla}\ \emph {et~al.}(2018)\citenamefont
  {Rezzolla}, \citenamefont {Most},\ and\ \citenamefont
  {Weih}}]{Rezzolla:2017aly}%
  \BibitemOpen
  \bibfield  {author} {\bibinfo {author} {\bibfnamefont {L.}~\bibnamefont
  {Rezzolla}}, \bibinfo {author} {\bibfnamefont {E.~R.}\ \bibnamefont {Most}},\
  and\ \bibinfo {author} {\bibfnamefont {L.~R.}\ \bibnamefont {Weih}},\
  }\bibfield  {title} {\bibinfo {title} {{Using gravitational-wave observations
  and quasi-universal relations to constrain the maximum mass of neutron
  stars}},\ }\href {https://doi.org/10.3847/2041-8213/aaa401} {\bibfield
  {journal} {\bibinfo  {journal} {Astrophys. J.}\ }\textbf {\bibinfo {volume}
  {852}},\ \bibinfo {pages} {L25} (\bibinfo {year} {2018})},\ \Eprint
  {https://arxiv.org/abs/1711.00314} {arXiv:1711.00314 [astro-ph.HE]}
  \BibitemShut {NoStop}%
\bibitem [{\citenamefont {Radice}\ \emph
  {et~al.}(2018{\natexlab{b}})\citenamefont {Radice}, \citenamefont {Perego},
  \citenamefont {Bernuzzi},\ and\ \citenamefont {Zhang}}]{Radice:2018xqa}%
  \BibitemOpen
  \bibfield  {author} {\bibinfo {author} {\bibfnamefont {D.}~\bibnamefont
  {Radice}}, \bibinfo {author} {\bibfnamefont {A.}~\bibnamefont {Perego}},
  \bibinfo {author} {\bibfnamefont {S.}~\bibnamefont {Bernuzzi}},\ and\
  \bibinfo {author} {\bibfnamefont {B.}~\bibnamefont {Zhang}},\ }\bibfield
  {title} {\bibinfo {title} {{Long-lived Remnants from Binary Neutron Star
  Mergers}},\ }\href {https://doi.org/10.1093/mnras/sty2531} {\bibfield
  {journal} {\bibinfo  {journal} {Mon. Not. Roy. Astron. Soc.}\ }\textbf
  {\bibinfo {volume} {481}},\ \bibinfo {pages} {3670} (\bibinfo {year}
  {2018}{\natexlab{b}})},\ \Eprint {https://arxiv.org/abs/1803.10865}
  {arXiv:1803.10865 [astro-ph.HE]} \BibitemShut {NoStop}%
\bibitem [{\citenamefont {Shibata}\ \emph {et~al.}(2019)\citenamefont
  {Shibata}, \citenamefont {Zhou}, \citenamefont {Kiuchi},\ and\ \citenamefont
  {Fujibayashi}}]{Shibata:2019ctb}%
  \BibitemOpen
  \bibfield  {author} {\bibinfo {author} {\bibfnamefont {M.}~\bibnamefont
  {Shibata}}, \bibinfo {author} {\bibfnamefont {E.}~\bibnamefont {Zhou}},
  \bibinfo {author} {\bibfnamefont {K.}~\bibnamefont {Kiuchi}},\ and\ \bibinfo
  {author} {\bibfnamefont {S.}~\bibnamefont {Fujibayashi}},\ }\bibfield
  {title} {\bibinfo {title} {{Constraint on the maximum mass of neutron stars
  using GW170817 event}},\ }\href {https://doi.org/10.1103/PhysRevD.100.023015}
  {\bibfield  {journal} {\bibinfo  {journal} {Phys. Rev. D}\ }\textbf {\bibinfo
  {volume} {100}},\ \bibinfo {pages} {023015} (\bibinfo {year} {2019})},\
  \Eprint {https://arxiv.org/abs/1905.03656} {arXiv:1905.03656 [astro-ph.HE]}
  \BibitemShut {NoStop}%
\bibitem [{\citenamefont {Abbott}\ \emph {et~al.}(2020)\citenamefont {Abbott}
  \emph {et~al.}}]{LIGOScientific:2020zkf}%
  \BibitemOpen
  \bibfield  {author} {\bibinfo {author} {\bibfnamefont {R.}~\bibnamefont
  {Abbott}} \emph {et~al.} (\bibinfo {collaboration} {LIGO Scientific,
  Virgo}),\ }\bibfield  {title} {\bibinfo {title} {{GW190814: Gravitational
  Waves from the Coalescence of a 23 Solar Mass Black Hole with a 2.6 Solar
  Mass Compact Object}},\ }\href {https://doi.org/10.3847/2041-8213/ab960f}
  {\bibfield  {journal} {\bibinfo  {journal} {Astrophys. J. Lett.}\ }\textbf
  {\bibinfo {volume} {896}},\ \bibinfo {pages} {L44} (\bibinfo {year}
  {2020})},\ \Eprint {https://arxiv.org/abs/2006.12611} {arXiv:2006.12611
  [astro-ph.HE]} \BibitemShut {NoStop}%
\bibitem [{\citenamefont {Weinberg}(1968)}]{Weinberg:1968de}%
  \BibitemOpen
  \bibfield  {author} {\bibinfo {author} {\bibfnamefont {S.}~\bibnamefont
  {Weinberg}},\ }\bibfield  {title} {\bibinfo {title} {{Nonlinear realizations
  of chiral symmetry}},\ }\href {https://doi.org/10.1103/PhysRev.166.1568}
  {\bibfield  {journal} {\bibinfo  {journal} {Phys. Rev.}\ }\textbf {\bibinfo
  {volume} {166}},\ \bibinfo {pages} {1568} (\bibinfo {year}
  {1968})}\BibitemShut {NoStop}%
\bibitem [{\citenamefont {Weinberg}(1990)}]{Weinberg:1990rz}%
  \BibitemOpen
  \bibfield  {author} {\bibinfo {author} {\bibfnamefont {S.}~\bibnamefont
  {Weinberg}},\ }\bibfield  {title} {\bibinfo {title} {{Nuclear forces from
  chiral Lagrangians}},\ }\href {https://doi.org/10.1016/0370-2693(90)90938-3}
  {\bibfield  {journal} {\bibinfo  {journal} {Phys. Lett. B}\ }\textbf
  {\bibinfo {volume} {251}},\ \bibinfo {pages} {288} (\bibinfo {year}
  {1990})}\BibitemShut {NoStop}%
\bibitem [{\citenamefont {Weinberg}(1991)}]{Weinberg:1991um}%
  \BibitemOpen
  \bibfield  {author} {\bibinfo {author} {\bibfnamefont {S.}~\bibnamefont
  {Weinberg}},\ }\bibfield  {title} {\bibinfo {title} {{Effective chiral
  Lagrangians for nucleon - pion interactions and nuclear forces}},\ }\href
  {https://doi.org/10.1016/0550-3213(91)90231-L} {\bibfield  {journal}
  {\bibinfo  {journal} {Nucl. Phys. B}\ }\textbf {\bibinfo {volume} {363}},\
  \bibinfo {pages} {3} (\bibinfo {year} {1991})}\BibitemShut {NoStop}%
\bibitem [{\citenamefont {Weinberg}(1992)}]{Weinberg:1992yk}%
  \BibitemOpen
  \bibfield  {author} {\bibinfo {author} {\bibfnamefont {S.}~\bibnamefont
  {Weinberg}},\ }\bibfield  {title} {\bibinfo {title} {{Three body interactions
  among nucleons and pions}},\ }\href
  {https://doi.org/10.1016/0370-2693(92)90099-P} {\bibfield  {journal}
  {\bibinfo  {journal} {Phys. Lett. B}\ }\textbf {\bibinfo {volume} {295}},\
  \bibinfo {pages} {114} (\bibinfo {year} {1992})},\ \Eprint
  {https://arxiv.org/abs/hep-ph/9209257} {arXiv:hep-ph/9209257} \BibitemShut
  {NoStop}%
\bibitem [{\citenamefont {Kaplan}\ \emph {et~al.}(1996)\citenamefont {Kaplan},
  \citenamefont {Savage},\ and\ \citenamefont {Wise}}]{Kaplan:1996xu}%
  \BibitemOpen
  \bibfield  {author} {\bibinfo {author} {\bibfnamefont {D.~B.}\ \bibnamefont
  {Kaplan}}, \bibinfo {author} {\bibfnamefont {M.~J.}\ \bibnamefont {Savage}},\
  and\ \bibinfo {author} {\bibfnamefont {M.~B.}\ \bibnamefont {Wise}},\
  }\bibfield  {title} {\bibinfo {title} {{Nucleon - nucleon scattering from
  effective field theory}},\ }\href
  {https://doi.org/10.1016/0550-3213(96)00357-4} {\bibfield  {journal}
  {\bibinfo  {journal} {Nucl. Phys. B}\ }\textbf {\bibinfo {volume} {478}},\
  \bibinfo {pages} {629} (\bibinfo {year} {1996})},\ \Eprint
  {https://arxiv.org/abs/nucl-th/9605002} {arXiv:nucl-th/9605002} \BibitemShut
  {NoStop}%
\bibitem [{\citenamefont {Kaplan}\ \emph
  {et~al.}(1998{\natexlab{a}})\citenamefont {Kaplan}, \citenamefont {Savage},\
  and\ \citenamefont {Wise}}]{Kaplan:1998tg}%
  \BibitemOpen
  \bibfield  {author} {\bibinfo {author} {\bibfnamefont {D.~B.}\ \bibnamefont
  {Kaplan}}, \bibinfo {author} {\bibfnamefont {M.~J.}\ \bibnamefont {Savage}},\
  and\ \bibinfo {author} {\bibfnamefont {M.~B.}\ \bibnamefont {Wise}},\
  }\bibfield  {title} {\bibinfo {title} {{A New expansion for nucleon-nucleon
  interactions}},\ }\href {https://doi.org/10.1016/S0370-2693(98)00210-X}
  {\bibfield  {journal} {\bibinfo  {journal} {Phys. Lett. B}\ }\textbf
  {\bibinfo {volume} {424}},\ \bibinfo {pages} {390} (\bibinfo {year}
  {1998}{\natexlab{a}})},\ \Eprint {https://arxiv.org/abs/nucl-th/9801034}
  {arXiv:nucl-th/9801034} \BibitemShut {NoStop}%
\bibitem [{\citenamefont {Kaplan}\ \emph
  {et~al.}(1998{\natexlab{b}})\citenamefont {Kaplan}, \citenamefont {Savage},\
  and\ \citenamefont {Wise}}]{Kaplan:1998we}%
  \BibitemOpen
  \bibfield  {author} {\bibinfo {author} {\bibfnamefont {D.~B.}\ \bibnamefont
  {Kaplan}}, \bibinfo {author} {\bibfnamefont {M.~J.}\ \bibnamefont {Savage}},\
  and\ \bibinfo {author} {\bibfnamefont {M.~B.}\ \bibnamefont {Wise}},\
  }\bibfield  {title} {\bibinfo {title} {{Two nucleon systems from effective
  field theory}},\ }\href {https://doi.org/10.1016/S0550-3213(98)00440-4}
  {\bibfield  {journal} {\bibinfo  {journal} {Nucl. Phys. B}\ }\textbf
  {\bibinfo {volume} {534}},\ \bibinfo {pages} {329} (\bibinfo {year}
  {1998}{\natexlab{b}})},\ \Eprint {https://arxiv.org/abs/nucl-th/9802075}
  {arXiv:nucl-th/9802075} \BibitemShut {NoStop}%
\bibitem [{\citenamefont {Beane}\ \emph {et~al.}(2002)\citenamefont {Beane},
  \citenamefont {Bedaque}, \citenamefont {Savage},\ and\ \citenamefont {van
  Kolck}}]{Beane:2001bc}%
  \BibitemOpen
  \bibfield  {author} {\bibinfo {author} {\bibfnamefont {S.~R.}\ \bibnamefont
  {Beane}}, \bibinfo {author} {\bibfnamefont {P.~F.}\ \bibnamefont {Bedaque}},
  \bibinfo {author} {\bibfnamefont {M.~J.}\ \bibnamefont {Savage}},\ and\
  \bibinfo {author} {\bibfnamefont {U.}~\bibnamefont {van Kolck}},\ }\bibfield
  {title} {\bibinfo {title} {{Towards a perturbative theory of nuclear
  forces}},\ }\href {https://doi.org/10.1016/S0375-9474(01)01324-0} {\bibfield
  {journal} {\bibinfo  {journal} {Nucl. Phys. A}\ }\textbf {\bibinfo {volume}
  {700}},\ \bibinfo {pages} {377} (\bibinfo {year} {2002})},\ \Eprint
  {https://arxiv.org/abs/nucl-th/0104030} {arXiv:nucl-th/0104030} \BibitemShut
  {NoStop}%
\bibitem [{\citenamefont {Rhoades}\ and\ \citenamefont
  {Ruffini}(1974)}]{Rhoades:1974fn}%
  \BibitemOpen
  \bibfield  {author} {\bibinfo {author} {\bibfnamefont {C.~E.}\ \bibnamefont
  {Rhoades}, \bibfnamefont {Jr.}}\ and\ \bibinfo {author} {\bibfnamefont
  {R.}~\bibnamefont {Ruffini}},\ }\bibfield  {title} {\bibinfo {title}
  {{Maximum mass of a neutron star}},\ }\href
  {https://doi.org/10.1103/PhysRevLett.32.324} {\bibfield  {journal} {\bibinfo
  {journal} {Phys. Rev. Lett.}\ }\textbf {\bibinfo {volume} {32}},\ \bibinfo
  {pages} {324} (\bibinfo {year} {1974})}\BibitemShut {NoStop}%
\bibitem [{\citenamefont {Koranda}\ \emph {et~al.}(1997)\citenamefont
  {Koranda}, \citenamefont {Stergioulas},\ and\ \citenamefont
  {Friedman}}]{Koranda:1996jm}%
  \BibitemOpen
  \bibfield  {author} {\bibinfo {author} {\bibfnamefont {S.}~\bibnamefont
  {Koranda}}, \bibinfo {author} {\bibfnamefont {N.}~\bibnamefont
  {Stergioulas}},\ and\ \bibinfo {author} {\bibfnamefont {J.~L.}\ \bibnamefont
  {Friedman}},\ }\bibfield  {title} {\bibinfo {title} {{Upper limit set by
  causality on the rotation and mass of uniformly rotating relativistic
  stars}},\ }\href {https://doi.org/10.1086/304714} {\bibfield  {journal}
  {\bibinfo  {journal} {Astrophys. J.}\ }\textbf {\bibinfo {volume} {488}},\
  \bibinfo {pages} {799} (\bibinfo {year} {1997})},\ \Eprint
  {https://arxiv.org/abs/astro-ph/9608179} {arXiv:astro-ph/9608179}
  \BibitemShut {NoStop}%
\bibitem [{\citenamefont {Forbes}\ \emph {et~al.}(2019)\citenamefont {Forbes},
  \citenamefont {Bose}, \citenamefont {Reddy}, \citenamefont {Zhou},
  \citenamefont {Mukherjee},\ and\ \citenamefont {De}}]{Forbes:2019xaz}%
  \BibitemOpen
  \bibfield  {author} {\bibinfo {author} {\bibfnamefont {M.~M.}\ \bibnamefont
  {Forbes}}, \bibinfo {author} {\bibfnamefont {S.}~\bibnamefont {Bose}},
  \bibinfo {author} {\bibfnamefont {S.}~\bibnamefont {Reddy}}, \bibinfo
  {author} {\bibfnamefont {D.}~\bibnamefont {Zhou}}, \bibinfo {author}
  {\bibfnamefont {A.}~\bibnamefont {Mukherjee}},\ and\ \bibinfo {author}
  {\bibfnamefont {S.}~\bibnamefont {De}},\ }\bibfield  {title} {\bibinfo
  {title} {{Constraining the neutron-matter equation of state with
  gravitational waves}},\ }\href {https://doi.org/10.1103/PhysRevD.100.083010}
  {\bibfield  {journal} {\bibinfo  {journal} {Phys. Rev. D}\ }\textbf {\bibinfo
  {volume} {100}},\ \bibinfo {pages} {083010} (\bibinfo {year} {2019})},\
  \Eprint {https://arxiv.org/abs/1904.04233} {arXiv:1904.04233 [astro-ph.HE]}
  \BibitemShut {NoStop}%
\bibitem [{\citenamefont {Drischler}\ \emph {et~al.}(2019)\citenamefont
  {Drischler}, \citenamefont {Hebeler},\ and\ \citenamefont
  {Schwenk}}]{Drischler:2017wtt}%
  \BibitemOpen
  \bibfield  {author} {\bibinfo {author} {\bibfnamefont {C.}~\bibnamefont
  {Drischler}}, \bibinfo {author} {\bibfnamefont {K.}~\bibnamefont {Hebeler}},\
  and\ \bibinfo {author} {\bibfnamefont {A.}~\bibnamefont {Schwenk}},\
  }\bibfield  {title} {\bibinfo {title} {{Chiral interactions up to
  next-to-next-to-next-to-leading order and nuclear saturation}},\ }\href
  {https://doi.org/10.1103/PhysRevLett.122.042501} {\bibfield  {journal}
  {\bibinfo  {journal} {Phys. Rev. Lett.}\ }\textbf {\bibinfo {volume} {122}},\
  \bibinfo {pages} {042501} (\bibinfo {year} {2019})},\ \Eprint
  {https://arxiv.org/abs/1710.08220} {arXiv:1710.08220 [nucl-th]} \BibitemShut
  {NoStop}%
\bibitem [{\citenamefont {Drischler}\ \emph {et~al.}(2020)\citenamefont
  {Drischler}, \citenamefont {Melendez}, \citenamefont {Furnstahl},\ and\
  \citenamefont {Phillips}}]{Drischler:2020yad}%
  \BibitemOpen
  \bibfield  {author} {\bibinfo {author} {\bibfnamefont {C.}~\bibnamefont
  {Drischler}}, \bibinfo {author} {\bibfnamefont {J.~A.}\ \bibnamefont
  {Melendez}}, \bibinfo {author} {\bibfnamefont {R.~J.}\ \bibnamefont
  {Furnstahl}},\ and\ \bibinfo {author} {\bibfnamefont {D.~R.}\ \bibnamefont
  {Phillips}},\ }\bibfield  {title} {\bibinfo {title} {{Quantifying
  uncertainties and correlations in the nuclear-matter equation of state}},\
  }\href {https://doi.org/10.1103/PhysRevC.102.054315} {\bibfield  {journal}
  {\bibinfo  {journal} {Phys. Rev. C}\ }\textbf {\bibinfo {volume} {102}},\
  \bibinfo {pages} {054315} (\bibinfo {year} {2020})},\ \Eprint
  {https://arxiv.org/abs/2004.07805} {arXiv:2004.07805 [nucl-th]} \BibitemShut
  {NoStop}%
\bibitem [{\citenamefont {Abbott}\ \emph {et~al.}(2023)\citenamefont {Abbott},
  \citenamefont {Detmold}, \citenamefont {Romero-L\'opez}, \citenamefont
  {Davoudi}, \citenamefont {Illa}, \citenamefont {Parre\~no}, \citenamefont
  {Perry}, \citenamefont {Shanahan},\ and\ \citenamefont
  {Wagman}}]{Abbott:2023coj}%
  \BibitemOpen
  \bibfield  {author} {\bibinfo {author} {\bibfnamefont {R.}~\bibnamefont
  {Abbott}}, \bibinfo {author} {\bibfnamefont {W.}~\bibnamefont {Detmold}},
  \bibinfo {author} {\bibfnamefont {F.}~\bibnamefont {Romero-L\'opez}},
  \bibinfo {author} {\bibfnamefont {Z.}~\bibnamefont {Davoudi}}, \bibinfo
  {author} {\bibfnamefont {M.}~\bibnamefont {Illa}}, \bibinfo {author}
  {\bibfnamefont {A.}~\bibnamefont {Parre\~no}}, \bibinfo {author}
  {\bibfnamefont {R.~J.}\ \bibnamefont {Perry}}, \bibinfo {author}
  {\bibfnamefont {P.~E.}\ \bibnamefont {Shanahan}},\ and\ \bibinfo {author}
  {\bibfnamefont {M.~L.}\ \bibnamefont {Wagman}} (\bibinfo {collaboration}
  {NPLQCD}),\ }\bibfield  {title} {\bibinfo {title} {{Lattice quantum
  chromodynamics at large isospin density}},\ }\href
  {https://doi.org/10.1103/PhysRevD.108.114506} {\bibfield  {journal} {\bibinfo
   {journal} {Phys. Rev. D}\ }\textbf {\bibinfo {volume} {108}},\ \bibinfo
  {pages} {114506} (\bibinfo {year} {2023})},\ \Eprint
  {https://arxiv.org/abs/2307.15014} {arXiv:2307.15014 [hep-lat]} \BibitemShut
  {NoStop}%
\bibitem [{\citenamefont {Sch\"afer}\ and\ \citenamefont
  {Wilczek}(1999{\natexlab{b}})}]{Schafer:1998ef}%
  \BibitemOpen
  \bibfield  {author} {\bibinfo {author} {\bibfnamefont {T.}~\bibnamefont
  {Sch\"afer}}\ and\ \bibinfo {author} {\bibfnamefont {F.}~\bibnamefont
  {Wilczek}},\ }\bibfield  {title} {\bibinfo {title} {{Continuity of quark and
  hadron matter}},\ }\href {https://doi.org/10.1103/PhysRevLett.82.3956}
  {\bibfield  {journal} {\bibinfo  {journal} {Phys. Rev. Lett.}\ }\textbf
  {\bibinfo {volume} {82}},\ \bibinfo {pages} {3956} (\bibinfo {year}
  {1999}{\natexlab{b}})},\ \Eprint {https://arxiv.org/abs/hep-ph/9811473}
  {arXiv:hep-ph/9811473} \BibitemShut {NoStop}%
\bibitem [{\citenamefont {Sch\"afer}\ and\ \citenamefont
  {Wilczek}(1999{\natexlab{c}})}]{Schafer:1999pb}%
  \BibitemOpen
  \bibfield  {author} {\bibinfo {author} {\bibfnamefont {T.}~\bibnamefont
  {Sch\"afer}}\ and\ \bibinfo {author} {\bibfnamefont {F.}~\bibnamefont
  {Wilczek}},\ }\bibfield  {title} {\bibinfo {title} {{Quark description of
  hadronic phases}},\ }\href {https://doi.org/10.1103/PhysRevD.60.074014}
  {\bibfield  {journal} {\bibinfo  {journal} {Phys. Rev. D}\ }\textbf {\bibinfo
  {volume} {60}},\ \bibinfo {pages} {074014} (\bibinfo {year}
  {1999}{\natexlab{c}})},\ \Eprint {https://arxiv.org/abs/hep-ph/9903503}
  {arXiv:hep-ph/9903503} \BibitemShut {NoStop}%
\bibitem [{\citenamefont {Sch\"afer}(2000)}]{Schafer:1999fe}%
  \BibitemOpen
  \bibfield  {author} {\bibinfo {author} {\bibfnamefont {T.}~\bibnamefont
  {Sch\"afer}},\ }\bibfield  {title} {\bibinfo {title} {{Patterns of symmetry
  breaking in QCD at high baryon density}},\ }\href
  {https://doi.org/10.1016/S0550-3213(00)00063-8} {\bibfield  {journal}
  {\bibinfo  {journal} {Nucl. Phys. B}\ }\textbf {\bibinfo {volume} {575}},\
  \bibinfo {pages} {269} (\bibinfo {year} {2000})},\ \Eprint
  {https://arxiv.org/abs/hep-ph/9909574} {arXiv:hep-ph/9909574} \BibitemShut
  {NoStop}%
\bibitem [{\citenamefont {Fujimoto}\ \emph {et~al.}(2022)\citenamefont
  {Fujimoto}, \citenamefont {Fukushima}, \citenamefont {McLerran},\ and\
  \citenamefont {Praszalowicz}}]{Fujimoto:2022ohj}%
  \BibitemOpen
  \bibfield  {author} {\bibinfo {author} {\bibfnamefont {Y.}~\bibnamefont
  {Fujimoto}}, \bibinfo {author} {\bibfnamefont {K.}~\bibnamefont {Fukushima}},
  \bibinfo {author} {\bibfnamefont {L.~D.}\ \bibnamefont {McLerran}},\ and\
  \bibinfo {author} {\bibfnamefont {M.}~\bibnamefont {Praszalowicz}},\
  }\bibfield  {title} {\bibinfo {title} {{Trace Anomaly as Signature of
  Conformality in Neutron Stars}},\ }\href
  {https://doi.org/10.1103/PhysRevLett.129.252702} {\bibfield  {journal}
  {\bibinfo  {journal} {Phys. Rev. Lett.}\ }\textbf {\bibinfo {volume} {129}},\
  \bibinfo {pages} {252702} (\bibinfo {year} {2022})},\ \Eprint
  {https://arxiv.org/abs/2207.06753} {arXiv:2207.06753 [nucl-th]} \BibitemShut
  {NoStop}%
\bibitem [{\citenamefont {Rapp}\ \emph {et~al.}(1998)\citenamefont {Rapp},
  \citenamefont {Sch\"afer}, \citenamefont {Shuryak},\ and\ \citenamefont
  {Velkovsky}}]{Rapp:1997zu}%
  \BibitemOpen
  \bibfield  {author} {\bibinfo {author} {\bibfnamefont {R.}~\bibnamefont
  {Rapp}}, \bibinfo {author} {\bibfnamefont {T.}~\bibnamefont {Sch\"afer}},
  \bibinfo {author} {\bibfnamefont {E.~V.}\ \bibnamefont {Shuryak}},\ and\
  \bibinfo {author} {\bibfnamefont {M.}~\bibnamefont {Velkovsky}},\ }\bibfield
  {title} {\bibinfo {title} {{Diquark Bose condensates in high density matter
  and instantons}},\ }\href {https://doi.org/10.1103/PhysRevLett.81.53}
  {\bibfield  {journal} {\bibinfo  {journal} {Phys. Rev. Lett.}\ }\textbf
  {\bibinfo {volume} {81}},\ \bibinfo {pages} {53} (\bibinfo {year} {1998})},\
  \Eprint {https://arxiv.org/abs/hep-ph/9711396} {arXiv:hep-ph/9711396}
  \BibitemShut {NoStop}%
\bibitem [{\citenamefont {Kaplan}\ and\ \citenamefont
  {Reddy}(2002)}]{Kaplan:2001qk}%
  \BibitemOpen
  \bibfield  {author} {\bibinfo {author} {\bibfnamefont {D.~B.}\ \bibnamefont
  {Kaplan}}\ and\ \bibinfo {author} {\bibfnamefont {S.}~\bibnamefont {Reddy}},\
  }\bibfield  {title} {\bibinfo {title} {{Novel phases and transitions in color
  flavor locked matter}},\ }\href {https://doi.org/10.1103/PhysRevD.65.054042}
  {\bibfield  {journal} {\bibinfo  {journal} {Phys. Rev. D}\ }\textbf {\bibinfo
  {volume} {65}},\ \bibinfo {pages} {054042} (\bibinfo {year} {2002})},\
  \Eprint {https://arxiv.org/abs/hep-ph/0107265} {arXiv:hep-ph/0107265}
  \BibitemShut {NoStop}%
\bibitem [{\citenamefont {Bedaque}\ and\ \citenamefont
  {Sch\"afer}(2002)}]{Bedaque:2001je}%
  \BibitemOpen
  \bibfield  {author} {\bibinfo {author} {\bibfnamefont {P.~F.}\ \bibnamefont
  {Bedaque}}\ and\ \bibinfo {author} {\bibfnamefont {T.}~\bibnamefont
  {Sch\"afer}},\ }\bibfield  {title} {\bibinfo {title} {{High density quark
  matter under stress}},\ }\href
  {https://doi.org/10.1016/S0375-9474(01)01272-6} {\bibfield  {journal}
  {\bibinfo  {journal} {Nucl. Phys. A}\ }\textbf {\bibinfo {volume} {697}},\
  \bibinfo {pages} {802} (\bibinfo {year} {2002})},\ \Eprint
  {https://arxiv.org/abs/hep-ph/0105150} {arXiv:hep-ph/0105150} \BibitemShut
  {NoStop}%
\bibitem [{\citenamefont {Cherman}\ \emph {et~al.}(2009)\citenamefont
  {Cherman}, \citenamefont {Cohen},\ and\ \citenamefont
  {Nellore}}]{Cherman:2009tw}%
  \BibitemOpen
  \bibfield  {author} {\bibinfo {author} {\bibfnamefont {A.}~\bibnamefont
  {Cherman}}, \bibinfo {author} {\bibfnamefont {T.~D.}\ \bibnamefont {Cohen}},\
  and\ \bibinfo {author} {\bibfnamefont {A.}~\bibnamefont {Nellore}},\
  }\bibfield  {title} {\bibinfo {title} {{A Bound on the speed of sound from
  holography}},\ }\href {https://doi.org/10.1103/PhysRevD.80.066003} {\bibfield
   {journal} {\bibinfo  {journal} {Phys. Rev. D}\ }\textbf {\bibinfo {volume}
  {80}},\ \bibinfo {pages} {066003} (\bibinfo {year} {2009})},\ \Eprint
  {https://arxiv.org/abs/0905.0903} {arXiv:0905.0903 [hep-th]} \BibitemShut
  {NoStop}%
\bibitem [{\citenamefont {Hohler}\ and\ \citenamefont
  {Stephanov}(2009)}]{Hohler:2009tv}%
  \BibitemOpen
  \bibfield  {author} {\bibinfo {author} {\bibfnamefont {P.~M.}\ \bibnamefont
  {Hohler}}\ and\ \bibinfo {author} {\bibfnamefont {M.~A.}\ \bibnamefont
  {Stephanov}},\ }\bibfield  {title} {\bibinfo {title} {{Holography and the
  speed of sound at high temperatures}},\ }\href
  {https://doi.org/10.1103/PhysRevD.80.066002} {\bibfield  {journal} {\bibinfo
  {journal} {Phys. Rev. D}\ }\textbf {\bibinfo {volume} {80}},\ \bibinfo
  {pages} {066002} (\bibinfo {year} {2009})},\ \Eprint
  {https://arxiv.org/abs/0905.0900} {arXiv:0905.0900 [hep-th]} \BibitemShut
  {NoStop}%
\bibitem [{\citenamefont {Hoyos}\ \emph {et~al.}(2016)\citenamefont {Hoyos},
  \citenamefont {Jokela}, \citenamefont {Rodr\'\i{}guez~Fern\'andez},\ and\
  \citenamefont {Vuorinen}}]{Hoyos:2016cob}%
  \BibitemOpen
  \bibfield  {author} {\bibinfo {author} {\bibfnamefont {C.}~\bibnamefont
  {Hoyos}}, \bibinfo {author} {\bibfnamefont {N.}~\bibnamefont {Jokela}},
  \bibinfo {author} {\bibfnamefont {D.}~\bibnamefont
  {Rodr\'\i{}guez~Fern\'andez}},\ and\ \bibinfo {author} {\bibfnamefont
  {A.}~\bibnamefont {Vuorinen}},\ }\bibfield  {title} {\bibinfo {title}
  {{Breaking the sound barrier in AdS/CFT}},\ }\href
  {https://doi.org/10.1103/PhysRevD.94.106008} {\bibfield  {journal} {\bibinfo
  {journal} {Phys. Rev. D}\ }\textbf {\bibinfo {volume} {94}},\ \bibinfo
  {pages} {106008} (\bibinfo {year} {2016})},\ \Eprint
  {https://arxiv.org/abs/1609.03480} {arXiv:1609.03480 [hep-th]} \BibitemShut
  {NoStop}%
\bibitem [{\citenamefont {Ecker}\ \emph {et~al.}(2017)\citenamefont {Ecker},
  \citenamefont {Hoyos}, \citenamefont {Jokela}, \citenamefont
  {Rodr\'\i{}guez~Fern\'andez},\ and\ \citenamefont
  {Vuorinen}}]{Ecker:2017fyh}%
  \BibitemOpen
  \bibfield  {author} {\bibinfo {author} {\bibfnamefont {C.}~\bibnamefont
  {Ecker}}, \bibinfo {author} {\bibfnamefont {C.}~\bibnamefont {Hoyos}},
  \bibinfo {author} {\bibfnamefont {N.}~\bibnamefont {Jokela}}, \bibinfo
  {author} {\bibfnamefont {D.}~\bibnamefont {Rodr\'\i{}guez~Fern\'andez}},\
  and\ \bibinfo {author} {\bibfnamefont {A.}~\bibnamefont {Vuorinen}},\
  }\bibfield  {title} {\bibinfo {title} {{Stiff phases in strongly coupled
  gauge theories with holographic duals}},\ }\href
  {https://doi.org/10.1007/JHEP11(2017)031} {\bibfield  {journal} {\bibinfo
  {journal} {JHEP}\ }\textbf {\bibinfo {volume} {11}},\ \bibinfo {pages}
  {031}},\ \Eprint {https://arxiv.org/abs/1707.00521} {arXiv:1707.00521
  [hep-th]} \BibitemShut {NoStop}%
\bibitem [{\citenamefont {Hoyos}\ \emph {et~al.}(2022)\citenamefont {Hoyos},
  \citenamefont {Jokela},\ and\ \citenamefont {Vuorinen}}]{Hoyos:2021uff}%
  \BibitemOpen
  \bibfield  {author} {\bibinfo {author} {\bibfnamefont {C.}~\bibnamefont
  {Hoyos}}, \bibinfo {author} {\bibfnamefont {N.}~\bibnamefont {Jokela}},\ and\
  \bibinfo {author} {\bibfnamefont {A.}~\bibnamefont {Vuorinen}},\ }\bibfield
  {title} {\bibinfo {title} {{Holographic approach to compact stars and their
  binary mergers}},\ }\href {https://doi.org/10.1016/j.ppnp.2022.103972}
  {\bibfield  {journal} {\bibinfo  {journal} {Prog. Part. Nucl. Phys.}\
  }\textbf {\bibinfo {volume} {126}},\ \bibinfo {pages} {103972} (\bibinfo
  {year} {2022})},\ \Eprint {https://arxiv.org/abs/2112.08422}
  {arXiv:2112.08422 [hep-th]} \BibitemShut {NoStop}%
\bibitem [{\citenamefont {Jokela}\ \emph {et~al.}(2021)\citenamefont {Jokela},
  \citenamefont {J\"arvinen}, \citenamefont {Nijs},\ and\ \citenamefont
  {Remes}}]{Jokela:2020piw}%
  \BibitemOpen
  \bibfield  {author} {\bibinfo {author} {\bibfnamefont {N.}~\bibnamefont
  {Jokela}}, \bibinfo {author} {\bibfnamefont {M.}~\bibnamefont {J\"arvinen}},
  \bibinfo {author} {\bibfnamefont {G.}~\bibnamefont {Nijs}},\ and\ \bibinfo
  {author} {\bibfnamefont {J.}~\bibnamefont {Remes}},\ }\bibfield  {title}
  {\bibinfo {title} {{Unified weak and strong coupling framework for nuclear
  matter and neutron stars}},\ }\href
  {https://doi.org/10.1103/PhysRevD.103.086004} {\bibfield  {journal} {\bibinfo
   {journal} {Phys. Rev. D}\ }\textbf {\bibinfo {volume} {103}},\ \bibinfo
  {pages} {086004} (\bibinfo {year} {2021})},\ \Eprint
  {https://arxiv.org/abs/2006.01141} {arXiv:2006.01141 [hep-ph]} \BibitemShut
  {NoStop}%
\bibitem [{\citenamefont {Gursoy}\ \emph {et~al.}(2018)\citenamefont {Gursoy},
  \citenamefont {Jarvinen},\ and\ \citenamefont {Nijs}}]{Gursoy:2017wzz}%
  \BibitemOpen
  \bibfield  {author} {\bibinfo {author} {\bibfnamefont {U.}~\bibnamefont
  {Gursoy}}, \bibinfo {author} {\bibfnamefont {M.}~\bibnamefont {Jarvinen}},\
  and\ \bibinfo {author} {\bibfnamefont {G.}~\bibnamefont {Nijs}},\ }\bibfield
  {title} {\bibinfo {title} {{Holographic QCD in the Veneziano Limit at a
  Finite Magnetic Field and Chemical Potential}},\ }\href
  {https://doi.org/10.1103/PhysRevLett.120.242002} {\bibfield  {journal}
  {\bibinfo  {journal} {Phys. Rev. Lett.}\ }\textbf {\bibinfo {volume} {120}},\
  \bibinfo {pages} {242002} (\bibinfo {year} {2018})},\ \Eprint
  {https://arxiv.org/abs/1707.00872} {arXiv:1707.00872 [hep-th]} \BibitemShut
  {NoStop}%
\bibitem [{\citenamefont {Anabalon}\ \emph {et~al.}(2018)\citenamefont
  {Anabalon}, \citenamefont {Andrade}, \citenamefont {Astefanesei},\ and\
  \citenamefont {Mann}}]{Anabalon:2017eri}%
  \BibitemOpen
  \bibfield  {author} {\bibinfo {author} {\bibfnamefont {A.}~\bibnamefont
  {Anabalon}}, \bibinfo {author} {\bibfnamefont {T.}~\bibnamefont {Andrade}},
  \bibinfo {author} {\bibfnamefont {D.}~\bibnamefont {Astefanesei}},\ and\
  \bibinfo {author} {\bibfnamefont {R.}~\bibnamefont {Mann}},\ }\bibfield
  {title} {\bibinfo {title} {{Universal Formula for the Holographic Speed of
  Sound}},\ }\href {https://doi.org/10.1016/j.physletb.2018.04.028} {\bibfield
  {journal} {\bibinfo  {journal} {Phys. Lett. B}\ }\textbf {\bibinfo {volume}
  {781}},\ \bibinfo {pages} {547} (\bibinfo {year} {2018})},\ \Eprint
  {https://arxiv.org/abs/1702.00017} {arXiv:1702.00017 [hep-th]} \BibitemShut
  {NoStop}%
\bibitem [{\citenamefont {Zhou}(2024{\natexlab{b}})}]{Zhou:mi1}%
  \BibitemOpen
  \bibfield  {author} {\bibinfo {author} {\bibfnamefont {D.}~\bibnamefont
  {Zhou}},\ }\bibfield  {title} {\bibinfo {title} {in prep},\ }\href@noop {} {\
   (\bibinfo {year} {2024}{\natexlab{b}})}\BibitemShut {NoStop}%
\bibitem [{\citenamefont {Zhou}\ and\ \citenamefont {Reddy}(2024)}]{Zhou:mi2}%
  \BibitemOpen
  \bibfield  {author} {\bibinfo {author} {\bibfnamefont {D.}~\bibnamefont
  {Zhou}}\ and\ \bibinfo {author} {\bibfnamefont {S.}~\bibnamefont {Reddy}},\
  }\bibfield  {title} {\bibinfo {title} {in prep},\ }\href@noop {} {\
  (\bibinfo {year} {2024})}\BibitemShut {NoStop}%
\bibitem [{\citenamefont {McKeen}\ \emph {et~al.}(2018)\citenamefont {McKeen},
  \citenamefont {Nelson}, \citenamefont {Reddy},\ and\ \citenamefont
  {Zhou}}]{mckeen:2018xwc}%
  \BibitemOpen
  \bibfield  {author} {\bibinfo {author} {\bibfnamefont {D.}~\bibnamefont
  {McKeen}}, \bibinfo {author} {\bibfnamefont {A.~E.}\ \bibnamefont {Nelson}},
  \bibinfo {author} {\bibfnamefont {S.}~\bibnamefont {Reddy}},\ and\ \bibinfo
  {author} {\bibfnamefont {D.}~\bibnamefont {Zhou}},\ }\bibfield  {title}
  {\bibinfo {title} {{Neutron stars exclude light dark baryons}},\ }\href
  {https://doi.org/10.1103/PhysRevLett.121.061802} {\bibfield  {journal}
  {\bibinfo  {journal} {Phys. Rev. Lett.}\ }\textbf {\bibinfo {volume} {121}},\
  \bibinfo {pages} {061802} (\bibinfo {year} {2018})},\ \Eprint
  {https://arxiv.org/abs/1802.08244} {arXiv:1802.08244 [hep-ph]} \BibitemShut
  {NoStop}%
\bibitem [{\citenamefont {Landry}\ and\ \citenamefont
  {Essick}(2019)}]{Landry:2018prl}%
  \BibitemOpen
  \bibfield  {author} {\bibinfo {author} {\bibfnamefont {P.}~\bibnamefont
  {Landry}}\ and\ \bibinfo {author} {\bibfnamefont {R.}~\bibnamefont
  {Essick}},\ }\bibfield  {title} {\bibinfo {title} {{Nonparametric inference
  of the neutron star equation of state from gravitational wave
  observations}},\ }\href {https://doi.org/10.1103/PhysRevD.99.084049}
  {\bibfield  {journal} {\bibinfo  {journal} {Phys. Rev. D}\ }\textbf {\bibinfo
  {volume} {99}},\ \bibinfo {pages} {084049} (\bibinfo {year} {2019})},\
  \Eprint {https://arxiv.org/abs/1811.12529} {arXiv:1811.12529 [gr-qc]}
  \BibitemShut {NoStop}%
\bibitem [{\citenamefont {Annala}\ \emph {et~al.}(2020)\citenamefont {Annala},
  \citenamefont {Gorda}, \citenamefont {Kurkela}, \citenamefont {N\"attil\"a},\
  and\ \citenamefont {Vuorinen}}]{Annala:2019puf}%
  \BibitemOpen
  \bibfield  {author} {\bibinfo {author} {\bibfnamefont {E.}~\bibnamefont
  {Annala}}, \bibinfo {author} {\bibfnamefont {T.}~\bibnamefont {Gorda}},
  \bibinfo {author} {\bibfnamefont {A.}~\bibnamefont {Kurkela}}, \bibinfo
  {author} {\bibfnamefont {J.}~\bibnamefont {N\"attil\"a}},\ and\ \bibinfo
  {author} {\bibfnamefont {A.}~\bibnamefont {Vuorinen}},\ }\bibfield  {title}
  {\bibinfo {title} {{Evidence for quark-matter cores in massive neutron
  stars}},\ }\href {https://doi.org/10.1038/s41567-020-0914-9} {\bibfield
  {journal} {\bibinfo  {journal} {Nature Phys.}\ }\textbf {\bibinfo {volume}
  {16}},\ \bibinfo {pages} {907} (\bibinfo {year} {2020})},\ \Eprint
  {https://arxiv.org/abs/1903.09121} {arXiv:1903.09121 [astro-ph.HE]}
  \BibitemShut {NoStop}%
\bibitem [{\citenamefont {Hairer}\ \emph {et~al.}(1993)\citenamefont {Hairer},
  \citenamefont {N\o{}rsett},\ and\ \citenamefont {Wanner}}]{10.5555/153158}%
  \BibitemOpen
  \bibfield  {author} {\bibinfo {author} {\bibfnamefont {E.}~\bibnamefont
  {Hairer}}, \bibinfo {author} {\bibfnamefont {S.~P.}\ \bibnamefont
  {N\o{}rsett}},\ and\ \bibinfo {author} {\bibfnamefont {G.}~\bibnamefont
  {Wanner}},\ }\href@noop {} {\emph {\bibinfo {title} {Solving ordinary
  differential equations I (2nd revised. ed.): nonstiff problems}}}\ (\bibinfo
  {publisher} {Springer-Verlag},\ \bibinfo {address} {Berlin, Heidelberg},\
  \bibinfo {year} {1993})\BibitemShut {NoStop}%
\bibitem [{\citenamefont {Tews}\ \emph
  {et~al.}(2018{\natexlab{c}})\citenamefont {Tews}, \citenamefont {Margueron},\
  and\ \citenamefont {Reddy}}]{Tews:2018iwm}%
  \BibitemOpen
  \bibfield  {author} {\bibinfo {author} {\bibfnamefont {I.}~\bibnamefont
  {Tews}}, \bibinfo {author} {\bibfnamefont {J.}~\bibnamefont {Margueron}},\
  and\ \bibinfo {author} {\bibfnamefont {S.}~\bibnamefont {Reddy}},\ }\bibfield
   {title} {\bibinfo {title} {How well does {GW170817} constrain the equation
  of state of dense matter?},\ }\bibfield  {journal} {\bibinfo  {journal}
  {Phys. Rev. C}\ }\textbf {\bibinfo {volume} {98}},\ \href
  {https://doi.org/10.1103/PhysRevC.98.04580} {10.1103/PhysRevC.98.04580}
  (\bibinfo {year} {2018}{\natexlab{c}}),\ \Eprint
  {https://arxiv.org/abs/1804.02783} {arXiv:1804.02783 [nucl-th]} \BibitemShut
  {NoStop}%
\bibitem [{\citenamefont {Hippert}\ \emph {et~al.}(2024)\citenamefont
  {Hippert}, \citenamefont {Noronha},\ and\ \citenamefont
  {Romatschke}}]{Hippert:2024hum}%
  \BibitemOpen
  \bibfield  {author} {\bibinfo {author} {\bibfnamefont {M.}~\bibnamefont
  {Hippert}}, \bibinfo {author} {\bibfnamefont {J.}~\bibnamefont {Noronha}},\
  and\ \bibinfo {author} {\bibfnamefont {P.}~\bibnamefont {Romatschke}},\
  }\bibfield  {title} {\bibinfo {title} {{Upper Bound on the Speed of Sound in
  Nuclear Matter from Transport}},\ }\href@noop {} {\  (\bibinfo {year}
  {2024})},\ \Eprint {https://arxiv.org/abs/2402.14085} {arXiv:2402.14085
  [nucl-th]} \BibitemShut {NoStop}%
\bibitem [{\citenamefont {Melendez}\ \emph {et~al.}(2019)\citenamefont
  {Melendez}, \citenamefont {Furnstahl}, \citenamefont {Phillips},
  \citenamefont {Pratola},\ and\ \citenamefont
  {Wesolowski}}]{Melendez:2019izc}%
  \BibitemOpen
  \bibfield  {author} {\bibinfo {author} {\bibfnamefont {J.~A.}\ \bibnamefont
  {Melendez}}, \bibinfo {author} {\bibfnamefont {R.~J.}\ \bibnamefont
  {Furnstahl}}, \bibinfo {author} {\bibfnamefont {D.~R.}\ \bibnamefont
  {Phillips}}, \bibinfo {author} {\bibfnamefont {M.~T.}\ \bibnamefont
  {Pratola}},\ and\ \bibinfo {author} {\bibfnamefont {S.}~\bibnamefont
  {Wesolowski}},\ }\bibfield  {title} {\bibinfo {title} {{Quantifying
  Correlated Truncation Errors in Effective Field Theory}},\ }\href
  {https://doi.org/10.1103/PhysRevC.100.044001} {\bibfield  {journal} {\bibinfo
   {journal} {Phys. Rev. C}\ }\textbf {\bibinfo {volume} {100}},\ \bibinfo
  {pages} {044001} (\bibinfo {year} {2019})},\ \Eprint
  {https://arxiv.org/abs/1904.10581} {arXiv:1904.10581 [nucl-th]} \BibitemShut
  {NoStop}%
\bibitem [{\citenamefont {Baade}\ and\ \citenamefont
  {Zwicky}(1934)}]{Baade_1934b}%
  \BibitemOpen
  \bibfield  {author} {\bibinfo {author} {\bibfnamefont {W.}~\bibnamefont
  {Baade}}\ and\ \bibinfo {author} {\bibfnamefont {F.}~\bibnamefont {Zwicky}},\
  }\bibfield  {title} {\bibinfo {title} {On super-novae},\ }\href
  {https://doi.org/10.1073/pnas.20.5.254} {\bibfield  {journal} {\bibinfo
  {journal} {Proceedings of the National Academy of Sciences}\ }\textbf
  {\bibinfo {volume} {20}},\ \bibinfo {pages} {254} (\bibinfo {year}
  {1934})}\BibitemShut {NoStop}%
\bibitem [{\citenamefont {Lattimer}\ and\ \citenamefont
  {Prakash}(2004)}]{Lattimer:2004pg}%
  \BibitemOpen
  \bibfield  {author} {\bibinfo {author} {\bibfnamefont {J.~M.}\ \bibnamefont
  {Lattimer}}\ and\ \bibinfo {author} {\bibfnamefont {M.}~\bibnamefont
  {Prakash}},\ }\bibfield  {title} {\bibinfo {title} {{The physics of neutron
  stars}},\ }\href {https://doi.org/10.1126/science.1090720} {\bibfield
  {journal} {\bibinfo  {journal} {Science}\ }\textbf {\bibinfo {volume}
  {304}},\ \bibinfo {pages} {536} (\bibinfo {year} {2004})},\ \Eprint
  {https://arxiv.org/abs/astro-ph/0405262} {arXiv:astro-ph/0405262 [astro-ph]}
  \BibitemShut {NoStop}%
\bibitem [{\citenamefont {Burrows}(2013)}]{Burrows:2012ew}%
  \BibitemOpen
  \bibfield  {author} {\bibinfo {author} {\bibfnamefont {A.}~\bibnamefont
  {Burrows}},\ }\bibfield  {title} {\bibinfo {title} {{Colloquium: Perspectives
  on core-collapse supernova theory}},\ }\href
  {https://doi.org/10.1103/RevModPhys.85.245} {\bibfield  {journal} {\bibinfo
  {journal} {Rev. Mod. Phys.}\ }\textbf {\bibinfo {volume} {85}},\ \bibinfo
  {pages} {245} (\bibinfo {year} {2013})},\ \Eprint
  {https://arxiv.org/abs/1210.4921} {arXiv:1210.4921 [astro-ph.SR]}
  \BibitemShut {NoStop}%
\bibitem [{\citenamefont {Cherman}\ \emph {et~al.}(2019)\citenamefont
  {Cherman}, \citenamefont {Sen},\ and\ \citenamefont
  {Yaffe}}]{Cherman:2018jir}%
  \BibitemOpen
  \bibfield  {author} {\bibinfo {author} {\bibfnamefont {A.}~\bibnamefont
  {Cherman}}, \bibinfo {author} {\bibfnamefont {S.}~\bibnamefont {Sen}},\ and\
  \bibinfo {author} {\bibfnamefont {L.~G.}\ \bibnamefont {Yaffe}},\ }\bibfield
  {title} {\bibinfo {title} {{Anyonic particle-vortex statistics and the nature
  of dense quark matter}},\ }\href
  {https://doi.org/10.1103/PhysRevD.100.034015} {\bibfield  {journal} {\bibinfo
   {journal} {Phys. Rev. D}\ }\textbf {\bibinfo {volume} {100}},\ \bibinfo
  {pages} {034015} (\bibinfo {year} {2019})},\ \Eprint
  {https://arxiv.org/abs/1808.04827} {arXiv:1808.04827 [hep-th]} \BibitemShut
  {NoStop}%
\bibitem [{\citenamefont {Hirono}\ and\ \citenamefont
  {Tanizaki}(2019)}]{Hirono:2018fjr}%
  \BibitemOpen
  \bibfield  {author} {\bibinfo {author} {\bibfnamefont {Y.}~\bibnamefont
  {Hirono}}\ and\ \bibinfo {author} {\bibfnamefont {Y.}~\bibnamefont
  {Tanizaki}},\ }\bibfield  {title} {\bibinfo {title} {{Quark-Hadron Continuity
  beyond the Ginzburg-Landau Paradigm}},\ }\href
  {https://doi.org/10.1103/PhysRevLett.122.212001} {\bibfield  {journal}
  {\bibinfo  {journal} {Phys. Rev. Lett.}\ }\textbf {\bibinfo {volume} {122}},\
  \bibinfo {pages} {212001} (\bibinfo {year} {2019})},\ \Eprint
  {https://arxiv.org/abs/1811.10608} {arXiv:1811.10608 [hep-th]} \BibitemShut
  {NoStop}%
\bibitem [{\citenamefont {Cherman}\ \emph {et~al.}(2020)\citenamefont
  {Cherman}, \citenamefont {Jacobson}, \citenamefont {Sen},\ and\ \citenamefont
  {Yaffe}}]{Cherman:2020hbe}%
  \BibitemOpen
  \bibfield  {author} {\bibinfo {author} {\bibfnamefont {A.}~\bibnamefont
  {Cherman}}, \bibinfo {author} {\bibfnamefont {T.}~\bibnamefont {Jacobson}},
  \bibinfo {author} {\bibfnamefont {S.}~\bibnamefont {Sen}},\ and\ \bibinfo
  {author} {\bibfnamefont {L.~G.}\ \bibnamefont {Yaffe}},\ }\bibfield  {title}
  {\bibinfo {title} {{Higgs-confinement phase transitions with fundamental
  representation matter}},\ }\href
  {https://doi.org/10.1103/PhysRevD.102.105021} {\bibfield  {journal} {\bibinfo
   {journal} {Phys. Rev. D}\ }\textbf {\bibinfo {volume} {102}},\ \bibinfo
  {pages} {105021} (\bibinfo {year} {2020})},\ \Eprint
  {https://arxiv.org/abs/2007.08539} {arXiv:2007.08539 [hep-th]} \BibitemShut
  {NoStop}%
\bibitem [{\citenamefont {Dumitrescu}\ and\ \citenamefont
  {Hsin}(2023)}]{Dumitrescu:2023hbe}%
  \BibitemOpen
  \bibfield  {author} {\bibinfo {author} {\bibfnamefont {T.~T.}\ \bibnamefont
  {Dumitrescu}}\ and\ \bibinfo {author} {\bibfnamefont {P.-S.}\ \bibnamefont
  {Hsin}},\ }\bibfield  {title} {\bibinfo {title} {{Higgs-Confinement
  Transitions in QCD from Symmetry Protected Topological Phases}},\ }\href@noop
  {} {\  (\bibinfo {year} {2023})},\ \Eprint {https://arxiv.org/abs/2312.16898}
  {arXiv:2312.16898 [hep-th]} \BibitemShut {NoStop}%
\bibitem [{\citenamefont {Cirigliano}\ \emph {et~al.}(2024)\citenamefont
  {Cirigliano}, \citenamefont {Dawid}, \citenamefont {Dekens},\ and\
  \citenamefont {Reddy}}]{Cirigliano:2024ocg}%
  \BibitemOpen
  \bibfield  {author} {\bibinfo {author} {\bibfnamefont {V.}~\bibnamefont
  {Cirigliano}}, \bibinfo {author} {\bibfnamefont {M.}~\bibnamefont {Dawid}},
  \bibinfo {author} {\bibfnamefont {W.}~\bibnamefont {Dekens}},\ and\ \bibinfo
  {author} {\bibfnamefont {S.}~\bibnamefont {Reddy}},\ }\bibfield  {title}
  {\bibinfo {title} {{A New Class of Three Nucleon Forces and their
  Implications}},\ }\href@noop {} {\  (\bibinfo {year} {2024})},\ \Eprint
  {https://arxiv.org/abs/2411.00097} {arXiv:2411.00097 [nucl-th]} \BibitemShut
  {NoStop}%
\end{thebibliography}
